\documentclass[iop,numberedappendix]{emulateapj}
\usepackage{graphicx}
\usepackage[usenames,dvipsnames]{color}

\def\pem/{$p_{\mathrm{em}}$} % polarization in emission
\def\pex/{$p_{\mathrm{ex}}$} % polarization in extinction
\def\p500/{$p_{500}$} % polarization at 500 microns
\def\thetai/{$\theta_{I}$} % polarization angle at I band
\def\thetab/{$\theta_{500}$} % polarization angle at I band
\def\av/{$A_{V}$} % visual extinction
\def\avst/{$A_{V}^\mathrm{st}$} % visual extinction
\def\eavst/{$\sigma A_{V}^\mathrm{st}$} % visual extinction
\def\avcl/{$A_{V}^\mathrm{cl}$} % visual extinction
\def\avall/{$A_{V}^\mathrm{tot}$} % visual extinction
\def\tauv/{$\tau_{V}$} % tau V
\def\pextauv/{$p_{\mathrm{ex}}/\tau_{V}$} 
\def\pempextauv/{$p_{\mathrm{em}}/(p_{\mathrm{ex}}/\tau_{V})$}
\def\plambdapitauv/{$p_{\mathrm{500}}/(p_{I}/\tau_{V})$} 
\def\per/{$R_{\mathrm{eff}}$} 
\def\p500pi/{$P_{\mathrm{500}}/p_{I}$} 

\newcommand{\specialcell}[2][t]{%
  \begin{tabular}[#1]{@{}c@{}}#2\end{tabular}}
\providecommand{\sorthelp}[1]{}

\shorttitle{Comparison of Sub-mm and Near-IR Polarization for Vela C}
\shortauthors{Santos et al.}

\begin{document}

\title{Comparing submillimeter polarized emission with near-infrared polarization \\ 
of background stars for the Vela C molecular cloud}

\author{Fabio P. Santos\altaffilmark{1},
Peter A. R. Ade\altaffilmark{2},
Francesco E. Angil\`e\altaffilmark{3},
Peter Ashton\altaffilmark{1},
Steven J. Benton\altaffilmark{4,5},
Mark J. Devlin\altaffilmark{3},
Bradley Dober\altaffilmark{3}, 
Laura M. Fissel\altaffilmark{1},
Yasuo Fukui\altaffilmark{6},
Nicholas Galitzki\altaffilmark{3}, 
Natalie N. Gandilo\altaffilmark{7,8},
Jeffrey Klein\altaffilmark{3},
Andrei L. Korotkov\altaffilmark{9},
Zhi-Yun Li\altaffilmark{10},
Peter G. Martin\altaffilmark{11},
Tristan G. Matthews\altaffilmark{1},
Lorenzo Moncelsi\altaffilmark{12},
Fumitaka Nakamura\altaffilmark{13},
Calvin B. Netterfield\altaffilmark{4,7},
Giles Novak\altaffilmark{1},
Enzo Pascale\altaffilmark{2},
Fr{\'e}d{\'e}rick Poidevin\altaffilmark{14,15},
Giorgio Savini\altaffilmark{16},
Douglas Scott\altaffilmark{17},
Jamil A. Shariff\altaffilmark{7,18},
Juan Diego Soler\altaffilmark{19},
Nicholas E. Thomas\altaffilmark{20},
Carole E. Tucker\altaffilmark{2},
Gregory S. Tucker\altaffilmark{9},
Derek Ward-Thompson\altaffilmark{21}} 

\altaffiltext{1}{Center for Interdisciplinary Exploration and Research in Astrophysics (CIERA) and Department\ of Physics \& Astronomy, Northwestern University, 2145 Sheridan Road, Evanston, IL 60208, U.S.A.}
\altaffiltext{2}{Cardiff University, School of Physics \& Astronomy, Queens Buildings, The Parade, Cardiff, CF24 3AA, U.K.} 
\altaffiltext{3}{Department of Physics \& Astronomy, University of Pennsylvania, 209 South 33rd Street, Philadelphia, PA, 19104, U.S.A.} 
\altaffiltext{4}{Department of Physics, University of Toronto, 60 St. George Street Toronto, ON M5S 1A7, Canada}
\altaffiltext{5}{Department of Physics, Princeton University, Jadwin Hall, Princeton, NJ 08544, U.S.A.}
\altaffiltext{6}{Department of Physics and Astrophysics, Nagoya University, Nagoya 464-8602, Japan}
\altaffiltext{7}{Department of Astronomy \& Astrophysics, University of Toronto, 50 St. George Street Toronto, ON M5S 3H4, Canada}
\altaffiltext{8}{Department of Physics and Astronomy, Johns Hopkins University, 3701 San Martin Drive, Baltimore, Maryland, U.S.A.}
\altaffiltext{9}{Department of Physics, Brown University, 182 Hope Street, Providence, RI, 02912, U.S.A.}
\altaffiltext{10}{Department of Astronomy, University of Virginia, 530 McCormick Rd, Charlottesville, VA 22904, U.S.A.}
\altaffiltext{11}{CITA, University of Toronto, 60 St. George St., Toronto, ON M5S 3H8, Canada}
\altaffiltext{12}{California Institute of Technology, 1200 E. California Blvd., Pasadena, CA, 91125, U.S.A.}
\altaffiltext{13}{National Astronomical Observatory, Mitaka, Tokyo 181-8588, Japan}
\altaffiltext{14}{Instituto de Astrofísica de Canarias, E-38200 La Laguna, Tenerife, Spain}
\altaffiltext{15}{Universidad de La Laguna, Dept. Astrof\'{i}sica, E-38206 La Laguna, Tenerife, Spain}
\altaffiltext{16}{Department of Physics \& Astronomy, University College London, Gower Street, London, WC1E 6BT, U.K.}
\altaffiltext{17}{Department of Physics \& Astronomy, University of British Columbia, 6224 Agricultural Road, Vancouver, BC V6T 1Z1, Canada}
\altaffiltext{18}{Department of Physics, Case Western Reserve University, 2076 Adelbert Road, Cleveland Ohio, 44106-7079, U.S.A.}
\altaffiltext{19}{Institute d'Astrophysique Spatiale, CNRS (UMR8617) Universit\'{e} Paris-Sud 11, B\^{a}timent 121, Orsay, France}
\altaffiltext{20}{NASA/Goddard Space Flight Center, Greenbelt , MD 20771, U.S.A.}
\altaffiltext{21}{Jeremiah Horrocks Institute, University of Central Lancashire, PR1 2HE, U.K.}

\begin{abstract} %250 words
We present a large-scale combination of near-infrared (near-IR) 
interstellar polarization data from background starlight with polarized emission data 
at submillimeter (sub-mm) wavelengths for the Vela C molecular cloud.
The near-IR data consist of more than 6700 detections probing a range of visual extinctions 
between $2$ and $20\,$mag in and around the cloud.
The sub-mm data was collected in Antartica by the Balloon-borne Large Aperture Submillimeter Telescope for Polarimetry 
(BLASTPol). 
This is the first direct combination of near-IR and sub-mm polarization data
for a molecular cloud aimed at measuring the ``polarization efficiency ratio" (\per/), a quantity that is expected 
to depend only on grain intrinsic physical properties. It is defined as
$p_{500}/(p_{I}/\tau_{V})$, where $p_{500}$ and $p_{I}$ are polarization fractions at $500\,\mu$m and
$I$-band, respectively, and $\tau_{V}$ is the optical depth.
To ensure that the same column density of material is producing 
both polarization from emission and from extinction, we conducted a careful selection
of near-background stars using 2MASS, {\it Herschel} and {\it Planck} data. 
This selection excludes objects contaminated by the Galactic diffuse background material
as well as objects located in the foreground.
Accounting for statistical and systematic uncertainties, we estimate an average \per/ 
value of $2.4\pm0.8$, which can be used to test the predictions of dust grain models designed 
for molecular clouds when such predictions become available. \per/ appears to be relatively flat as a 
function of the cloud depth for the range of visual extinctions probed.
%suggesting that significant grain modification might occur only at higher densities. 
\end{abstract}


\section{Introduction}
\label{introduction}

Astronomers have known about the existence of magnetic fields in the interstellar medium (ISM) for 
over 60 years, as initially revealed by observations of starlight polarization 
\citep{hall1949,hiltner1949,mathewson1970,davis1951,serkowski1975}.
From the diffuse neutral material to molecular clouds and dense cores, 
polarimetry of starlight and polarized thermal emission from dust
have historically proved to be the best tracers of the sky-projected component of 
the magnetic field. 
Despite extensive efforts to understand the role of magnetic fields in the ISM, 
many open questions remain.
For example, although molecular clouds are widely known to be sites
of star formation, the role of magnetic fields in this process is not entirely understood.
Molecular clouds exhibit intricate patterns of filaments 
and striations, but the relation of these structures to magnetic fields is still under debate 
\citep{goldsmith2008,andre2010,arzou2011}. Furthermore, 
we do not know whether magnetic fields are able to
%it is generally not known for which specific conditions magnetic fields are relevant to
support clouds against gravitational collapse thereby affecting the efficiency for forming new 
stars \citep{mou1981,mckee2007}.
%\citep{mou1981}, and therefore they represent an important variable
%in models of star formation, possibly interfering with the efficiency to form new 
%stars \citep{mckee2007}.

Although the above mentioned magnetic field mapping technique is now widely used,
the detailed mechanisms regulating polarized emission and extinction
by dust are not entirely understood. Starlight
of background objects becomes linearly polarized after passing through an interstellar
cloud in which a subset population of non-spherical grains 
have their long axis preferentially
aligned perpendicular
to the magnetic field \citep{hall1949,hiltner1949,davis1951,mathewson1970}. 
The observed polarization orientation will be parallel to the sky-projected magnetic field.
The degree of polarization of background starlight is
detectable in
the ultraviolet, peaks in the optical ($\lambda\approx0.55\,\mu$m), and falls off in the near-infrared 
(near-IR) spectral bands \citep{serkowski1975}. 
This wavelength dependence
gives clues regarding the 
size distribution of aligned particles \citep{kim1994,kim1995}. 
Aligned dust grains radiate thermally at wavelengths longer than the mid-infrared 
spectral bands (according to their typical temperatures of $\sim 10 - 100\,$K),
and this emission is polarized perpendicularly to the magnetic 
field \citep{1983hildebrand,1988hildebrand}.

The limitations in interpreting polarization data from extinction or emission 
are usually related to uncertainties regarding the alignment mechanism or the physical 
properties of the dust grains. 
The most promising grain alignment theory, known as radiative torques 
\citep[RATs, ][]{1976dolginov,draine_wein_1996,draine_wein_1997,lazarian2000,lazarian2007}, requires an
anisotropic radiation field having $\lambda \sim a$, where $a$ is the grain size. 
This is consistent with evidence that inside starless cores
there is a depth beyond which no alignment takes place \citep{whittet2008,alves2014,jones2015}.
Other observations consistent with RATs' predictions include the poor alignment 
of small-sized grains \citep{kim1995} and the angular dependence of polarization efficiency 
around sources of radiation, relative to the magnetic field direction \citep{andersson2011,vaillancourt2015}.
However, the classical alignment mechanism \citep[paramagnetic relaxation;][]{davis1951} may 
still be significant for a subset of smaller sized grains \citep{hoang2014}, 
suggesting a balance between both effects \citep[for a review see][]{andersson2015}.

The most basic observational constraint on dust properties that can be derived from interstellar polarimetry is that
a fraction of the grain population must be non-spherical, a necessary condition
to produce polarization. The grain composition is primarily silicates 
and carbonaceous material \citep[for a review see][]{draine2003}.
Spectropolarimetry of silicate spectral features show that
silicate grains are subject to alignment mechanisms \citep{smith2000}.
By contrast, the non-detection of polarization levels in spectral features associated with carbonaceous grains 
suggests that these are generally not aligned \citep{chiar2006}, but more study is needed. 
In general, it is possible to draw conclusions regarding 
grain properties by comparing observations of the polarization spectra to predictions 
based on physical grain models \citep[e.g.,][]{bethell2007,draine2009}. The predictions can be adjusted by varying 
a range of input parameters.

The fractional polarization levels detected in extinction and emission 
(\pex/ and \pem/, respectively) are strongly affected by the grain alignment conditions
(i.e., the alignment efficiency), grain intrinsic properties (shapes, sizes, and chemical 
compositions), and the inclination of magnetic fields to the line-of-sight (LOS).
For polarization by extinction, the polarization degree generally increases linearly with 
the amount of material distributed along 
the LOS \citep{jones1989}, so normalizing this quantity by the visual optical depth 
(\pextauv/) makes it a suitable probe of the polarization efficiency, analogous to \pem/. 
In view of all the variables that can affect \pem/ and \pextauv/, it is useful to 
find quantities that are invariant with respect to one or another of these physical parameters, 
allowing their
combined effect to be disentangled. One of these quantities is the ``polarization
efficiency ratio", defined as \pempextauv/. Assuming a situation in which 
the same population of dust grains distributed along the LOS produces both polarization
by emission and by extinction, \pem/ and \pextauv/ are expected to be 
equally dependent on alignment efficiencies and inclinations of magnetic fields to the 
LOS. Therefore, their ratio should depend only on properties inherent to the grains 
themselves, such as emission and extinction cross-sections, which in turn depend on 
their shapes and dielectric functions \citep{1988hildebrand,martin2007,jones2015,jones2015b}. 
Therefore the polarization efficiency ratio is
a powerful probe of dust properties over a wide range of densities and temperatures, 
and is particularly interesting to compare against grain models that relate 
\pem/ to \pextauv/ using a range of adjustable parameters \citep{draine2009}. 

The main goal of the work presented here is to determine \pempextauv/ for the 
Vela C molecular cloud, which is a portion of a larger complex of clouds known 
as the Vela Molecular Ridge \citep[VMR,][]{murphy1991,netterfield2009}. 
Vela C is located at a distance of
$700\pm200\,$pc \citep{liseau1992}. This cloud was the main observational target of BLASTPol 
(the Balloon-borne Large-Aperture Sub-millimeter Telescope for Polarimetry), which 
conducted deep submillimeter (sub-mm) observations of the polarized thermal emission 
from the cloud \citep[][]{2016fissel}. We report the results of an extensive observational 
survey of near-IR stellar polarization for a wide portion of the cloud, providing
over 6700 detections. This enabled us to carry out a large-scale combination of polarization from extinction and
emission, in which complementary data from 2MASS, {\it Herschel} and {\it Planck} were also utilized.
Section \ref{s:obsdata} describes the observations and data reduction scheme for both the sub-mm and 
near-IR samples. 
In Section \ref{s:results1} we compare the magnetic field angles inferred respectively from 
sub-mm and near-IR data. In Section \ref{s:results2} we introduce a major challenge in the analysis
which is our lack of prior knowledge concerning stellar distances. We show how the above-mentioned 
complementary data can give us a handle on this problem. Section \ref{s:results3} describes the computation
of the polarization efficiency ratio \pempextauv/, for which we adopt the symbol \per/.
A discussion of the results is given in Section \ref{s:discussion} and our main conclusions
are summarized in Section \ref{s:conclusions}. 

\section{Observational data}
\label{s:obsdata}

\subsection{Polarized thermal emission from BLASTPol}

BLASTPol is a high-altitude balloon-borne experiment that was launched on 26 December 2012 from Antartica.
It was equipped with a $1.8$m diameter primary mirror and a series of dichroic filters that allowed 
us to carry out simultaneous observations of total intensity $I$ in three spectral bands centered at $250$, $350$ and 
$500\,\mu$m. Additionally, using a polarizing grid mounted in front of the detector arrays, 
together with an achromatic half-wave plate \citep{moncelsi2014}, BLASTPol was able to measure the linear 
polarization Stokes parameters
$Q$ and $U$. A thorough description of the instrument and the observational strategy adopted, as well as the data reduction, 
beam analysis, and null tests for data quality assurance may be found in \citet{2016fissel}.

Although BLASTPol targeted various Galactic molecular clouds, Vela C 
was its highest priority science target. We carried out a ``deep" $43\,$h integration toward the densest portions of the 
cloud \citep[more specifically, covering four of the five Vela C subregions defined by][]{hill2011}. In addition,  
an extra $11\,$h of integration were spent on a wider area around the cloud ($\sim10\,$deg$^2$), 
consisting of more diffuse interstellar material. The observing mode involved a series of raster 
scans, using four different half-wave plate angles.

For the purposes of this work, we are focusing only on the $500\,\mu$m data set. 
Polarimetry at 250 and 350$\,\mu$m and its relationship with polarimetry at 
500$\,\mu$m are discussed in a separate work \citep{2016gandilo}.
As described by \citet{2016fissel},
for this particular set of observations our beam FWHM was larger than had been predicted by our
optics model, containing significant structure with non-Gaussian shape. The data
were smoothed in order to obtain an approximately round beam having a FWHM  of $2.5'$.

%-------------------------------------------------------------
   \begin{figure}
   \centering
   \includegraphics[width=0.48\textwidth]{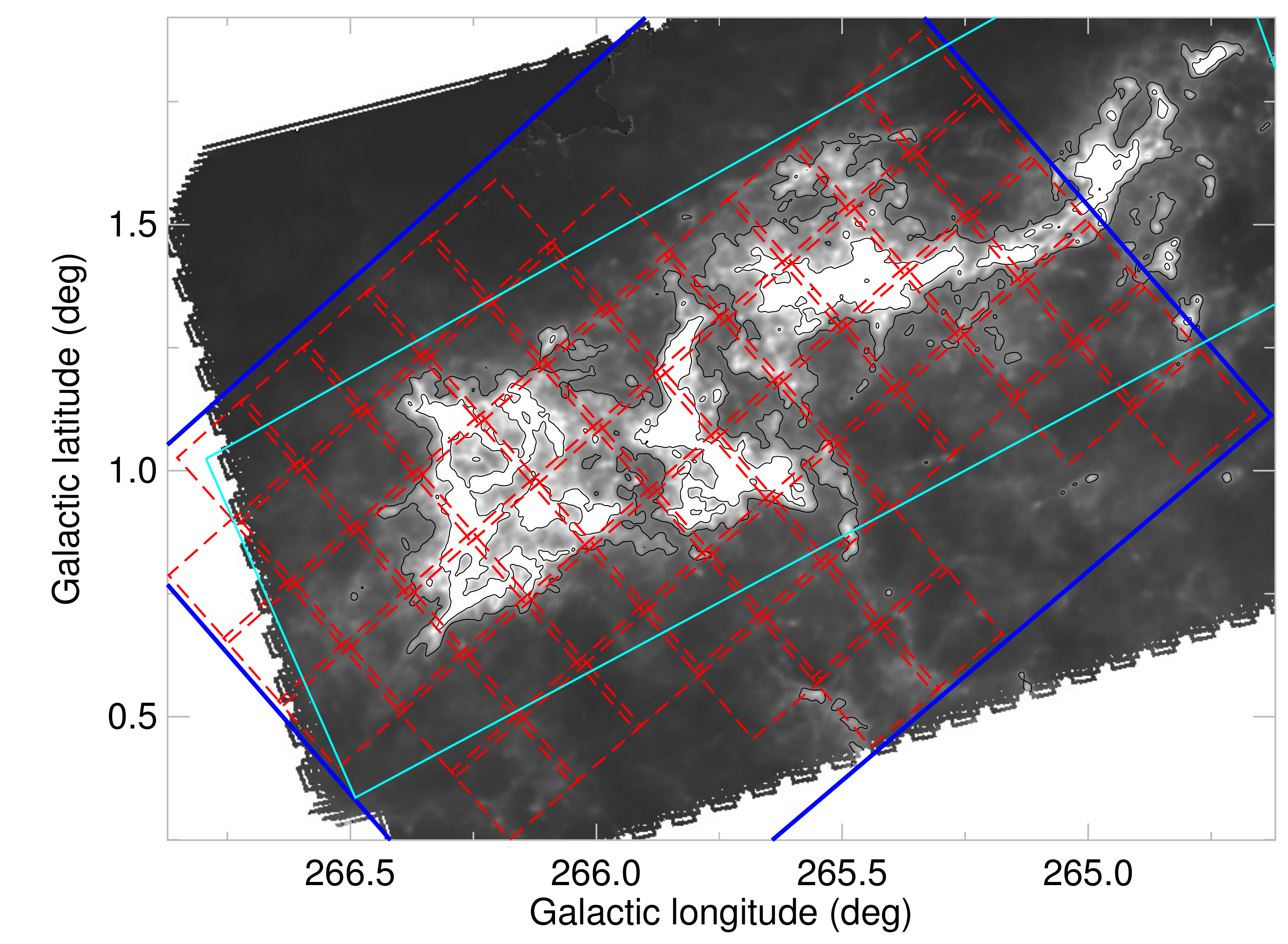} \\
   \caption{Column density image of Vela C (obtained from {\it Herschel} data, 
	    as described in Section \ref{s:avcl}),
               with contours representing the cloud's visual extinction (\avcl/)
               at levels of $10$ and $25$ mag. The red dashed-line
               mosaic shows observation fields used in the 
               $I$-band survey, and the cyan box represents the ``validity region"
               for the BLASTPol data set (see Section \ref{s:backsub_datasel}). 
               The blue box is the area used to retrieve 
               near-IR photometric data from 2MASS (Section \ref{s:avst}). Stellar objects located 
               both in this box and within the boundaries of the {\it Herschel} map define the 
	       ``wide photometric field" (see Table \ref{tab}).
              }
         \label{f:mosaic}
   \end{figure}
%-------------------------------------------------------------

\begin{table*}[!t]
\centering
\caption{\label{tab} Selection criteria for the data sets used in the analysis}
\begin{tabular}{cccccc}
\tableline\tableline
\specialcell{Data set \\ denomination} & Source & Selection ID & Selection criteria & $N$ & Figures \\ \tableline

Basic BLASTPol & BLASTPol & S1 & \specialcell{De-biasing ($p_{500}\rightarrow\sqrt{p_{500}^{2}-\sigma_{p500}^{2}}$), \\ consistency between aggressive and \\ conservative diffuse background \\ subtraction methods (see Section \ref{s:dataselection}), \\ data inside the validity region (cyan, Figure \ref{f:mosaic}) \\ $p_{500}/\sigma_{p500} > 3$ } & 3157$^{a}$ & \specialcell{\ref{f:polmap} (red \\ pseudo-vectors} \\ \tableline 

Basic $I$-band & OPD & S2 & \specialcell{De-biasing ($p_{I}\rightarrow\sqrt{p_{I}^{2}-\sigma_{pI}^{2}}$), \\ $p_{I}/\sigma_{pI} > 3$} & 6740 & \specialcell{\ref{f:polmap} (cyan \\ pseudo-vectors)} \\ \tableline

\specialcell{Basic polarization \\ combination} & \specialcell{BLASTPol,\\ OPD} & S3 & \specialcell{Selections S1 and S2, \\ areas of overlap between \\ sub-mm and $I$-band pseudo-vectors} & 1355$^{a}$ & \ref{f:angle}  \\ \tableline

Wide photometric field & \specialcell{2MASS, \\ Herschel} & S4 & \specialcell{stars inside blue box (Figure \ref{f:mosaic}) and \\ within the boundaries of the Herschel map, \\ 2MASS ``AAA" quality, \\ points inside reddening band (blue, Figure \ref{f:ccdiag})} & 20348 & \specialcell{\ref{f:ccdiag}, \ref{f:avavdiag},\\ \ref{f:avavdiag2} (black dots), \\ \ref{f:hist}} \\ \tableline

\specialcell{Planck-2MASS \\ combination} & \specialcell{Planck, \\ 2MASS, \\ Herschel} & S5 & \specialcell{Same objects from the wide photometric \\ field (selected using S4) combined to \\ Planck $\tau_{353}$ data}   & 20348 & \ref{f:compavall}  \\ \tableline

\specialcell{$I$-band-2MASS \\ combination} & \specialcell{OPD, \\ 2MASS}  & S6 & \specialcell{Selection S2, 2MASS ``AAA" quality, \\ points inside reddening band (blue, Figure \ref{f:ccdiag})}  & 5980 & \ref{f:fore} \\ \tableline

\specialcell{Corrected polarization \\ combination} & \specialcell{BLASTPol,\\ OPD, \\ 2MASS, \\ Herschel}  & S7 & \specialcell{Selections S3 and S4, \\ magnetic field orientation consistency ($\Delta\theta < 15\degr$) \\ and removal of RCW36 area (see Section \ref{s:dataselection_ang}), \\ foreground correction (see Section \ref{s:foreground}) \\ with  $p_{I}/\sigma_{pI} > 3$ re-applied, \\ \avst/$/$\eavst/$>3$} & 834$^{a}$ & \specialcell{\ref{f:avavdiag2} (red \\ crosses), \ref{f:reffback}} \\ \tableline

\specialcell{Ideal stellar \\ sample} & \specialcell{BLASTPol,\\ OPD, \\ 2MASS, \\ Herschel} & S8 & \specialcell{Selections S7, \\ points within the ideal stellar locus \\ (see Figure \ref{f:avavdiag2})} & 87$^{a,b}$ & \ref{f:reff1}, \ref{f:perhist} \\ \tableline

\end{tabular}
\tablecomments{The table shows a list of selection criteria for each data set used in this work. The columns 
respectively represent the adopted denomination of the data set, the source of the data set itself and data products
used to apply the selection, an identifier (ID) to specify the list of selections, the selection criteria
applied to each data set, the number of data points ($N$) obtained after selection, and the figures where each 
data set is used for analysis.
\\ $^{(a)}$ Valid for the intermediate diffuse emission subtraction method.
\\ $^{(b)}$ Average number considering systematic variations of the GL method (see Appendix \ref{ap:gl}).
}
\end{table*}

\subsection{Near-IR polarization from OPD}

The near-IR polarization data were acquired at the 
Pico dos Dias Observatory (OPD\footnote{The Pico dos Dias Observatory
is operated by the Brazilian National Laboratory for Astrophysics (LNA), 
a research institute of the Ministry of Science, Technology and Innovation (MCTI).}, 
Brazil), in a series of observations between November 2013 and February 2014.
Both the 1.6m and the 0.6m telescopes were used in alternating night shifts, 
together with the IAGPOL polarimeter with the $I$-band near-IR filter 
($0.79\,\mu$m, Cousins) and the optical CCD detector. 
In both telescopes, the detector covers a field-of-view of approximately 
$11'\times11'$, and therefore a careful mosaic-mapping was needed in order 
to cover a large portion of the molecular cloud. In Figure \ref{f:mosaic}, 
the red dashed boxes represent each of the $62$ areas observed in the $I$-band.

The polarimeter \citep{magalhaes1996} consists of a sequence of optical elements positioned in 
the optical path before.
The incident light first passes through an achromatic half-wave plate 
(HWP, with an optical axis orientation of $\psi$), which is made to rotate in discrete
steps of $22.5\degr$. 
%Consecutive HWP positions represent rotations
%of $45\degr$ in the orientation of the polarization plane. 
Next, a Savart analyser 
splits the beam into two orthogonally polarized components. These components 
then pass through a spectral filter and the duplicated 
stellar images are simultaneously detected by the CCD. 
Sequential rotations of the HWP cause flux variations in the orthogonally polarized components, 
so that the flux ratios can be fit to a modulation function proportional 
to $\bar{Q}\cos{4\psi}+\bar{U}\sin{4\psi}$ ($\bar{Q}=Q/I$ and $\bar{U}=U/I$ are the
flux-normalized Stokes parameters, where $I$ is the total intensity). 
Since the polarimetric quantities are 
derived from flux ratios, the observational strategy is essentially analogous to 
differential photometry, and any atmospheric variations are canceled through 
this operation.

For all $62$ observational fields, two independent sets of observations were carried
out respectively using short ($10$ to $20\,$s) and long ($60$ to $100\,$s) exposure times, at each
of the eight positions of the half-wave plate. 
In cases where a single object was observed multiple times, the measurement with the highest
S/N was selected.
%Multiple observations 
%of individual objects were later selected based on the highest polarization degree
%S/N achieved. 
At least three polarimetric standard stars 
were observed each night \citep{hsu1982,clemens1990,turnshek1990,larson1996}, 
in order to calibrate the polarization position angles and check the consistency of the 
instrumental polarization (which can be safely neglected for this instrumental set, 
considering that it is much smaller than the typical uncertainties of $\approx 0.1\%$).

The data reduction process consisted of standard image treatment (bias, flat-fielding
and bad-pixel correction), followed by aperture photometry of all duplicated images
of point-like sources. The resulting fluxes were subsequently used to build modulation 
functions for each object using a set of specifically designed IRAF\footnote{IRAF is distributed by the 
National Optical Astronomy Observatories, which are operated by the Association of 
Universities for Research in Astronomy, Inc., under cooperative agreement with the National 
Science Foundation \citep{tody1986}.} routines 
\citep[PCCDPACK, ][]{pereyra2000}. The polarization degree $p_{I}$ and orientation
\thetai/ as well as their respective uncertainties were calculated for each object based
on the corresponding normalized Stokes parameters. A detailed description of the data processing 
can be found in \citet{santos2012}.

%-------------------------------------------------------------
   \begin{figure*}
   \centering
   \includegraphics[width=\textwidth]{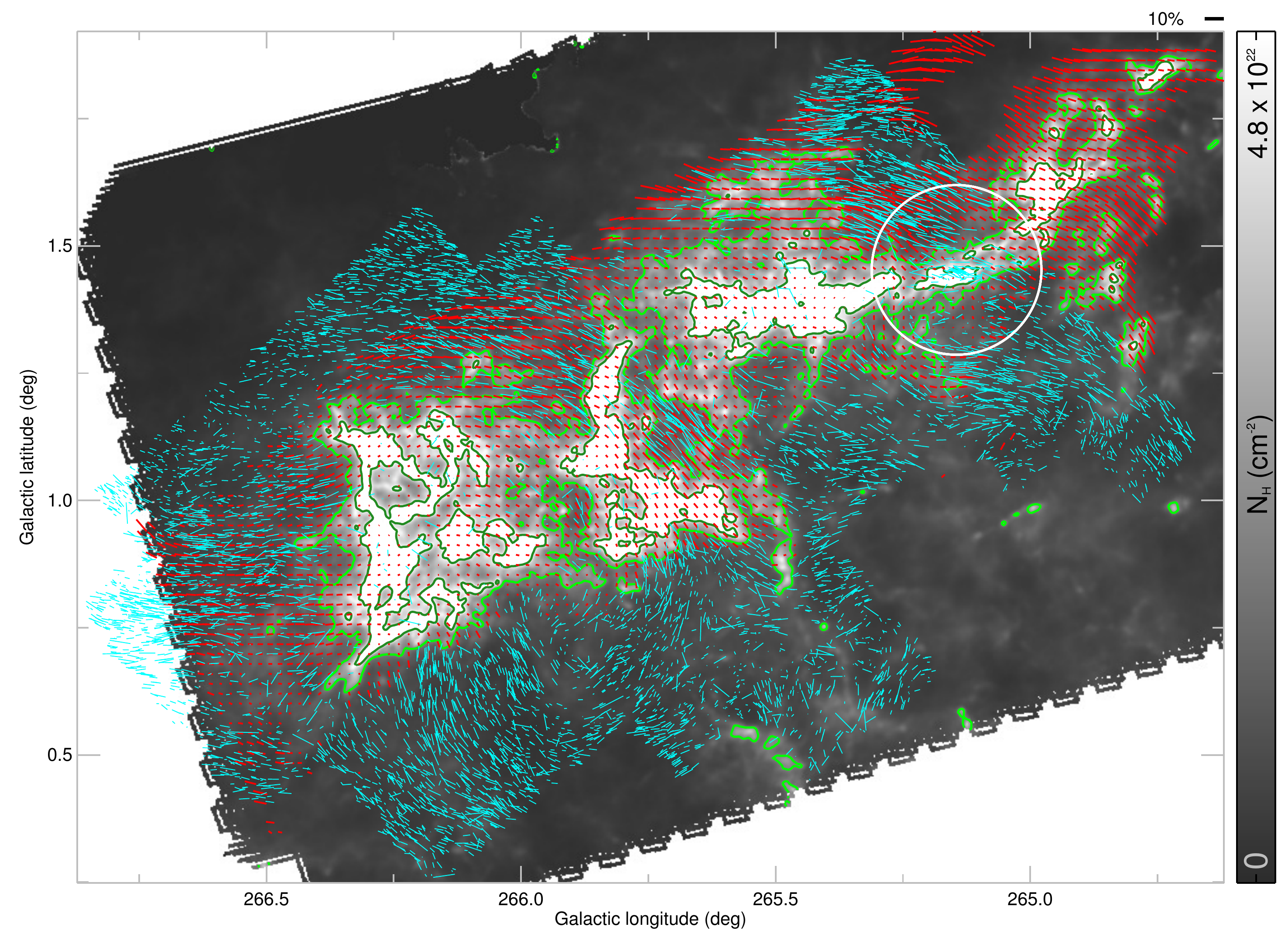} \\
      \caption{Polarization map of Vela C, including both the $I$-band (cyan)
              and the BLASTPol (red, $90\degr$-rotated) basic data sets as defined in Table \ref{tab}. 
              The background image is described in Section \ref{s:combinedpolmap} and is the same as shown in Figure \ref{f:mosaic}. 
	       It shows column density ($N_\mathrm{H}$), with outer and inner 
               contours representing visual extinction levels of $10$ (light green) and 
               $25$ mag (dark green) respectively. The sizes of pseudo-vectors are proportional 
               to the polarization degree, 
               with a reference $10\%$ pseudo-vector shown at the top right. The white circle has $10'$ radius and 
               is centered on RCW36.
              }
         \label{f:polmap}
   \end{figure*}
%-------------------------------------------------------------

\subsection{Corrections applied to BLASTPol and $I$-band polarization data}
\label{s:backsub_datasel}

Before carrying out a comparison between near-IR and sub-mm polarimetric data, 
a set of data corrections and selections must be carried out in order to make sure that only a 
high quality sub-sample is used for the comparison. 
Table 1 summarizes the corrections and selections for the various data sets that are used 
in this work.
The corrections are described 
in the present sub-section.

To define both the basic BLASTPol and the basic $I$-band data sets, 
we first apply the debiasing correction according to the prescription
$p\rightarrow\sqrt{p^{2}-\sigma_p^{2}}$ \citep{wardle1974}, which is not valid for lower S/N data
(the low S/N data will later be rejected, as discussed in Section \ref{s:dataselection}).
Secondly, it is necessary 
to remove from the BLASTPol data set the contribution from the diffuse Galactic 
emission (foreground and background), thereby isolating only the polarized dust 
emission from the Vela C molecular cloud itself. This process is described in detail by \citet{2016fissel}. 
It was carried out using two different methods. In the first method, 
which we refer to as ``conservative", we assume that most of the diffuse emission 
near Vela C is actually associated with the cloud, and therefore the goal was to avoid 
subtracting such emission. This was achieved by using a well-separated nearby low-flux 
region (also mapped by BLASTPol) as a representation
for the background/foreground dust emission. In this subtraction method, 
we are effectively assuming that the background/foreground
emission is uniform across the region.
In the second method, referred to as ``aggressive", the diffuse emission near the cloud is assumed
to be not associated with it. 
It was subtracted by defining two reference regions along the cloud's
North and South edges, and then using the $I$, $Q$, and $U$ measurements
in these regions to fit a linear emission profile, which
was then subtracted. The use of the two reference 
regions effectively 
defines a ``validity area" between them (cyan box in Figure \ref{f:mosaic}). 

Finally, following \citet{2016fissel} we proceed under the assumption that the most suitable 
diffuse emission subtraction method probably 
corresponds to an intermediate
version, lying between the aggressive and conservative methods. Accordingly, an ``intermediate" diffuse emission
subtraction method is introduced, which involves averaging the respective
I, Q, and U maps corresponding to the two extreme methods. 
In this work, unless otherwise explicitly stated, all results and analysis employ 
intermediate diffuse emission subtraction. However, our final analysis of the polarization efficiency ratio
(Section \ref{s:reff1}) takes into account the systematic uncertainties 
associated with the diffuse emission subtraction process.

\subsection{Data selections applied to BLASTPol and $I$-band polarization measurements}
\label{s:dataselection}

Similarly to the procedure adopted by \citet{2016fissel}, we select for analysis only the data
that do not present strong variations between the results obtained from the various diffuse subtraction methods.
Representing polarization degrees and position angles for the intermediate, conservative, and aggressive
diffuse emission subtraction methods respectively as $(p_\mathrm{int}$, $\phi_\mathrm{int})$,
$(p_\mathrm{con}$, $\phi_\mathrm{con})$, and $(p_\mathrm{agg}$, $\phi_\mathrm{agg})$,
we require that $p_\mathrm{int} > 3|p_\mathrm{int}-p_\mathrm{agg}|$ and $p_\mathrm{int} > 3|p_\mathrm{int}-p_\mathrm{con}|$, 
and also that $|\phi_\mathrm{int}-\phi_\mathrm{agg}| < 10\degr$ and $|\phi_\mathrm{int}-\phi_\mathrm{con}| < 10\degr$.
Finally, for both the $I$-band and BLASTPol data sets, we require that the S/N in polarization fraction 
satisfies $p/\sigma_{p} > 3$, completing the definitions of the 
basic BLASTPol and basic $I$-band data sets (Table \ref{tab}). 
After applying these selection criteria, 6740 stars remain in the basic $I$-band data set.

\section{Consistency between sub-mm and near-IR magnetic field angles}
\label{s:results1}

\subsection{Combined polarization map}
\label{s:combinedpolmap}

Figure \ref{f:polmap} shows the combined polarization map, in which cyan pseudo-vectors represent $I$-band 
observations and red pseudo-vectors are the $500\,\mu$m polarimetric measurements (rotated by $90\degr$
in order to match with the orientation of sky-projected magnetic field), using the basic data sets in both cases. 
Pseudo-vectors lengths are proportional to polarization degree. The background
image (see also Figure \ref{f:mosaic}) is a map of hydrogen column density estimated from 
{\it Herschel} dust emission data (see Section \ref{s:avcl}).
The {\it Herschel} data also provide an estimate of the cloud visual extinction levels, as shown 
by the inner (\av/$=25\,$mag, dark green) and outer (\av/$=10\,$mag, light green) contours in Figure \ref{f:polmap}.

$I$-band pseudo-vectors surround the denser portions of the cloud, tracing the more diffuse molecular material, 
with far fewer detections at \av/$>10$ mag (in this work we will generally 
limit the analysis to cloud extinction levels below $20\,$mag). An exception 
occurs in the vicinity of the RCW 36 H{\sc ii} 
region, as indicated by a white circle ($10'$ radius). Here the 
presence of bright OB-type stars in the star-forming cluster \citep{baba2004} allowed many 
$I$-band polarization detections even at higher extinction levels. Good-quality sub-mm 
detections, on the other hand, are mainly found in the denser regions of the cloud (see Section \ref{s:dataselection_ang}), 
where higher fluxes give better sensitivity. Large areas of overlap
between sub-mm and near-IR pseudo-vectors 
are seen in Figure \ref{f:polmap}. 
These areas are used to define the ``basic polarization combination" data set (see Table \ref{tab}).
We select sub-mm
polarization values corresponding to each stellar position, using a finely gridded BLASTPol map 
with $10''$ pixel size. The basic polarization combination data set is comprised of
$1355$ individual lines-of-sight containing both sub-mm and $I$-band data. However,
before correlating polarization data from extinction and emission directly,
a careful selection of the suitable stars for this comparison needs to be done, 
as discussed in the next sections.

\subsection{Agreement between magnetic field orientation and exclusion of the RCW 36 region}
\label{s:dataselection_ang}

As discussed in detail in Section \ref{s:results2}, the
main challenge to be dealt with before directly comparing polarization from emission and
extinction is to make sure both methods are probing the same interstellar
material along the LOS. 
The polarized emission data traces only the molecular cloud (see Section \ref{s:backsub_datasel}),
while polarization from stars, which are distributed in a range of distances along the line-of-sight
(see Appendix \ref{ap:avstuncertainties}), could be contaminated by the foreground/background material.
A first step is to select 
data for which respective 
polarization angles 
from the two data sets agree, thereby ensuring that the 
set of sky-projected magnetic field orientations sampled along the LOS coincides.
This procedure could exclude, for example, stars in the foreground or far away in the
background, tracing magnetic field structures not associated with the 
cloud itself.
Note, however, that if there are no strong changes in field orientation 
along the LOS, similar angles will be found
even when probing different columns of interstellar material
(e.g., see discussion in Section \ref{s:results2}). Therefore, the 
angle requirement is necessary but not sufficient.

%-------------------------------------------------------------
   \begin{figure}
   \centering
   \includegraphics[width=0.48\textwidth]{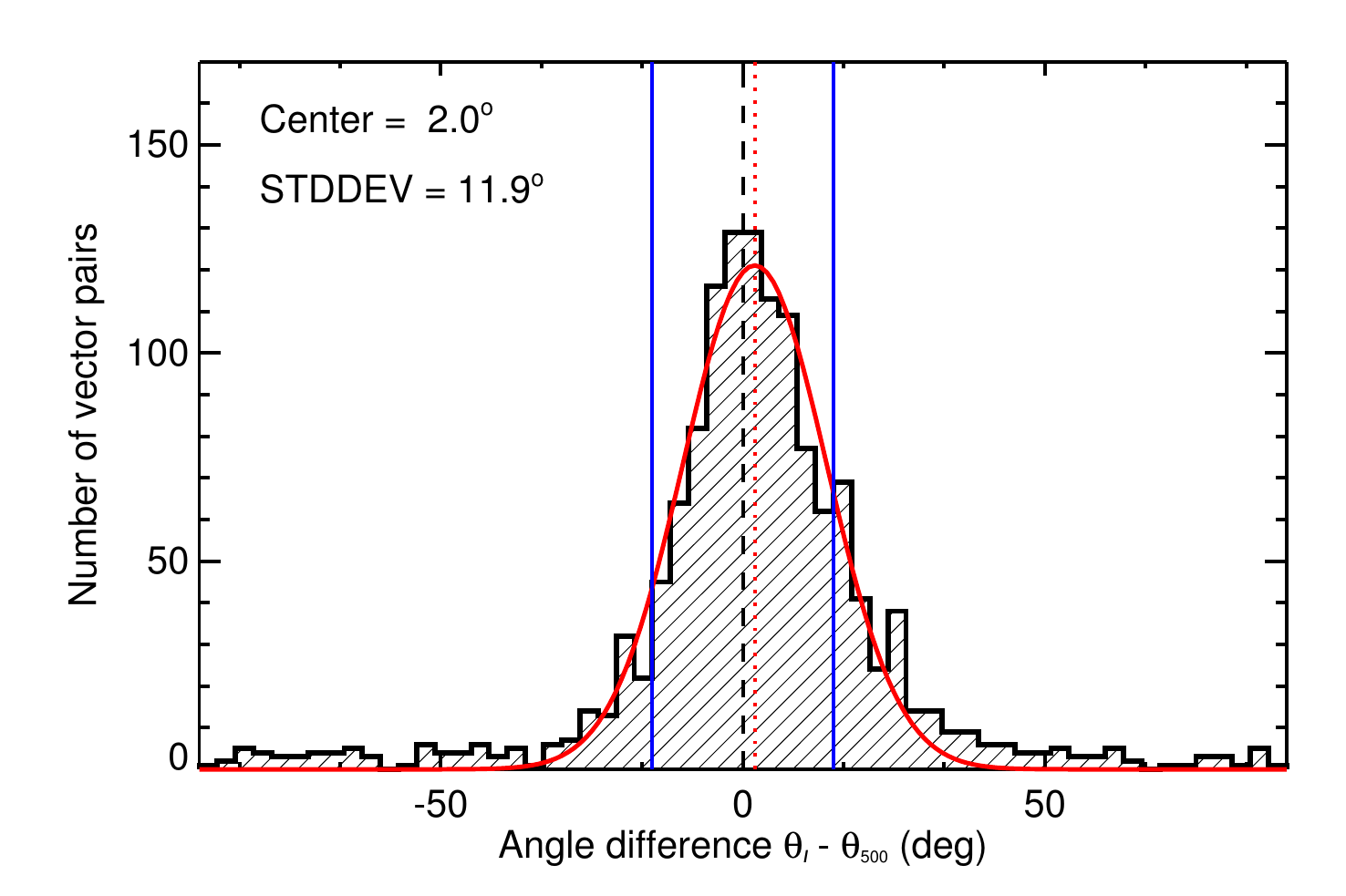} \\
   \includegraphics[width=0.48\textwidth]{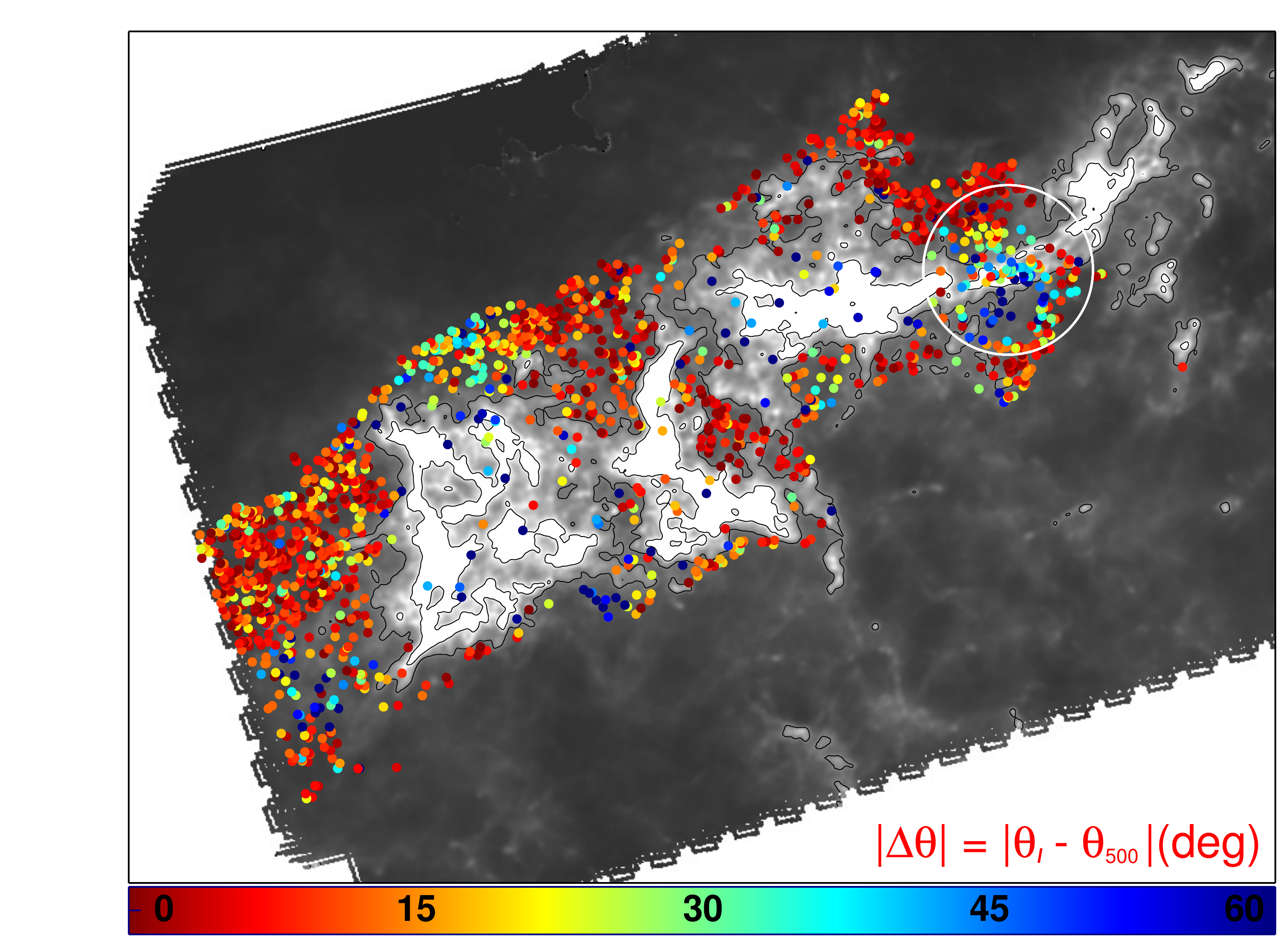} \\
      \caption{{\it Top:} histogram of the difference in magnetic field orientation between the 
	      $I$-band and $500\,\mu$m, data from the basic polarization combination data set 
              of 1355 stars (see Table \ref{tab}). 
              The red curve represents a Gaussian
              fit to the distribution (its center and standard deviation are specified in the figure). 
              Vertical blue lines represent the $\Delta\theta < 15\degr$ data cut (see Section \ref{s:dataselection_ang}).
	       {\it Bottom:} Estimated column density map of Vela C 
	       (same as shown in Figure \ref{f:mosaic}), with colored 
               dots representing the absolute value of the magnetic field angle 
               difference ($|\Delta\theta|=|\theta_{I}-\theta_{500\mu\mathrm{m}}|$).
	       The white circle is centered on RCW 36.
              }
         \label{f:angle}
   \end{figure}
%-------------------------------------------------------------

   Figure \ref{f:polmap} allows a visual comparison of the sky-projected 
   magnetic field lines traced by the two data sets, showing good agreement over 
most of the map. Representing the $I$-band and 
$500\mu$m magnetic field angles respectively as \thetai/ and \thetab/, in 
Figure \ref{f:angle} we show a histogram of the difference $\Delta\theta = $\thetai/$-$\thetab/ (top)
as well as a map where the color scale represents the absolute value of those differences (bottom).
The distribution is closely centered near $\Delta\theta = 0\degr$ (the Gaussian fit is peaked
at $2.0\degr$ with a standard deviation of $12\degr$). Since background stars at different distances
map different interstellar background components, one might expect a large 
discrepancy when comparing near-IR and sub-mm polarization angles. The good 
correlation seen in Figure \ref{f:angle} suggests that among all interstellar components along the LOS, 
the Vela C cloud
itself has a dominant effect in determining the polarization angle.
Nevertheless, to be prudent,
we will restrict our sample to $\Delta\theta < 15\degr$, which corresponds approximately to 
half of the distribution's FWHM. This criterion removes the outliers 
for which the two data sets could be probing different interstellar 
components. This is a conservative choice, given that even if 
no restriction to $\Delta\theta$ is applied, although the
number of data samples available for the analysis increases slightly, it does not 
significantly affect the final results that are presented in Sections
\ref{s:reff1} and \ref{s:finalper}.

BLASTPol data from the RCW 36 area suffer from systematic uncertainties that 
are typically larger than the statistical errors. Null tests carried out by \citet{2016fissel} show
significant structures around RCW 36 in the residual $Q$ and $U$ maps.
Furthermore, analysing the map in Figure \ref{f:angle}, we notice that around RCW 36 (white circle)
$\Delta\theta$ is systematically higher. Since it is known that many of the stars detected 
in that area are part of the star-forming cluster in the H{\sc ii} region 
\citep{baba2004} and therefore embedded in the cloud,
we believe that the discrepancy might be explained as follows: while the $500\,\mu$m 
polarization integrates the emission
along the entire cloud, the near-IR pseudo-vectors trace magnetic fields only up to the position of the corresponding embedded star.
In view of this possibility, we adopt a conservative approach by excluding all stars located inside
the white circle. 

\section{The stellar distance problem and \av/ estimates}
\label{s:results2}

%-------------------------------------------------------------
   \begin{figure}
   \centering
   \includegraphics[width=0.5\textwidth]{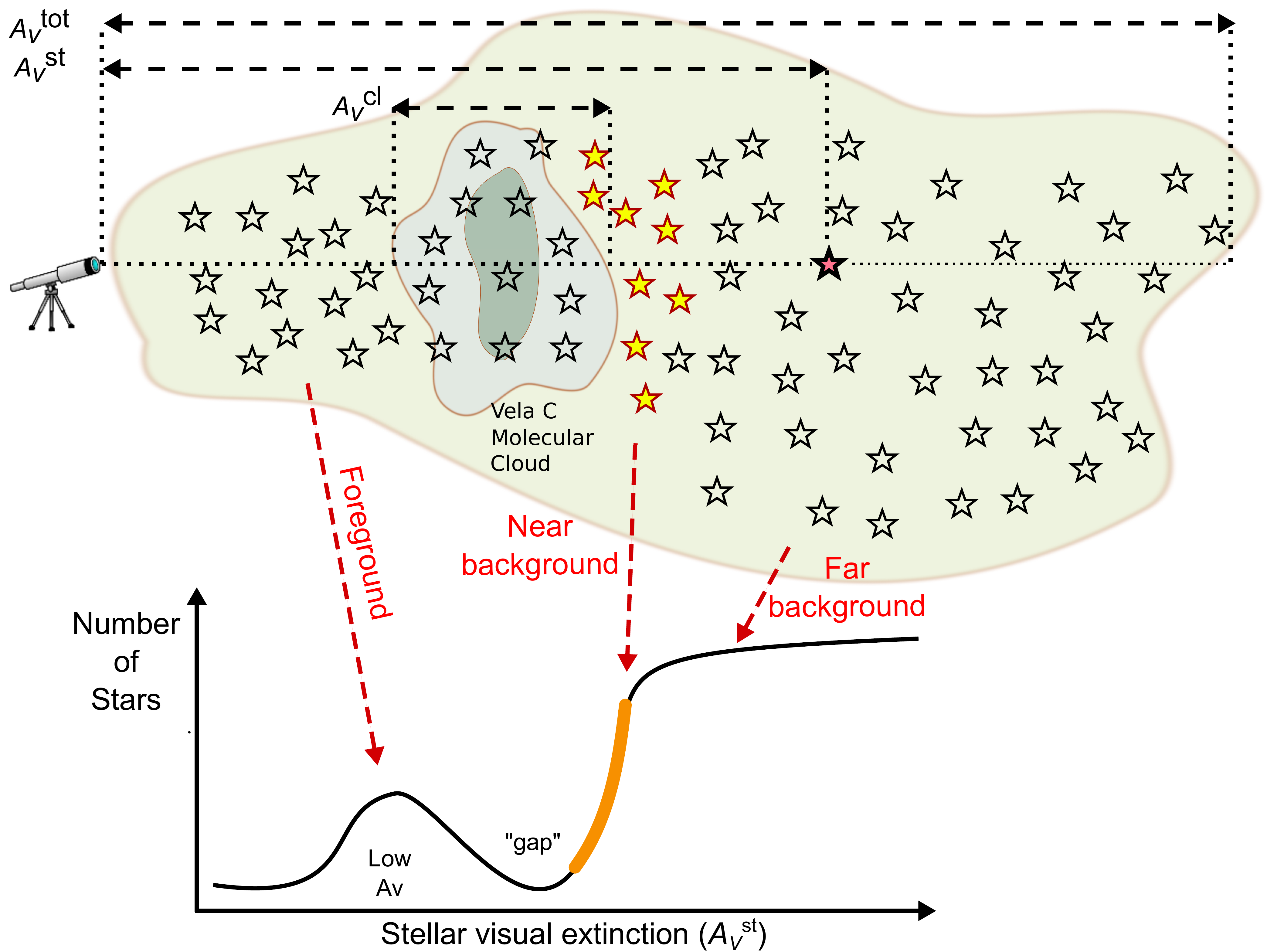} \\
      \caption{Schematic image showing the distribution of stellar objects in the
               the direction of Vela C, consisting of foreground, embedded, 
	       near-background (a.k.a. ideal; yellow) and far-background stars. The expected distribution  
               of stellar extinctions is also shown (bottom); this includes a Gaussian-like
               population of foreground stars at low extinctions and a steep rise 
               corresponding to objects located in the near-background (ideal stars).
	       For the LOS of a particular example object (denoted by the red star), 
               three types of extinction measurements are defined (top). These are the stellar extinction \avst/,
	       the cloud extinction \avcl/, and the total LOS extinction \avall/.
              }
         \label{f:scheme}
   \end{figure}
%%%%%%%%% use gs -o /dev/null -sDEVICE=bbox fig_stellar_distance.pdf
%%%%%%%%% to find bounding box
%-------------------------------------------------------------

As described in Section \ref{s:backsub_datasel} above, 
after subtraction of the foreground/background contribution, 
the polarized emission data traces only the molecular cloud.
Therefore, the stars with $I$-band polarization data that are suitable for 
comparison with polarized emission data are the ones immediately behind the molecular cloud, in the near-background.
The situation is illustrated in Figure \ref{f:scheme}.
%(even so, a small contribution
%from the cloud's foreground still needs to be subtracted from the stellar light data, 
%as will be discussed in Section \ref{s:foreground}). 
The important issue is that stars 
detected via our near-IR polarimetry observations are distributed at a range of distances 
in the cloud's LOS, but only a small subset of objects located in the near-background 
of the cloud should be selected, avoiding the inclusion of foreground stars and 
also far background stars contaminated by material from the Galactic disk.
For the purpose of adopting a clear nomenclature, near-background 
objects are henceforth referred to as ``ideal" stars, and objects located sufficiently
far away in the background (so that the additive extinction from the diffuse Galactic
ISM is non-negligible) are referred to as ``far-background" stars (see Figure \ref{f:scheme}).
%In this section we will define three types of visual extinctions for each stellar 
%sight line. These types are summarized at the top of Figure \ref{f:scheme}
%and will be discussed in detail below.

\subsection{The basic method for choosing near-background stars}
\label{s:avdef}

%The exclusion of far-background stars in this case is particularly important, 
%because, since Vela C is close to the Galactic Plane ($0.5\degr < b < 2.0\degr$), 
%a significant amount of background material from the disk is expected to exist
%(see Section \ref{s:compavs}).

Since individual stellar distances are typically not known through photometric or trigonometric
parallax techniques, we will identify ideal stars by analysing the distribution
of stellar visual extinctions (\avst/), as illustrated in Figure \ref{f:scheme}. 
Considering a specific LOS in the direction of the cloud, the distribution will typically exhibit a 
distinguishable population of foreground stars with low extinctions,
illustrated in Figure \ref{f:scheme} as a low-\av/ peak.
As we move to larger
distances going through the cloud, near-background (ideal) stars will have higher extinctions, 
therefore defining a rise in the distribution. Continuing to even larger distances, far-background
stars might have additional extinction from some background material in the Galactic disk.
Therefore, stars located at the rise in the distribution (after the ``gap" located just beyond the foreground
stars), should be ideal stars that are suitable for use in computing the polarization efficiency ratio.

In order to carry out the analysis described above, first it is necessary to define three different 
types of visual extinction measurement (see Figure \ref{f:scheme}): 
(1) the stellar extinction, \avst/, defined by the column of 
material extending as far as the stellar location, which can be estimated through near-IR photometry; 
(2) the cloud visual extinction, \avcl/, which accounts only for the molecular cloud column,
therefore being foreground and background subtracted; and (3) a visual extinction 
accounting for the entire column of interstellar material along the LOS, 
defined as \avall/. In Sections \ref{s:avst}, \ref{s:avcl} and \ref{s:avtot} we describe how these three types of 
visual extinction measurement
are determined. These measurements will subsequently be used to select ideal stars.

\subsection{Determining stellar visual extinction (\avst/) from 2MASS}
\label{s:avst}

%-------------------------------------------------------------
   \begin{figure}
   \centering
   \includegraphics[width=0.48\textwidth]{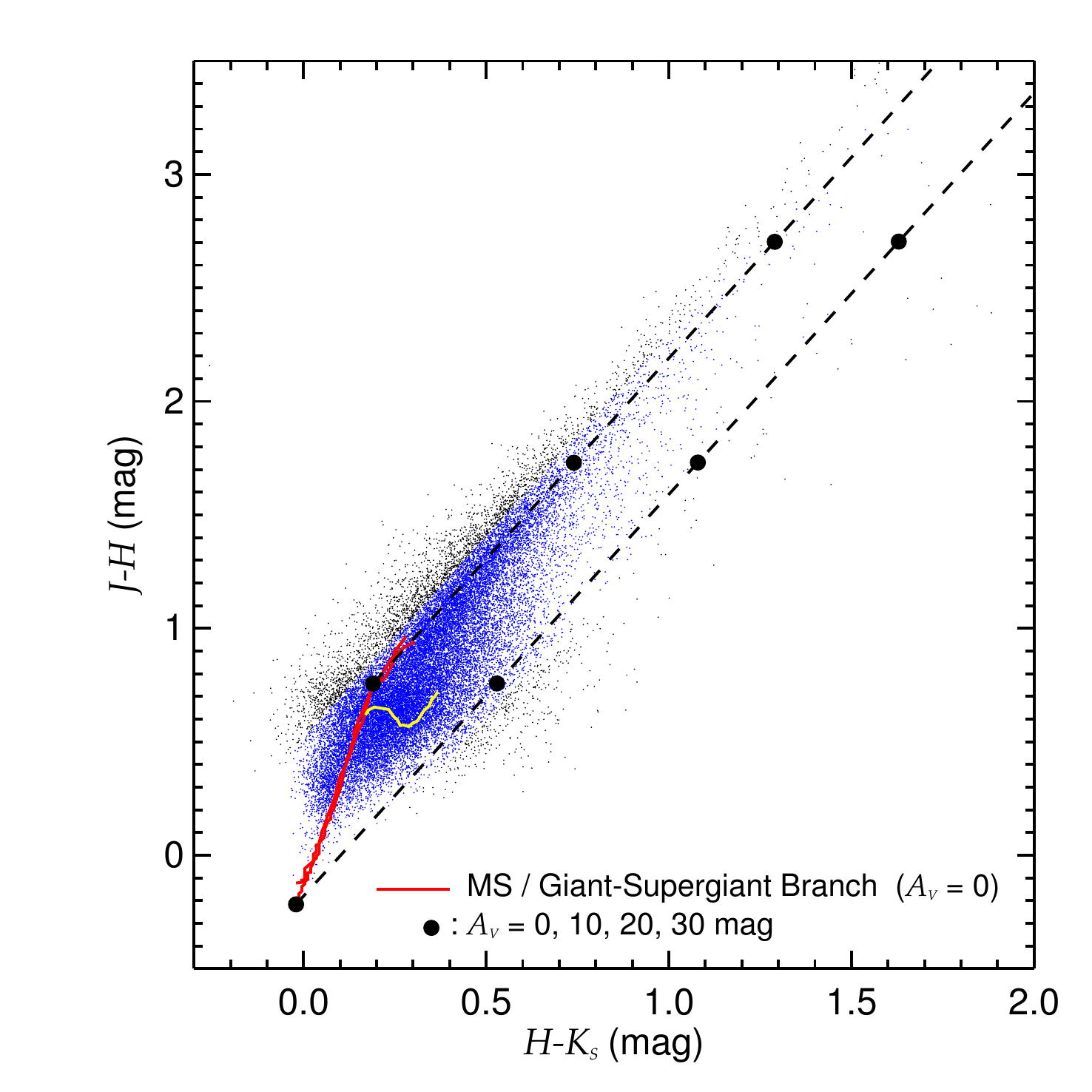} \\
   \includegraphics[width=0.50\textwidth]{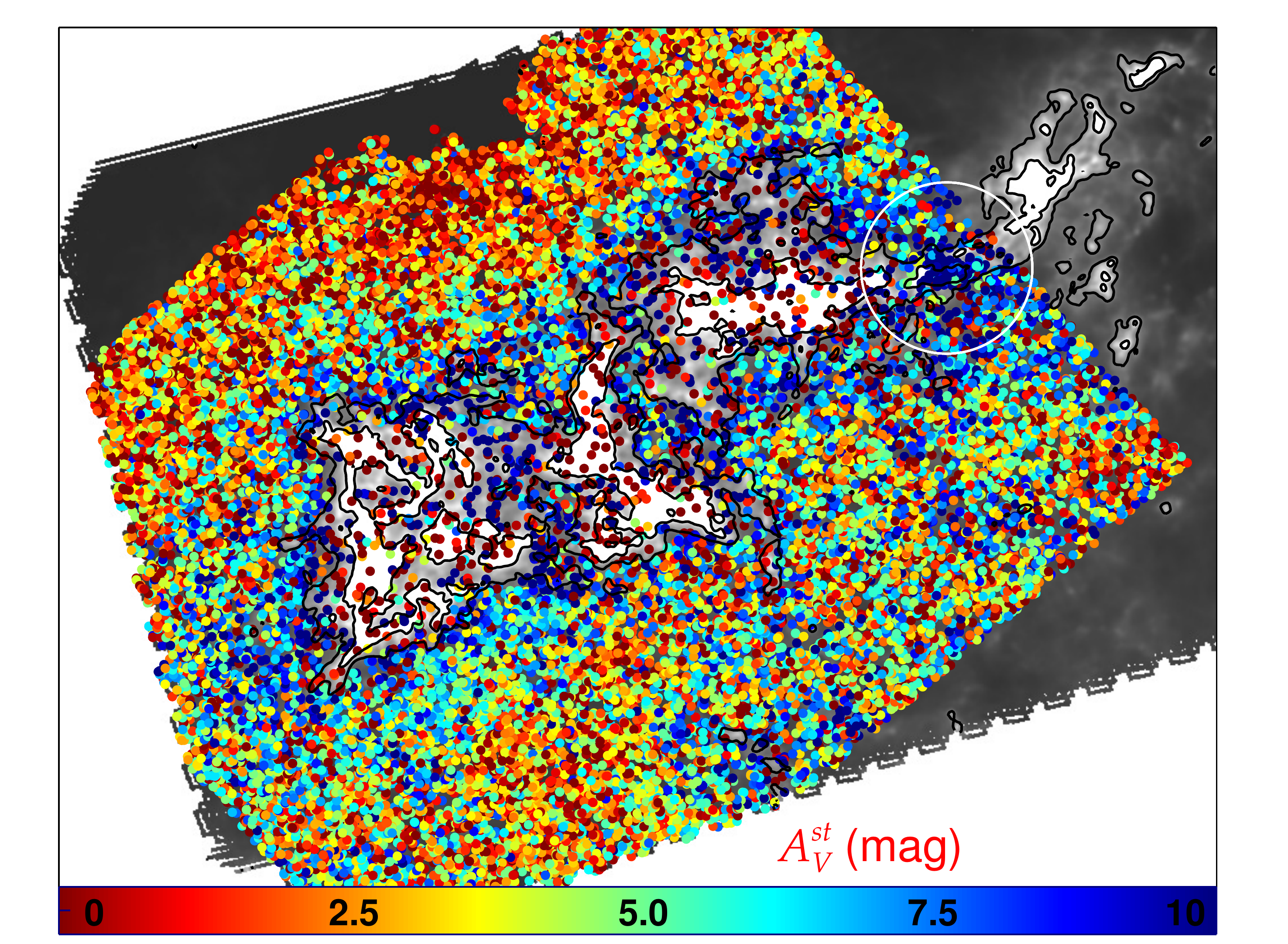} \\
      \caption{{\it Top:} Color-color diagram ($J-H \mathrm{\ vs.\ } H-K_{s}$) 
              for 2MASS stars in the wide photometric field (see Section \ref{s:avst} and Table \ref{tab}). 
	       Stellar extinctions (\avst/) are estimated for 
               all objects located inside the reddening band (the area between the parallel black dashed 
               lines). \avst/ is proportional to the distance between the object and the main-sequence/giant 
               locus (the red lines at the bottom left). Objects outside and inside the 
               reddening band are shown as black and blue points, respectively.
               The yellow line corresponds to the locus of main-sequence stars with 
               spectral types later than K7.
	       {\it Bottom:} Estimated column density map of Vela C (same as shown 
	       in Figure \ref{f:mosaic}), with colored dots overlaid
               representing stellar extinctions \avst/ for 2MASS stars in the wide photometric field.             
	       The white circle is centered on RCW 36.
              }
         \label{f:ccdiag}
   \end{figure}
%-------------------------------------------------------------

Even considering that the individual spectral types for each object are not known, an approximate
estimate of \avst/ may be obtained using the stars' $J-H$ and $H-K_s$ colors from the 
2MASS catalog \citep{skru2006}. This method has been used, for example, 
by \citet{whittet2008}, in their analysis of optical polarization efficiencies ($p/A_\mathrm{V}$)
in the Taurus and Ophiuchus dark clouds.
Each observed star defines a position in a color-color diagram (Figure \ref{f:ccdiag}, top), and  
the color excess values ($E(J-H)$ and $E(H-K_s)$) may be obtained through extrapolation 
along the reddening band (black dashed lines) onto the intrinsic color lines\footnote{Intrinsic 
colors are obtained from \citet{koornneef1983} and further corrected to the 
2MASS photometric system through transformation relations provided by \citet{carpenter2001}.} (superposed solid red lines). 
The method discussed here was applied to all objects located within both the blue 
box of Figure \ref{f:mosaic} and the boundaries of the {\it Herschel} map. We 
define this sample as the ``wide photometric field" (Table \ref{tab}). It encompasses most stars with 
$I$-band polarimetric detections, in addition to a vast sample of the stellar population
in the direction of Vela C. This wider set of photometric data will be useful for 
the analysis of Sections \ref{s:compavs} and \ref{s:gl}. 

\citet{martin2012} showed that for the ISM around Vela C, the slope of the 
reddening band is $1.77\pm0.01$, which is evident here in the elongated distribution of
points along the black dashed lines in Figure \ref{f:ccdiag}. 
De-reddening each point along the reddening band generally provides unambiguous results,
since the main sequence, giant, and super-giant loci all correspond to superposed lines in this diagram, 
except for a subset of
late-type main-sequence stars (the yellow line in Figure \ref{f:ccdiag}). However, taking into account the 
2MASS photometric completeness limits in the $J$, $H$, and $K_{s}$ bands, it is straightforward to 
show that at distances of $700\,$pc or greater, main sequence stars with spectral types 
later than approximately K7 would not be bright enough to be detected, 
and therefore the portion of the main sequence indicated by a thin yellow line may be ignored
(foreground objects are an obvious exception, but these will be removed from the analysis later; 
see Section \ref{s:gl}).

The conversion from color excess to visual extinction can be carried out in several different ways. 
For instance, canonical relations can be obtained from \citet{rieke1985} or
\citet{fitzpatrick1999}, provided that some value for the total-to-selective 
extinction is adopted. Another option is to take advantage of the updated 
relation between $E(J-K_{s})$ and total hydrogen column density $N_\mathrm{H}$
obtained by \citet[][Equation 9]{martin2012} by reassessing previously 
published ultraviolet stellar spectroscopic data and comparing it with 2MASS data. 
Since in the present work we are using the same catalog 
of near-IR photometry (2MASS) as was used by \citet{martin2012}, this last method seems most appropriate.
By using it, we avoid any conversion errors due to mismatch in the photometric system employed.
%for making sure that small differences due to slightly different photometric 
%systems (used by \citealt{rieke1985} or \citealt{fitzpatrick1999}) 
%will not affect the conversion. 
We obtain $E(J-K_{s})$ for each star by summing $E(J-H)$ and $E(H-K_{s})$.
Then, we combine Equation 9 from \citet{martin2012} (which relates $E(J-K_{s})$ to $N_\mathrm{H}$) 
with the gas-to-dust relation $N_\mathrm{H} = 1.9\times10^{21}$cm$^{-2}$\avst/ 
\citep[][]{savage1977,bohlin1978,rachford2009}.
The resulting relation between $E(J-K_{s})$ and \avst/ is: 

\begin{equation}
A_\mathrm{V}^\mathrm{st} = 6.05 E(J-K_{s}) - 0.04.
\end{equation}

\noindent It is important to point out that the above-mentioned gas-to-dust relation 
includes the assumption that $A_{V}=3.1 E(B-V)$ \citep[e.g.,][]{draine2003}, and that the total-to-selective
extinction ratio depends on grain properties, thus
providing a source of systematic uncertainties (see discussion at the end of this subsection).

%As noted by \citet{martin2012}, it is important to point out that the range of $A_{V}$
%in which the above stated gas-to-dust relation has been tested to be valid is somewhat low 
%(up to 4 mag), and therefore the application above this limit is obviously an extrapolation.
%If an increased extinction (and column density) is simply an effect of large path lengths
%along the LOS, then there is no reason why the grain properties would change
%and therefore the same relations would probably still be valid. However, they might be different 
%if large extinctions are actually due to increased volume density.

To define the wide photometric field sample, we keep only stars
with 2MASS photometric quality ``AAA", signifying a photometric detection with S/N $> 10$ and 
uncertainties in $J$, $H$ and $K_{s}$ below $0.1\,$mag. 
Furthermore, stars located well outside the reddening band (black dots in Figure \ref{f:ccdiag}, top)
are excluded in order to avoid extragalactic sources
and young stars with circumstellar disks, which are known to exhibit infrared excess
and sometimes intrinsic polarization. 

%Uncertainties in \avst/
%obtained solely through propagation of the photometric uncertainties typically range 
%between $0.3$ and $0.8\,$mag. 
%However, one should 
%note that the fractional error is lower for higher extinction stars. 
%For the purpose of studying the polarization efficiency 
%ratio (Sections \ref{s:reff1}), only stars with \avst/$/\Delta$\avst/$>3$ were kept. 

The distribution of stellar extinctions \avst/ is shown in the bottom panel of Figure \ref{f:ccdiag}.
Note that closer to the cloud the average extinctions are generally higher. 
Also note that many stars with very low extinctions may be found within the cloud contours, 
and objects with high extinctions may be found well off the cloud.
These objects have properties consistent with being, respectively, foreground 
and far-background stars. 
In Section \ref{s:compavs} and Appendix \ref{ap:avstuncertainties} we show that the wide
distribution of \avst/ values seen for off-cloud positions is primarily due to distance,
with stars located at large distances behind the cloud having higher extinctions
due to the presence of diffuse ISM in the far-background.

%{\bf 
%Section \ref{s:compavs} and Appendix \ref{ap:avstuncertainties} discusses the main factors 
%that can affect the \avst/ distribution and its comparison with cloud extinctions.
%It is clear that stars are detected over a wide range of distances 
%toward Vela C, so that interstellar material from the background is also being 
%probed. The wide stellar distance distribution along the line-of-sight is the main reason 
%why in the off-cloud regions of Figure \ref{f:ccdiag} (bottom), for instance, there is a significant variation in the \avst/ values.
%}

\subsection{Determining cloud visual extinction (\avcl/) from {\it Herschel}}
\label{s:avcl}

The foreground and background subtracted molecular cloud extinction \avcl/ 
was estimated from dust emission maps made by 
{\it Herschel} SPIRE at $250$, $350$ and $500\,\mu$m. The technique used is 
similar though not identical to the one described by 
\citet{2016fissel}. One difference is that we did not smooth the {\it Herschel}
maps to the BLASTPol resolution. Another difference is that we did not make use 
of the $160\mu$m PACS map, since it 
covers a smaller sky area in comparison with the SPIRE maps. 
In brief, the technique consists of, firstly, 
using previously selected ``diffuse emission regions" surrounding the cloud
(containing little or no emission from the cloud itself) to calculate the 
diffuse Galactic contribution for each waveband. 
These are then subtracted from each corresponding
SPIRE map. Modified blackbody SED fits were then constructed for each pixel,
assuming the dust opacity law of \citet{1983hildebrand} with a dust spectral
index of $\beta  = 2$, thus generating column density ($N_\mathrm{H}$) and temperature maps ($T$). 
Finally, the relation $N_\mathrm{H} = 1.9\times10^{21}$ cm$^{-2}$\avcl/ was used to obtain the 
cloud extinction map. 
It is important to point out that 
the assumptions used above are strictly valid only for diffuse lines-of-sight.
There is evidence in the literature that 
the sub-mm optical depth per unit column density increases somewhat for higher density 
molecular clouds due to grain processing \citep{planck2011-XXV}, leading to some uncertainty

As previously mentioned, \avcl/ contours corresponding to $10$ and $25\,$mag 
are shown in Figure \ref{f:polmap}. Note that for sky regions covered by the
$160\,\mu$m maps, we found very little difference between \avcl/ values derived with and without
the $160\,\mu$m data.

\subsection{Determining total visual extinction for the entire line-of-sight (\avall/) from {\it Planck}}
\label{s:avtot}

In order to obtain the visual extinction for the entire LOS corresponding to each
individual star in our sample, including the entire column up to and beyond the star,
we use the $353\,$GHz optical depth from {\it Planck} all-sky mapping.  
\cite{2014planckxi} correlated their $353\,$GHz optical depth ($\tau_{353}$) 
with estimates of color excess $E(B-V)$ for quasars, based on photometric measurements 
from the Sloan Digital Sky Survey (SDSS). Using extragalactic objects rather than Galactic stars ensured that 
the entire Galactic column in the direction of each quasar was probed, 
avoiding biases that could arise from background contamination. 
They found $E(B-V)/\tau_{353} = (1.49\pm0.03)\times10^4\,$mag.
By assuming that $A_{V}=3.1E(B-V)$, we converted the {\it Planck}-derived selective 
extinction measurements to \avall/.

\subsection{Comparisons between \avst/, \avcl/, and \avall/}
\label{s:compavs}

\begin{figure}
   \centering
   \includegraphics[width=0.48\textwidth]{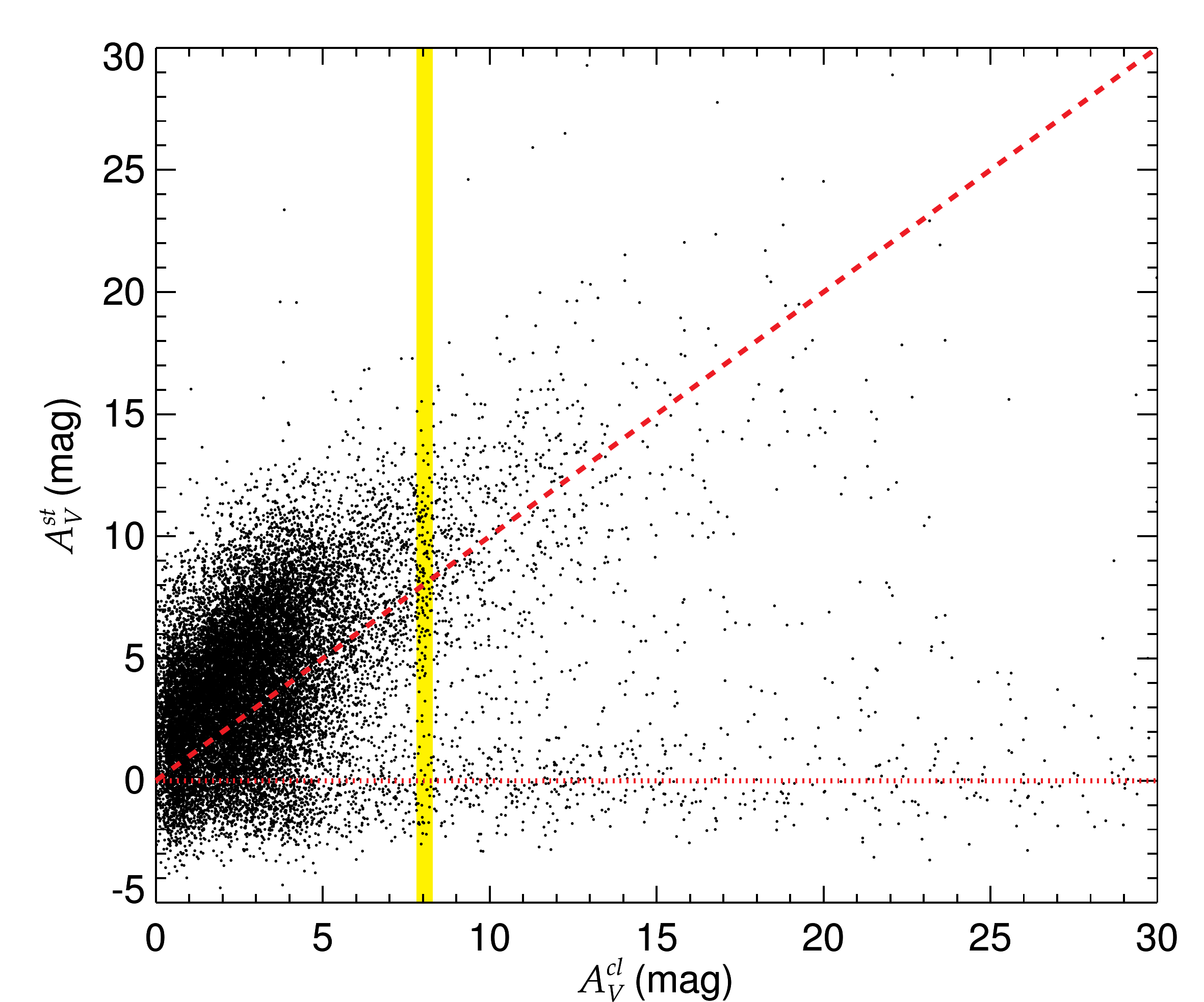} \\
   \includegraphics[width=0.48\textwidth]{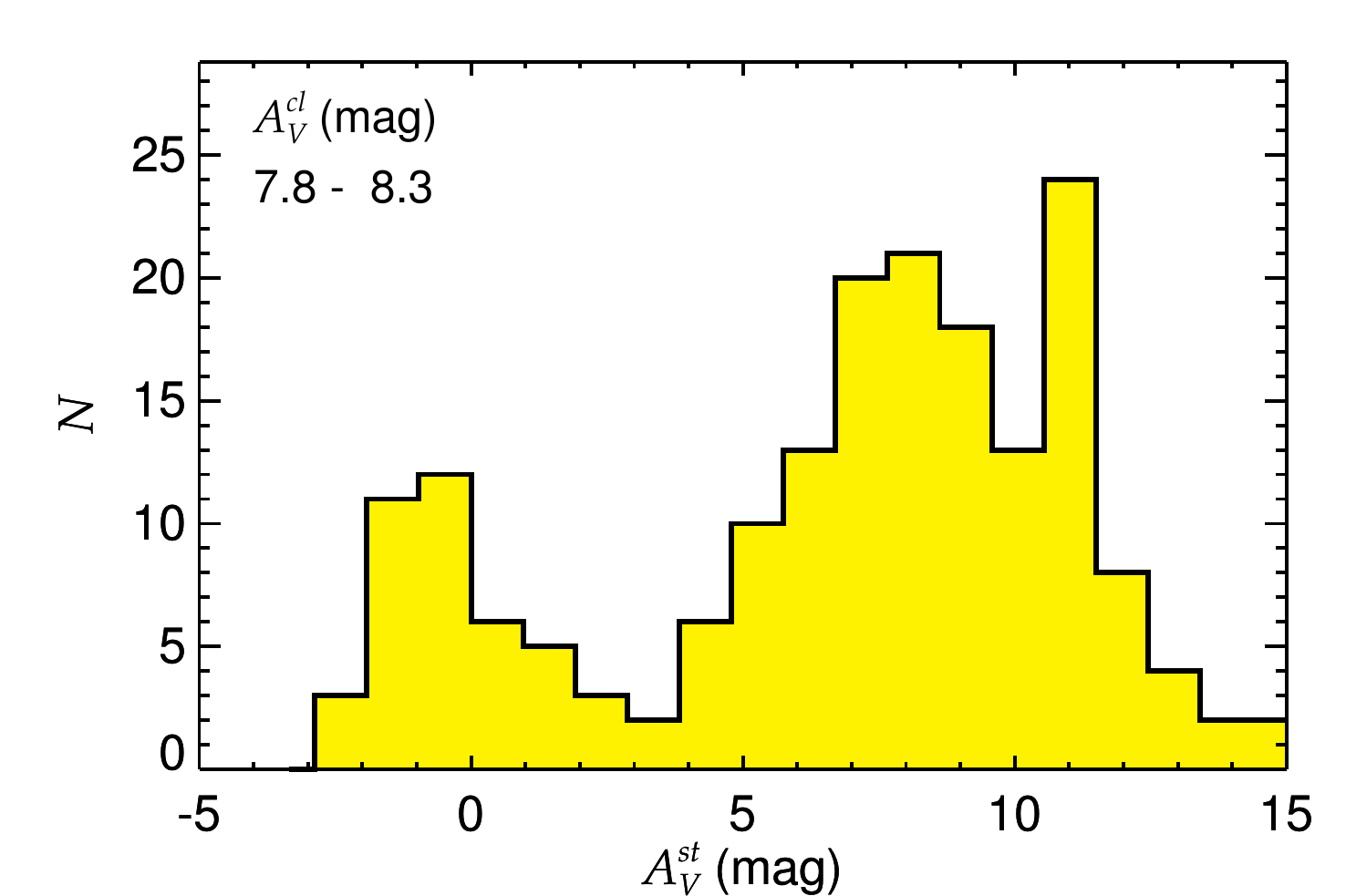} \\
   \caption{Top: Diagram comparing stellar extinctions (\avst/) and cloud extinctions (\avcl/) for objects 
	      located within the wide photometric field (see Table \ref{tab}). 
               Dashed and dotted lines representing the 
               equality \avst/=\avcl/ and \avst/$=0$, respectively, are shown for reference.
	       Bottom: Example \avst/ histogram for a bin of 
	       \avcl/ centered on \avcl/$ = 8\,$mag (corresponding to the yellow vertical band in the top panel), 
	       showing the gap between the
	       foreground and background stellar extinction distributions.
              }
         \label{f:avavdiag}
   \end{figure}

A comparison between \avst/ and \avcl/ is shown in Figure~\ref{f:avavdiag} (top). This diagram 
includes all the stars from the wide photometric field (Section \ref{s:avst} and Table \ref{tab}).
Note that most of the points are located somewhat
above the equality line (dashed red line), which suggests that many of these objects are 
affected by extinction from the background ISM.
In Appendix \ref{ap:avstuncertainties} we consider
Galactic models for stellar and dust distribution 
together with the sensitivity of the 2MASS survey, and we conclude that the wide photometric 
field is expected to include large numbers of stars located in
the far-background ($\approx2-10\,$kpc), 
behind several magnitudes of additional extinction caused by diffuse Galactic ISM
behind the cloud.
Although several other factors may affect
comparisons between \avst/ and \avcl/ (Appendix \ref{ap:avstuncertainties}), the primary cause for the wide spread 
of points above the equality line 
is this population of far-background stars contaminated by background extinction.
For a fixed \avcl/ value 
one can see that there is a population of foreground objects near \avst/$=0$ (dotted line). 
As one moves further up in stellar extinction \avst/, a ``gap" region is found, followed by 
a rise in the number of stars. 
For example, for \avcl/$\approx8\,$mag (vertical yellow band in Figure~\ref{f:avavdiag}, top), 
we see a cluster of points near \avst/$=0$, 
another cluster around \avst/$= 6 - 12\,$mag, and very few points in the ``gap" near
\avst/$= 2 - 4\,$mag. This can be seen in Figure~\ref{f:avavdiag} (bottom),
which is a histogram of \avst/ for a small \avcl/ bin centered on \avcl/$=8\,$mag.
The histogram clearly exhibits a gap between the foreground and background stellar populations.
This is consistent with the 
expectation described above (Figure \ref{f:scheme}, Section \ref{s:avdef}).

%-------------------------------------------------------------
%-------------------------------------------------------------

\begin{figure*}
   \centering
   \includegraphics[width=0.75\textwidth]{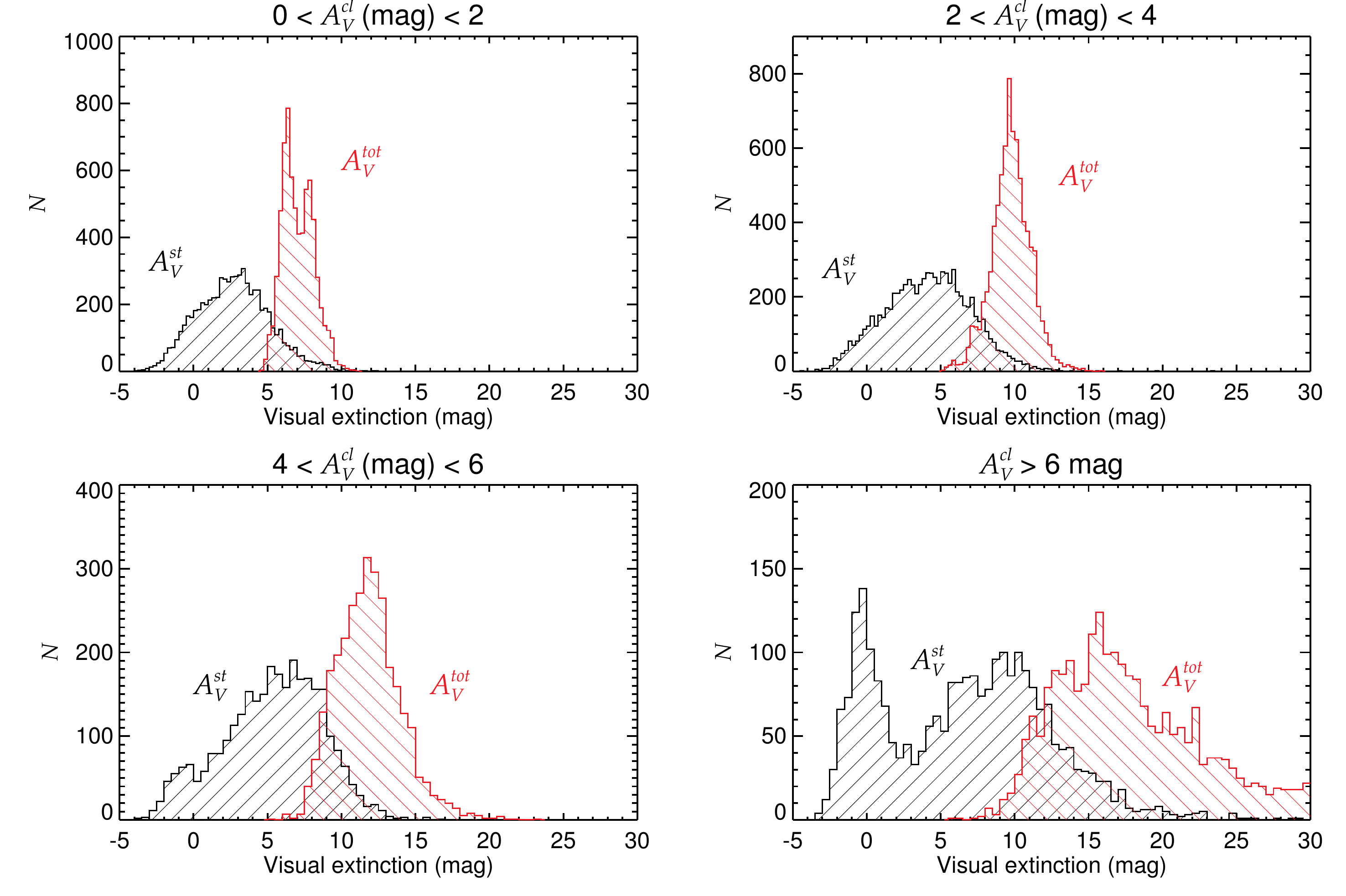} \\
      \caption{Histograms of stellar extinction 
	      \avst/ and total LOS extinction \avall/)
               for different ranges of cloud extinction \avcl/, including all objects
	       from the Planck-2MASS combination data set (see Table \ref{tab}).
              }
         \label{f:compavall}
   \end{figure*}

   We have argued that many stars are contaminated by a
background Galactic extinction component (and therefore are located further away from the cloud, 
in the far-background). An independent way of testing this is to compare with 
visual extinction estimates that account for the entire line-of-sight (\avall/), using
the Planck-2MASS combination data set (see Table \ref{tab}). Figure \ref{f:compavall}
shows visual extinction histograms for different \avcl/ ranges, including the distributions 
for both \avst/ (black) and \avall/ (red). It is clear that regardless of which depth through the 
cloud one uses (i.e., for all \avcl/ ranges), the \avall/ distribution always extends to higher
levels than \avst/. In particular, that is true even when the contribution 
from the cloud itself is small (see the first histogram of Figure \ref{f:compavall}, 
for which $0 <$\avcl/$<2\,$mag). This histogram shows that even for the relatively 
diffuse areas surrounding the cloud, the visual extinction integrated along the entire LOS is typically
between $5$ and $10\,$mag, while stellar extinctions have a broader distribution, but centered
at $\approx3\,$mag. 

%-------------------------------------------------------------
%-------------------------------------------------------------

Our analysis of Figure \ref{f:compavall} supports the existence
of a significant column density of background 
ISM. This hypothesis is also consistent with the distribution of points in Figure \ref{f:avavdiag} (top), 
and with the analysis of Appendix \ref{ap:avstuncertainties},
as noted earlier.
We conclude that many stellar objects are contaminated by background ISM
and will have to be removed from the sample.
Appendix \ref{ap:backevidence} shows an independent set of evidence for the existence 
of this contaminating far-background material, based on a separate analysis of stellar 
extinction as a function of distance.

\section{Computing the polarization efficiency ratio \per/}
\label{s:results3}

\subsection{Foreground correction for stellar extinction and polarization}
\label{s:foreground}

Our qualitative analysis of Figure \ref{f:avavdiag} (top) revealed that there is 
a group of stars forming a ``band" approximately following the line \avst/ $=0\,$mag (red dotted line). 
For these stars, independently of the cloud's extinction along the LOS, stellar 
extinctions are very low. This is a characteristic feature of foreground stars.
Such stars must be removed from this analysis. Furthermore, it is expected that 
diffuse material in front of the Vela C cloud contributes a small fraction of the extinction measured
for background stars (although the \avst/ values derived from 2MASS do not 
provide the necessary sensitivity to estimate this small component). Additionally, 
a foreground polarization is also imposed on the stellar light from background 
stars. Both the extinction and polarization components 
originating from the foreground ISM must be subtracted.

Estimates of the foreground extinction and polarization in the direction 
of Vela C are obtained in Appendix \ref{ap:fore}. We find that the foreground ISM towards
Vela C is in general very diffuse, with an extinction level of approximately $0.15\pm0.09\,$mag.
For the purposes of defining the corrected polarization combination data set that will 
be used in the analysis of Section \ref{s:reff1} (see also Table \ref{tab}), this foreground extinction value is subtracted 
from \avst/, and additionally, only stars with \avst/$/$\eavst/$>3$ are used (where \eavst/ are the statistical 
uncertainties derived from propagation of 2MASS photometric errors). 
The foreground polarization is estimated as $p_{I}=0.4\%$ and \thetai/$=132\degr$. 
This component is subtracted from the measured polarization values of our sample, 
using standard techniques \citep[e.g.,][]{santos2014},
%which consists in 
%calculating the Stokes parameters $\bar{Q}$ and $\bar{U}$ for the foreground polarization,
%and subsequently subtract it from each stellar polarization Stokes values.
and then we re-apply the $p_{I}/\sigma_{pI}>3$ criterion.
These selections complete the definition of the corrected polarization
combination data set as specified in Table \ref{tab} (where we also summarize all the 
additional selection criteria
described in Sections \ref{s:dataselection}, \ref{s:dataselection_ang}, and \ref{s:avst}).

\subsection{The Gaussian-logistic method of selecting ideal stars}
\label{s:gl}

%-------------------------------------------------------------
   \begin{figure*}
   \centering
   \includegraphics[width=0.90\textwidth]{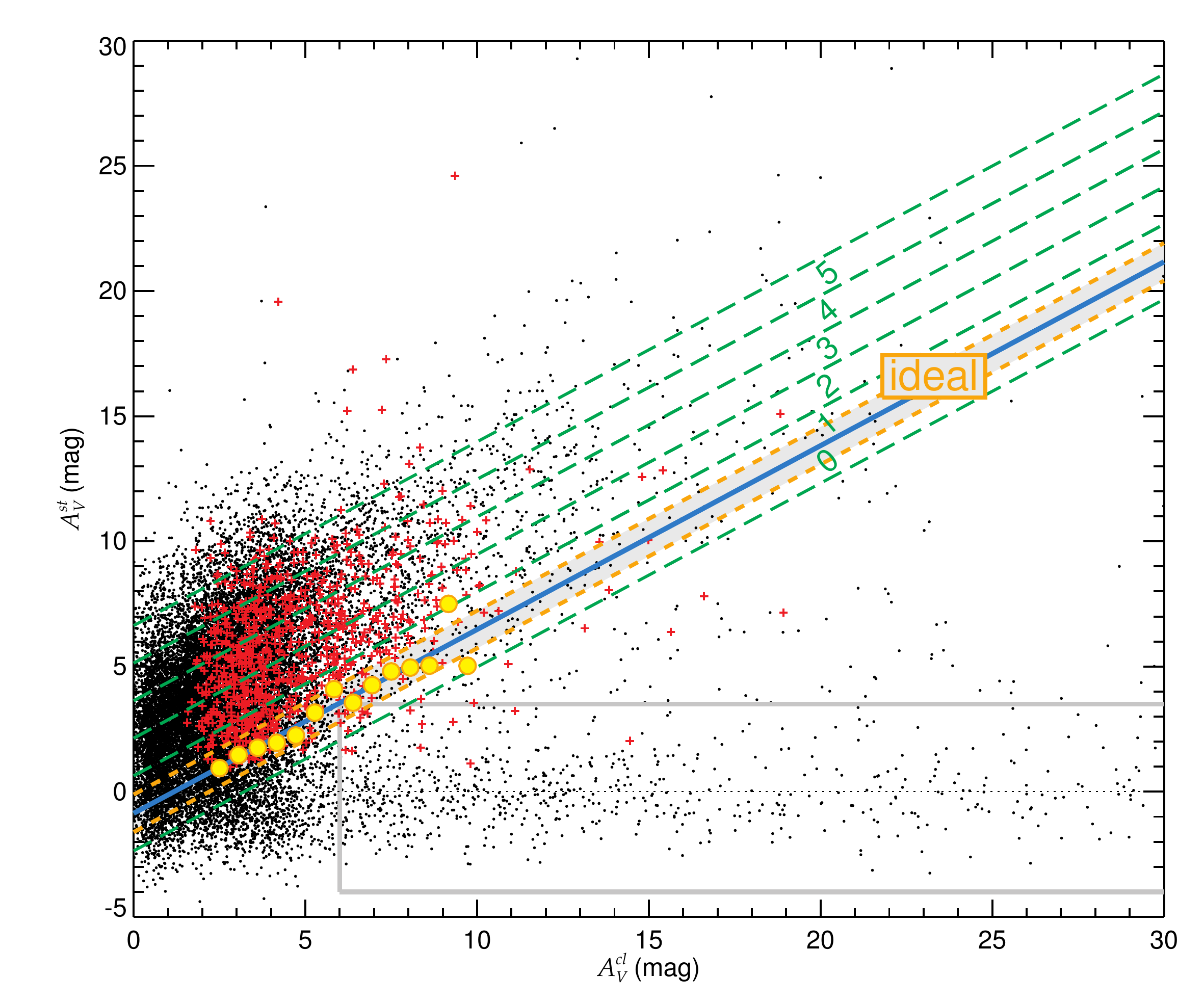}
      \caption{Stellar extinctions \avst/ vs. cloud extinction \avcl/ for stars in the wide photometric 
               field data set (black dots), and stars in the corrected polarization combination data set 
	       (red crosses). Data sets are defined in Table \ref{tab}.
               The gray box is used to the define the \avst/ distribution of foreground stars 
               as an input to the GL method, and the yellow circles are the mid-points $A_{V}^\mathrm{0}$
               of the logistic function for different \avcl/ bins, obtained
               as outputs from the same method (see Section \ref{s:gl} and Appendix \ref{ap:gl}). The blue line is a fit 
               to the yellow circles, and allow us to define the ``ideal stellar sample" (points in the
	       grayed area between the two dashed orange lines), which is the set of objects located in the
               near-background (Section \ref{s:gl}). Strips 0 to 5 (bounded by the green lines) are defined 
               parallel to the ideal stellar locus
               and are used to study the effects of background contamination (see Sections \ref{s:gl} and \ref{s:contamback}). 
               %using configuration 18 5 2 
              }
         \label{f:avavdiag2}
   \end{figure*}
%-------------------------------------------------------------

As discussed in Section \ref{s:avdef} above, only ideal stars (located in the
near-background) are suitable for a quantitative comparison between polarization data sets
obtained from extinction and emission. 
Notice that these objects cannot be unbiasedly selected simply through 
        a direct comparison such as \avst/ = \avcl/, because of the uncertainties associated 
with the derivation of \avcl/ from {\it Herschel} data (see Section \ref{s:avcl}).
Instead, we apply an empirical method that does not rely on a direct
comparison between \avst/ and \avcl/.
Figure \ref{f:scheme} showed a schematic profile of 
the stellar extinction distribution expected toward a given LOS, composed of a foreground 
population at low extinction, followed by a steep rise in the number of stars, corresponding
to the ``ideal" stars.
Furthermore, we observed this expected profile in the data from the wide photometric 
field (Figure \ref{f:avavdiag}).
Accordingly, 
we model the \avst/ distribution within different bins of cloud extinction \avcl/ using a Gaussian-logistic (GL)
function, defined as a Gaussian function added to a logistic profile (which can be 
described as a smoothed step-function):

\begin{equation}
N^\mathrm{st} (A_{V}^\mathrm{st}) =\alpha e^{\frac{(A_{V}^\mathrm{st}-\beta)^{2}}{2\sigma^{2}}} + \frac{a}{1+e^{-b(A_{V}^\mathrm{st}-A_{V}^\mathrm{0})}}.
\label{e:gl}
\end{equation}
 
\noindent In this equation, $N^\mathrm{st}$ is the number of stars (within a certain 
bin of cloud extinctions \avcl/), given as a function of the stellar extinction
\avst/. In the first term, which represents the foreground stellar population, parameters 
$\alpha$, $\beta$, and $\sigma$ are the height, displacement, and 
width (the standard deviation) of the Gaussian fit, respectively. In the second term, 
which represents the background stellar population, parameters $a$, $b$, and $A_{V}^\mathrm{0}$ 
represent the height of the logistic function, its steepness, and the mid-point
of the logistic curve, respectively. 

To specify the subsets of the data that are used in the GL fits, 
in Figure \ref{f:avavdiag2} we show again the comparison between stellar and cloud extinction.
The black dots represent the wide photometric field sample, 
identical to what is shown in 
Figure \ref{f:avavdiag} (top). The GL method 
does not require any polarization measurements, and therefore
should be applied to the maximum number of stars available. 
For this reason we apply it to the wide photometric field.
After the application of the GL method, however, stars that will be available for the 
computation of the polarization efficiency ratio are only those
in the corrected polarization combination data set (containing both $I$-band and 
$500\mu$m polarization data; see Table \ref{tab}). These are indicated 
in Figure \ref{f:avavdiag2} by the red crosses; a subset of these stars
selected as ideal objects via the GL method will be used to compute the 
polarization efficiency ratio.

A detailed description of the GL method is given in Appendix \ref{ap:gl}. 
The basic idea is to fit equation \ref{e:gl} to different distributions representing
different bins of \avcl/ (one can imagine this as a series of vertical slices in
Figure \ref{f:avavdiag2}; see also Figure \ref{f:avavdiag}).
The important quantity here is $A_{V}^\mathrm{0}$, which represents the position
of the rise in the number of stars (the mid-point of the logistic function), and therefore defines the locus of ideal stars for 
each bin of cloud extinctions. 
In Figure \ref{f:avavdiag2}, the yellow 
circles show the position of $A_{V}^\mathrm{0}$ for each \avcl/ bin (using the center of each 
bin). These points are fit to a straight line (the blue line), 
which represents the positions where ideal stars are found. We define a series of ``strips" parallel
to this line, labeled from $0$ to $5$ (delimited by green dashed lines), and a special strip defined 
as the ``ideal stellar locus" (grayed area between the orange dashed lines). In the figure, strip-0 is below the ideal stellar 
locus, and higher strips represent increasingly distant far-background stars whose
\avst/ values (and $I$-band polarization) are increasingly contaminated by the Galactic ISM. 
Red crosses inside the ideal locus define the ``ideal stellar sample" (Table \ref{tab}), 
which will be used to compute the polarization efficiency ratio.
The vertical separation between consecutive strips is $1.5\,$mag, which is on the order of 
the typical uncertainty in \avst/ (see Appendix \ref{ap:fore}). Since the separation 
between lower and higher strips is larger than the typical \avst/ uncertainties, 
we expect that higher-numbered strips will clearly show increasing levels of
background contamination in their measured $I$-band polarizations. We return to this 
point in Section \ref{s:contamback}.

\subsection{Polarization efficiency ratio and analysis of systematic uncertainties}
\label{s:reff1}

%-------------------------------------------------------------
   \begin{figure*}
   \centering
   \includegraphics[width=0.48\textwidth]{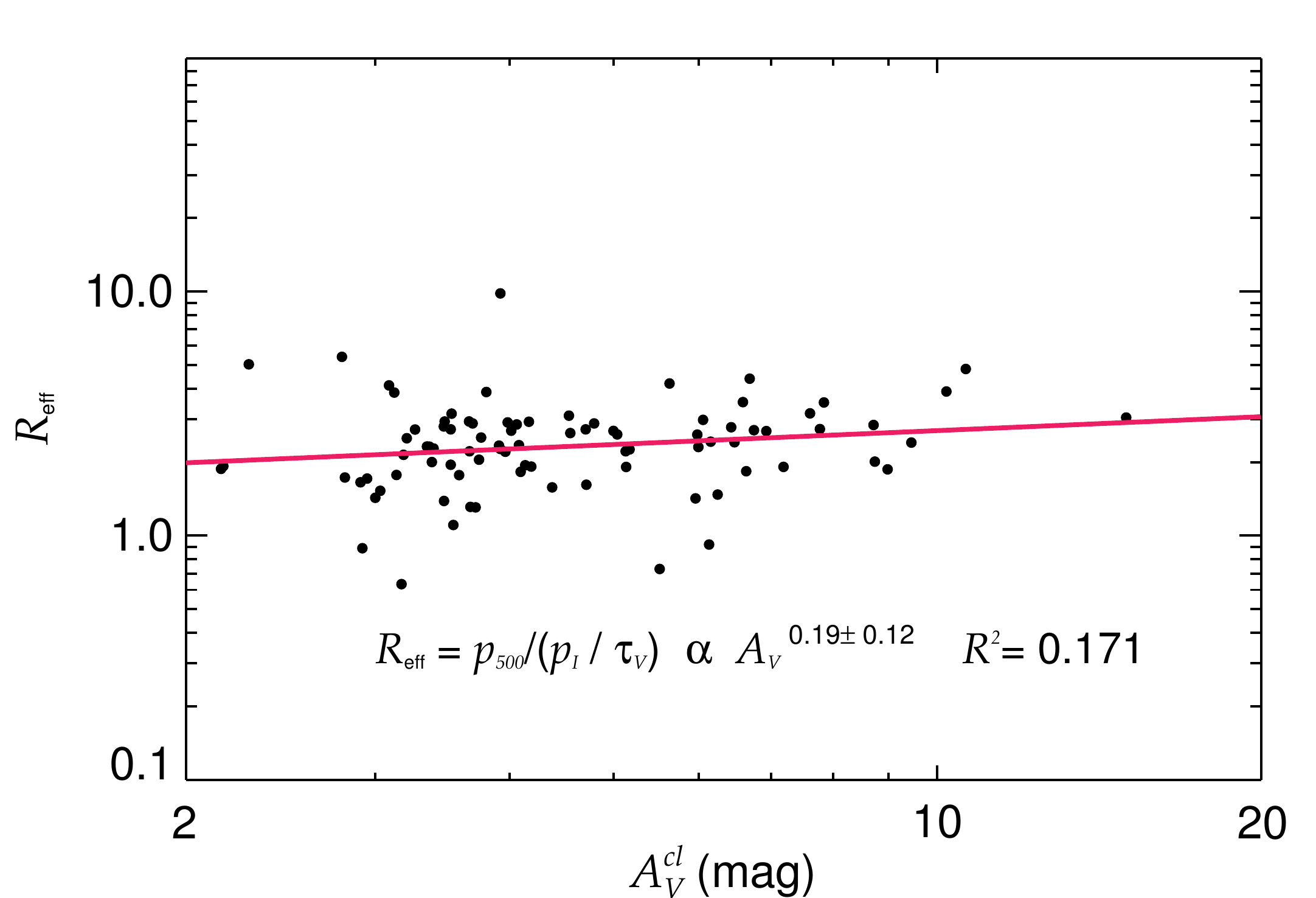} 
   \includegraphics[width=0.48\textwidth]{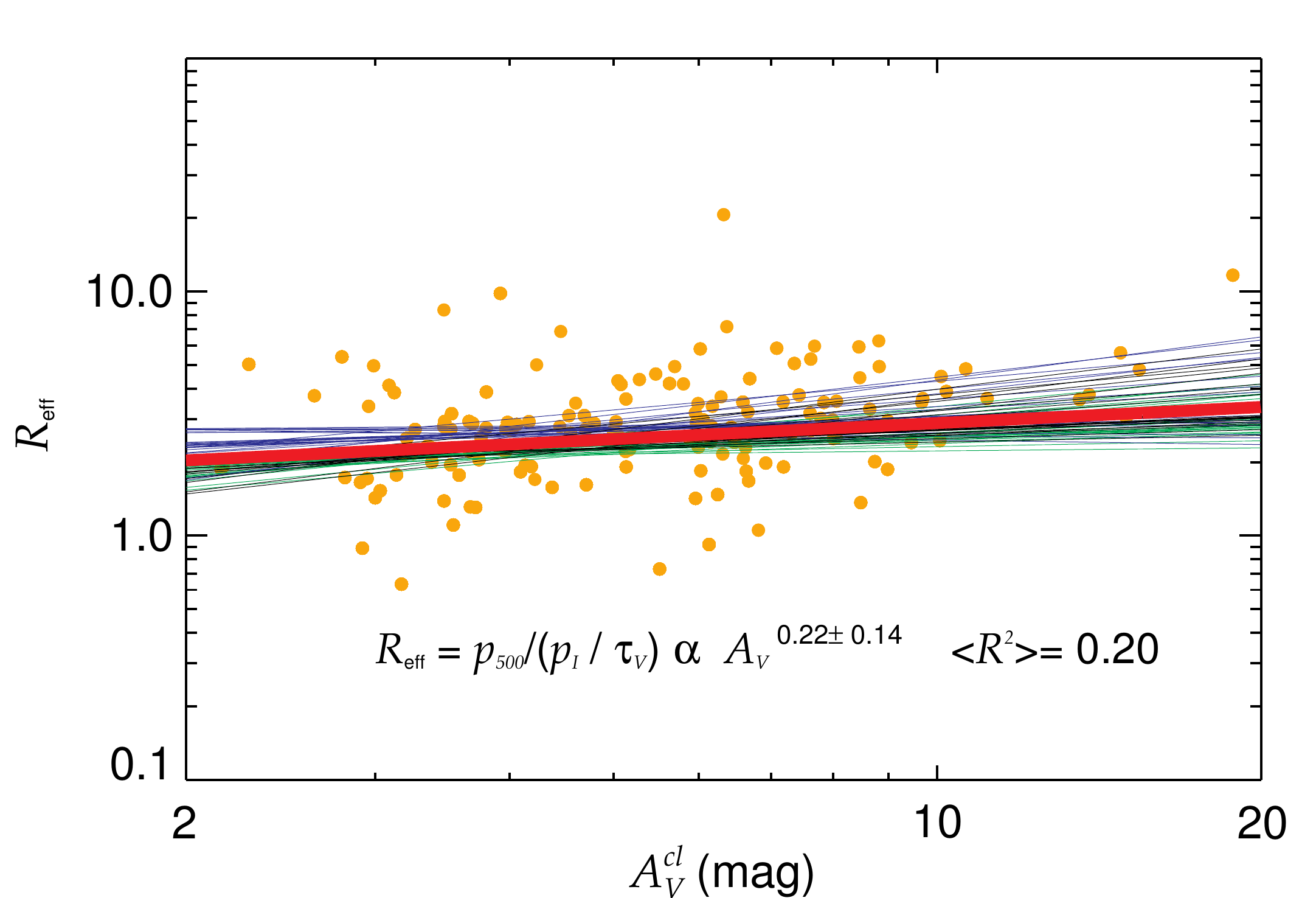} \\
   \includegraphics[width=0.48\textwidth]{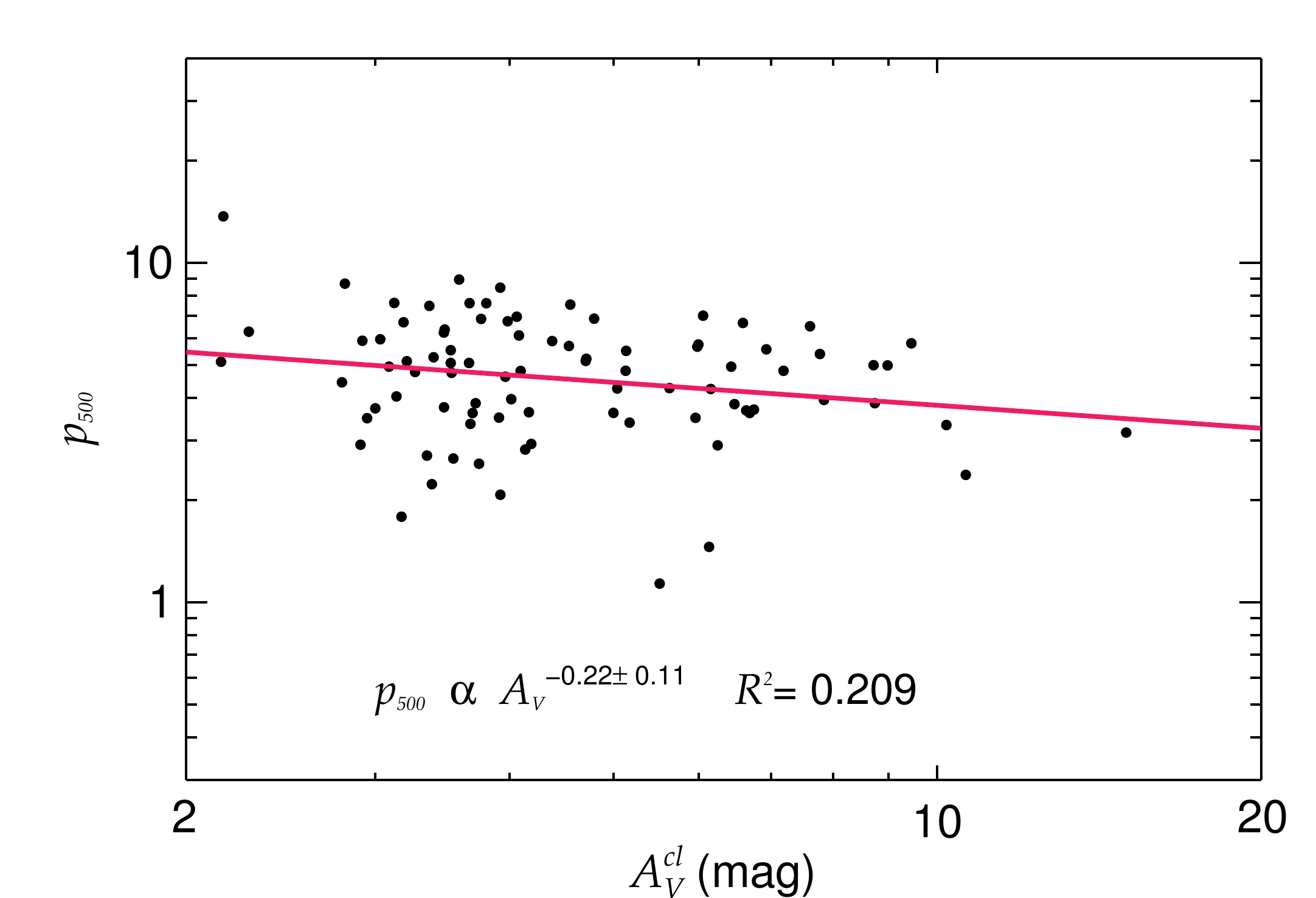}
   \includegraphics[width=0.48\textwidth]{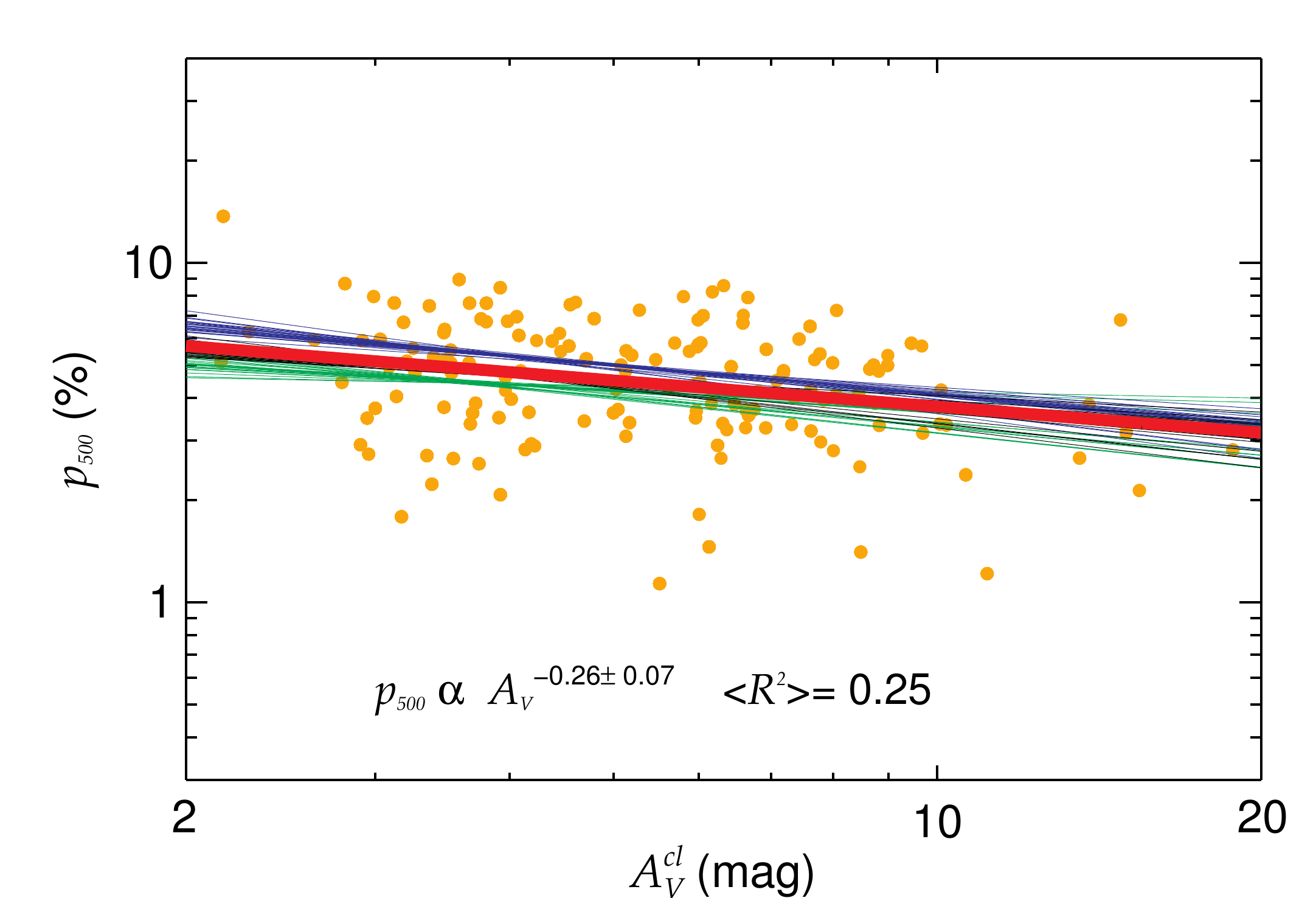} \\
   \includegraphics[width=0.48\textwidth]{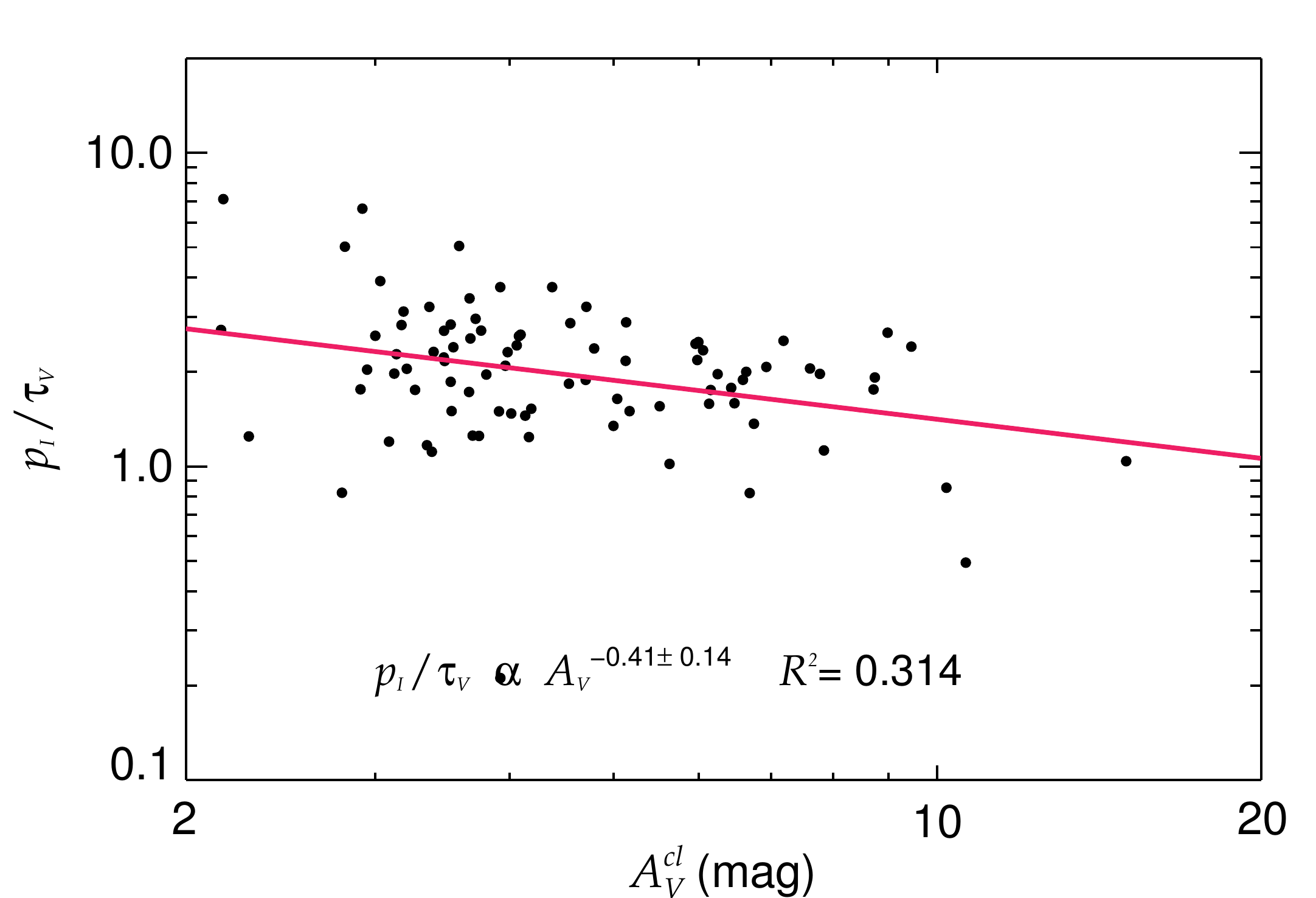}
   \includegraphics[width=0.48\textwidth]{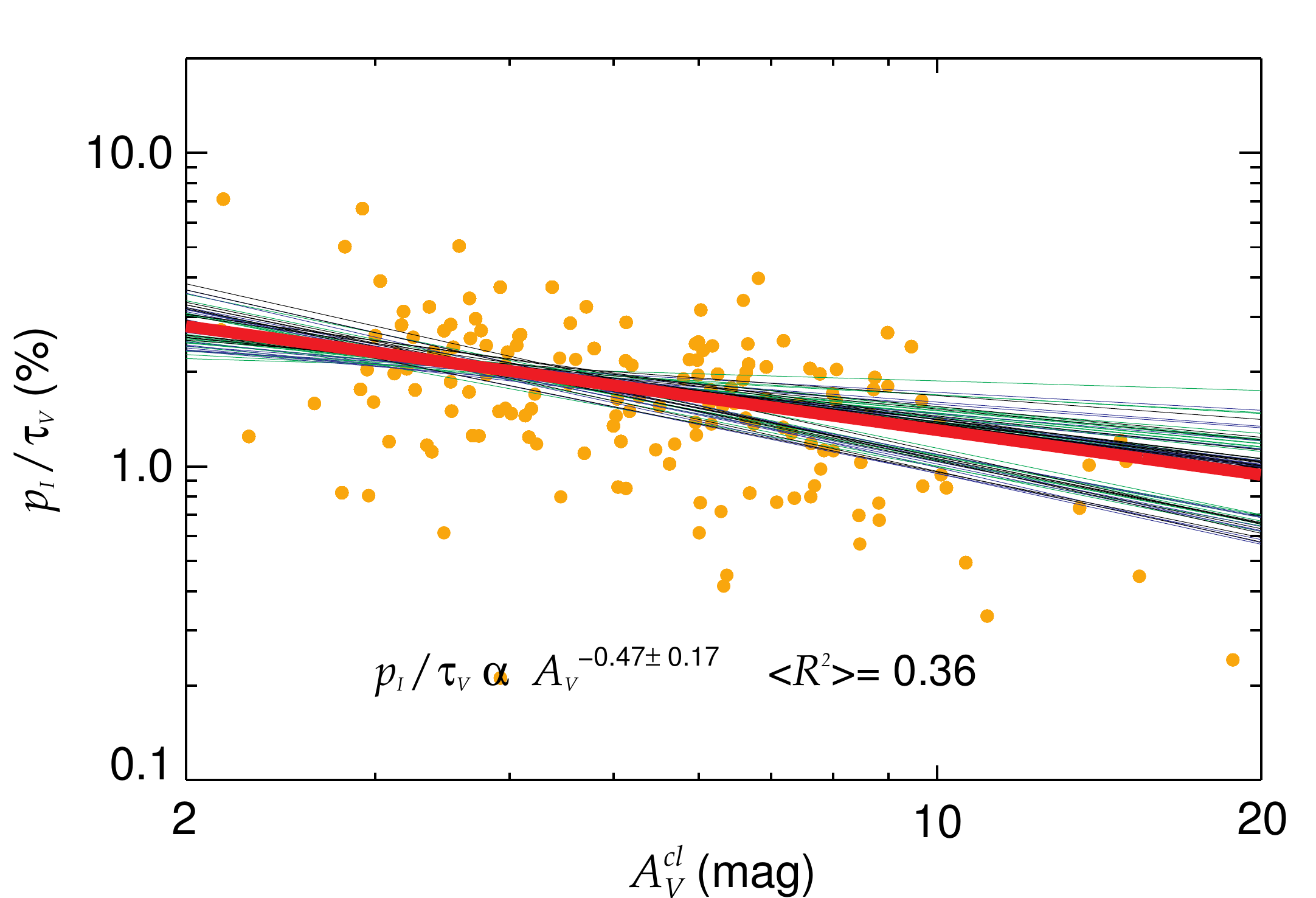} \\
      \caption{\scriptsize Diagrams of \per/$\mathrm{\ vs.\ }$\avcl/ (top, the polarization efficiency ratio 
               $p_{500}/(p_{I}/\tau_{V})$ as a function of cloud extinction),
               $p_{500}$$\mathrm{\ vs.\ }$\avcl/ (middle) and 
               $p_{I}/\tau_{V}$$\mathrm{\ vs.\ }$\avcl/ (bottom), using only the ideal stellar sample 
               in each case. The ideal stellar sample is selected using the GL method. 
               Diagrams on the left are for the standard example. Power-law fits are shown in each case, together with Pearson 
               correlation coefficients ($R^2$). Diagrams on the right account for systematic uncertainties through 
               a series of variations of the GL method input parameters ($N_{\mathrm{d}}, R_{\mathrm{bin}}$, and $N_{\mathrm{max}}$), 
               resulting in slightly different ideal stellar loci and consequently different 
	       fitted curves. The differences between diffuse emission subtraction methods are also accounted for; black,
               blue and green curves are, respectively, for intermediate, aggressive and conservative subtraction methods.
               The red curves are obtained taking the mean (and standard deviation) of the 
               individual power-law exponents for the various individual fitted curves shown. 
               Orange dots represent the ``extended" ideal stellar sample, showing only the points for 
               the intermediate diffuse subtraction method. Details are given in Section \ref{s:reff1} and Appendix \ref{ap:gl}
              }
         \label{f:reff1}
%%%% SHOWING 18 5 2
   \end{figure*}
%-------------------------------------------------------------

Having determined the locus of ideal stars in the \avst/ $\mathrm{\ versus\ }$ \avcl/ diagram, we 
are now in a position to study the polarization efficiency ratio (\per/).
We will refer to the method of data analysis described in this sub-section 
as the ``standard analysis" of polarization properties.
\per/ is here defined as the ratio between polarization fraction at $500\,\mu$m ($p_{500}$) and 
polarization efficiency in the $I$-band, $p_{I}/\tau_{V}$ (where $\tau_{V} =$\avst/$/1.086$ is the 
optical depth):

\begin{equation}
R_\mathrm{eff} = \frac{p_{500}}{p_{I}/\tau_{V}}
\end{equation}

In order to understand how the various relevant quantities varies as we move toward higher 
cloud depths, in Figure \ref{f:reff1} ({\it left}) we show \per/, $p_{500}$ and $p_{I}/\tau_{V}$, respectively,
as a function of \avcl/, using only objects from the ideal stellar sample (Table \ref{tab}). 
For each of these profiles, we also fit a power law (red curve), 
together with a calculation of $R^2$, the Pearson correlation coefficient. 
All points are given equal weight, and the fits are limited to the range $2\,$mag $<$\avcl/$<20\,$mag, 
where most of the data are distributed. 

Although curves in Figure \ref{f:reff1} ({\it left}) might seem sufficient to analyse the polarization 
efficiency ratio and its dependence on \avcl/, the analysis is affected by systematic uncertainties 
that depend on the various choices of input parameters for the GL method and also on the choice of diffuse emission subtraction method
(Section \ref{s:backsub_datasel}). 
The example shown in Figure \ref{f:reff1} ({\it left}) corresponds to a single choice of 
input parameters which we refer to as the ``standard example" (see Appendix \ref{ap:gl}).
As described in Section \ref{s:backsub_datasel}, three alternate types of diffuse emission subtraction were used 
(conservative, aggressive, and intermediate). The choice of
method affects the calculation of \per/, and the resulting uncertainty should be accounted for in the analysis.
A detailed description of our treatment of these systematic uncertainties is given in Appendix \ref{ap:gl}. 
Basically, the GL method is re-applied a number of times, in each case varying a set of input parameters that 
slightly change the resulting locus of ideal stars.
Using slightly different ideal stellar loci in turn
changes the resulting parameters of the power-law fits. 
The diffuse emission subtraction method is also varied.

The results are shown in Figure \ref{f:reff1} ({\it right}), in which each curve is obtained using 
the ideal stellar sample that corresponds to one particular parameter set, with black, blue, and green curves 
corresponding to intermediate, aggressive, and conservative diffuse emission 
subtraction, respectively. Identically to Figure \ref{f:reff1} ({\it left}), we show 
\per/, $p_{500\mu\mathrm{m}}$, and $p_{I}/\tau_{V}$ as a function of \avcl/.
The points in these diagrams (orange dots), show all stars that were 
found inside the ideal stellar locus at least once. We define this combination of points as the 
``extended" ideal stellar sample.
The power-law exponents (and errors) shown in each panel are the averages (and standard
deviations) of the set of exponents obtained for the various individual fits
corresponding to the various choices of input parameters and diffuse emission subtraction
method.
The same 
procedure is applied to obtain the displayed correlation coefficients. 
Using the average exponent values, we draw the average curve (red), which may be viewed as
the best solution, with an uncertainty represented by the range of individual curves.
Notice that 
for each of the three plots on the right side of Figure \ref{f:reff1},
the dispersion in the fitted exponents 
(the standard deviation) is similar to the statistical 
uncertainty in the power law exponents obtained using the standard example fit (left panels).

The GL method was applied in order to avoid including far-background stars 
whose polarization measurements are significantly contaminated by the interstellar material 
of the Galactic disc. For completeness, it is also instructive to understand the effect 
of including far-background objects, by applying the ``standard analysis" of polarimetric 
properties to all strips defined in Figure \ref{f:avavdiag2}.
Section \ref{s:contamback} shows the results 
of this analysis and provides further discussion on background contamination.

\subsection{Determination of the mean polarization efficiency ratio \per/ in the $2<$\av/$<20$~mag range}
\label{s:finalper}

%-------------------------------------------------------------
   \begin{figure}[!t]
   \centering
   \includegraphics[width=0.48\textwidth]{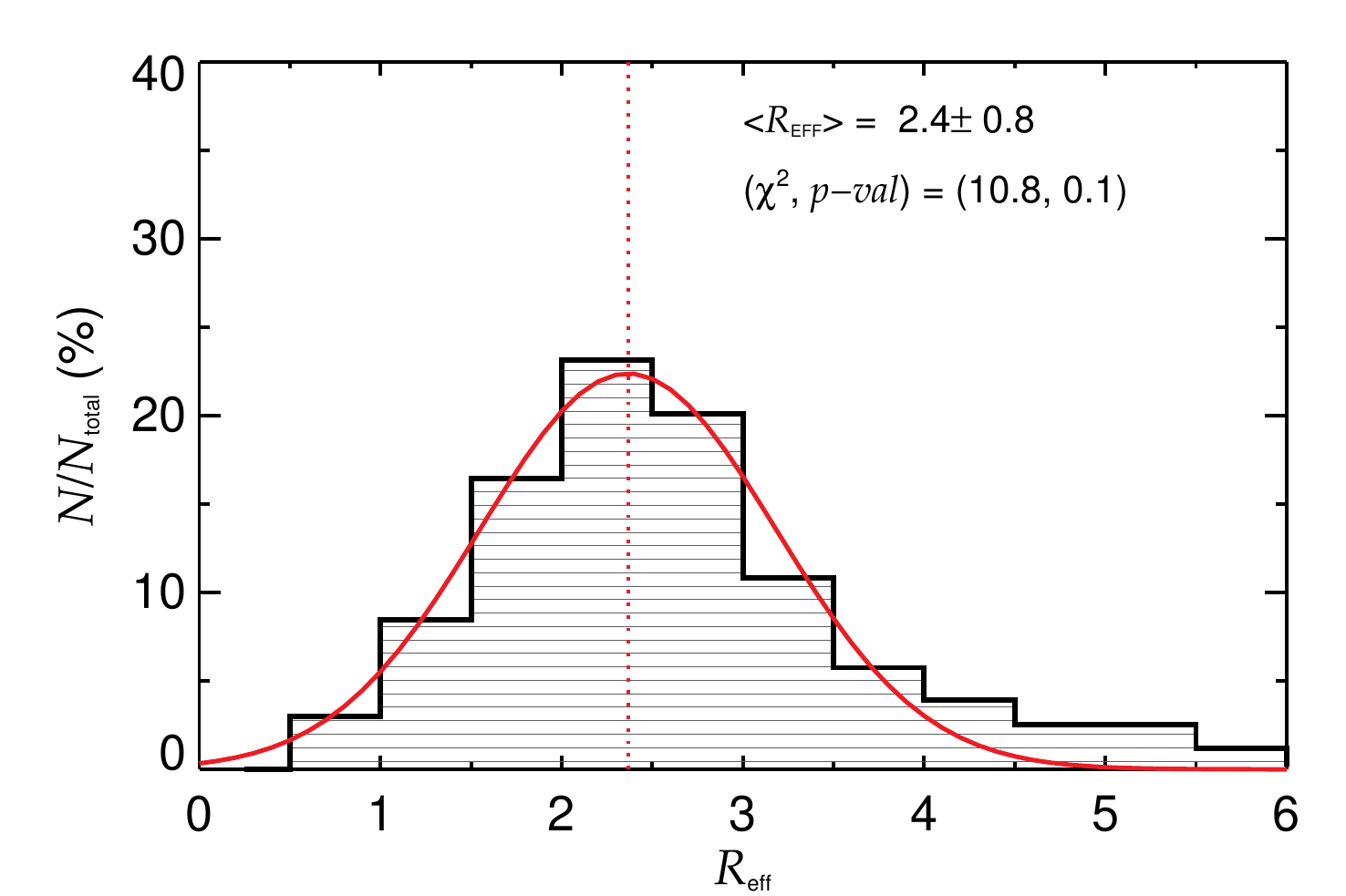} \\
   \includegraphics[width=0.48\textwidth]{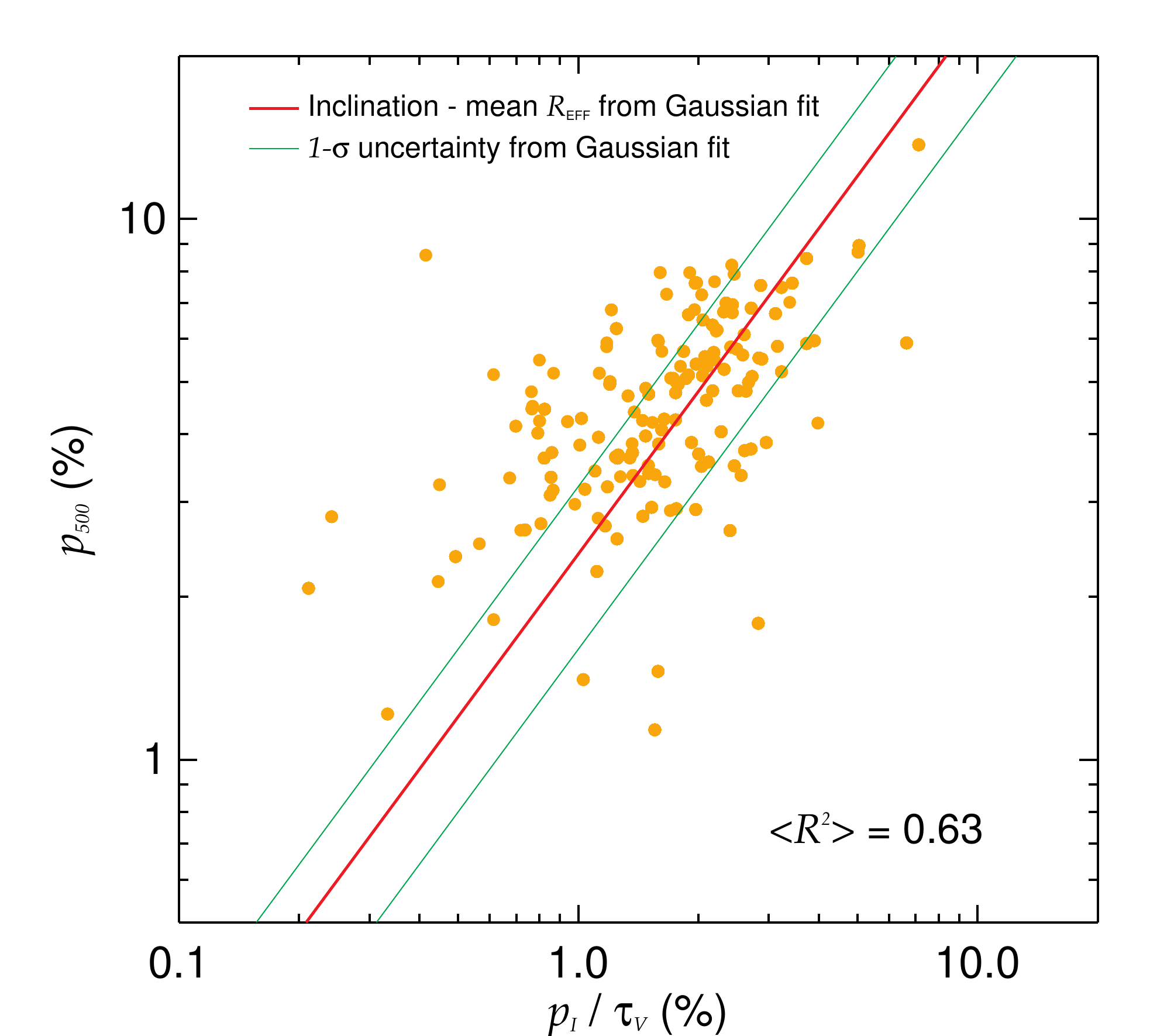} \\
      \caption{Normalized histogram of the polarization efficiency ratio \per/=$p_{500}/(p_{I}/\tau_{V})$ (top) 
	       and a direct comparison between 
	       $p_{500}$ and $p_{I}/\tau_{V}$ (bottom). Both diagrams include objects
	       from the ``extended" ideal stellar sample, but the bottom one is showing 
	       only points for the intermediate diffuse emission subtraction method. The Gaussian fit 
               to the \per/ histogram gives a best estimate of \per/=$2.4\pm0.8$. The 
               chi-square ($\chi^2$) and associated p-value for the Gaussian fit are also shown.
               The positions of the red and green lines in the bottom diagram 
               represent this best value estimate and its uncertainty, respectively, as obtained 
               from the Gaussian fit in the top panel (these lines are not obtained from a
               linear fit, see Section \ref{s:finalper}). 
              }
         \label{f:perhist}
   \end{figure}
%-------------------------------------------------------------

The \per/$\mathrm{\ vs.\ }$\avcl/ curve 
shown in Figure \ref{f:reff1} (top right) has a positive slope, with a power-law 
exponent given by $0.22\pm0.14$. The correlation coefficient is low ($\left\langle R^{2}\right\rangle = 0.20$).
Nominally, our best estimate for the power-law exponent
implies \per/(20 mag)/\per/(2 mag) = 1.7, where
\per/$(x)$ is the value of \per/ at \avcl/$=x$. However, the estimated uncertainty
in the power-law exponent is comparable to the value of this exponent, so the 
positive slope seen in the \per/ vs. \avcl/ curve is not statistically significant. Thus, we will instead interpret
our result as an upper limit on the steepness of this curve, conservatively 
setting bounds of $-0.06$ and $0.50$ on the value of the exponent by respectively
subtracting and adding twice the uncertainty to the nominal value.
We can then express corresponding limits on the overall steepness of the \per/ vs \avcl/
curve as $0.9 < $\per/(20 mag)/\per/(2 mag)$ < 3.2$.

Given the lack of statistically significant changes in \per/, we next proceed to derive
a best estimate for the characteristic mean value of \per/ for the \avcl/
range that we have studied, along with an estimated uncertainty in this
\per/ value. We do this by fitting a Gaussian function to the distribution
of \per/ values shown in Figure \ref{f:perhist} (top), which
%The trend in \per/ is practically flat, in the $2 <$\avcl/(mag)$< 20$ range.
%Given that \per/ is practically constant, we determine its mean value 
%and standard deviation by 
%adjusting a Gaussian fit to the distribution of \per/ values 
%shown in Figure~\ref{f:perhist}~(top), which 
includes all objects
from the ``extended" ideal stellar sample. 
In this histogram, higher weights 
are given to points proportionally to the number of times each star was found 
inside the ideal stellar locus, considering all the systematic variations discussed in Section \ref{s:reff1}.
With that calculation we find \per/=$2.4\pm0.8$, where the uncertainty is here estimated to be equal
to the 1-$\sigma$ width of the distribution. 

Figure \ref{f:perhist} (bottom) shows a direct comparison between  $p_{500}$ and $p_{I}/\tau_{V}$
using the same ``extended" ideal stellar sample. The red and green lines in that diagram
represent the mean \per/ value as well as its uncertainty, respectively, 
as determined from the Gaussian fit in Figure \ref{f:perhist} (top). 
Since the \per/ vs. \avcl/ curve is not perfectly flat (Figure \ref{f:reff1}, top right), 
the distribution of points in the $p_{500}$ vs. $p_{I}/\tau_{V}$ diagram is not expected to 
exactly match the slope of the red line which assumes a direct proportionality
between the two quantities.
Computing the correlation coefficient between 
$p_{500}$ and $p_{I}/\tau_{V}$ for each variation of the GL method, 
and taking the average value, we obtain $\left\langle R^{2} \right\rangle = 0.63$, 
demonstrating that a significant correlation exists between these quantities. 
   
\section{Discussion}
\label{s:discussion}

\subsection{Polarization properties in the $2 <$\av/$< 20$ mag range}

The main goal of this work was to compute the polarization efficiency ratio,
\per/.
In addition to \per/, in Section \ref{s:reff1}, we investigated how 
$p_{500}$ and $p_{I}/\tau_{V}$ vary as a function of \avcl/, in order 
to understand how these two quantities 
separately affect \per/. We found that, in the range $2\,$mag$<$\avcl/$< 20\,$mag, 
both $p_{500}$ and $p_{I}/\tau_{V}$ show decreasing trends, with 
power-law exponents of $-0.26\pm0.07$ and $-0.47\pm0.17$, 
respectively (see Figure \ref{f:reff1}, right panels). 

\citet{2016fissel} presented a detailed analysis of the BLASTPol data set for Vela C, 
including studies of $p_{500}$ as a function of \avcl/. These authors found a decreasing 
trend, corresponding to a power-law exponent
of $-0.45$. The apparent discrepancy between this value and our $-0.26$ exponent probably arises from 
the fact that \citet{2016fissel} studied the entire range of cloud depths between 
\avcl/$\sim2\,$mag and $\sim50\,$mag, whereas in this work we only used 
the subset of BLASTPol data for which correlation with ideal stars was possible. This subset covers
$2\,$mag$ <$\avcl/$< 20\,$mag. It is clear from Figure 12 of 
\citet{2016fissel} that for low column density sight lines, the $p_{500}$ vs. \avcl/ curve is
relatively flatter in comparison to what is seen deeper in the cloud, consistent with 
our observation of a shallower exponent for our lower-density sight lines.

The power law exponents we found for both the submillimeter ($-$0.26) and near-IR ($-$0.47) data are 
comparable to what has been found in similar studies that have been carried out for other 
molecular clouds  
\citep{goodman1995,gerakines1995,matthews2002,whittet2008,chapman2011,cashman2014,alves2014,jones2015}.
These studies used either near-IR or submillimeter data, or sometimes a 
combination of the two, and found exponents generally lying in the range $-$0.3 to $-$1.0. 
\citet{jones2015} show that for the largest column densities the exponents tend to be more 
negative, consistent with our qualitative observations concerning Figure 12 of \citet{2016fissel}.

The well-known tendency for \pem/ and \pextauv/ to decrease with column density has been 
modeled in several papers.  For example, it has been interpreted as an effect of 
turbulence \citep{falceta2008}, as a loss of grain alignment toward well-shielded 
regions \citep{whittet2008}, or as a combination of both effects \citep{jones2015}.

In Section \ref{s:finalper}, we showed that \per/ shows 
no statistically significant changes with \avcl/, over the sampled range of $2\,$mag$ <$\avcl/$< 20\,$mag.
As previously discussed, if we assume that the same population of dust grains distributed 
along the LOS is producing both polarized emission and polarization by extinction, \per/
should depend only on intrinsic grain properties. 
In quiescent molecular clouds, grain processing effects, such as growth due to coagulation, 
may take place as one goes deeper into the cloud \citep{draine2003,jones2004}.
Potentially, these could affect \per/.
The relatively flat profile of \per/ vs. \avcl/ that we have found for Vela C, 
implies that any changes in \per/ that are occurring in the cloud due to changes
in dust grain properties with increasing \avcl/ must be too small for us to detect
given our sensitivity limitations. From the discussion in Section \ref{s:finalper}, 
we conclude that no change in \per/ larger than about a factor of three
is occurring over the \avcl/ range sampled.
%the following explanations may be 
%given: (a) no significant modifications in grain morphology or chemical composition occurs 
%in this density range, but perhaps more substantial changes in dust grains would occur 
%deeper into the cloud, at larger density ranges; or (b) changes in dust grain properties 
%are taking place in the extinction range probed by our study, but the effect of these 
%changes on \per/ is too small for us to detect, given our systematic and statistical uncertainties.
We cannot rule out the possibility that much larger changes in \per/ occur
for \avcl/$ > 20\,$mag.

\subsection{Observations and predictions of \per/ for the diffuse ISM}

Our value of $2.4\pm0.8$ for $p_{500}/(p_{I}/\tau_{V})$ can in principle
be compared against the prediction of dust grain models that are able to link the 
polarization extinction and emission spectra. 
%by adjusting different 
%parameters such as grain shapes and composition.
\citet{draine2009} present four such models, in which
observed polarization spectra at optical/near-IR wavelengths are used as
inputs \citep{serkowski1975,martin1992}. One of the output products of the model 
is the polarized emission spectrum, allowing a direct comparison of \pem/
with \pextauv/ for specific wavelengths. 
However, these models were designed to reproduce 
the conditions of the diffuse atomic ISM, so a direct comparison with our 
results is problematic.
The {\it Planck} collaboration carried out a
comparison between polarized emission at $850\,\mu$m and published $V$-band starlight polarization 
for diffuse-emission sight-lines \citep{planckxxi2015}.
A polarization efficiency ratio of $p_{850}/(p_{V}/\tau_{V}) = 4.2\pm0.3$ was obtained,
which may be compared with \citet{draine2009} predictions of
$2.9 - 4.1$ for the same quantity.

\citet{planckxxi2015} also determined $P_{850}/p_{V}$ (where $P_{850}$ is the 
polarized {\it flux} at $850\,\mu$m), which also may be compared
against models. As pointed out by \citet{planckxxi2015}, this quantity is easier
to measure since it is independent of the typical systematic uncertainties that 
affect $\tau_{V}$. In the case of molecular clouds, however, the emission depends
on grain temperature \citep[which usually decreases for higher densities;][]{2016fissel}, so
$P_{500}/p_{I}$ is also expected to vary. Therefore, in this 
work we focused only on \plambdapitauv/.

Despite the above-mentioned mismatch between our observations and the model of
\citet{draine2009}, we will compare our value for $p_{500}/(p_{I}/\tau_{V})$,
which is $2.4\pm0.8$, to the 
\citet{draine2009} predictions, which are $\sim3.3$ for diffuse ISM models 
in which both carbonaceous and silicate grains are aligned, and $\sim4.6$ 
when only alignment by silicate grains are considered. This range of values 
was obtained by combining Figures 8 and 6 from \citet{draine2009}.
%which
%provide prediction values of $p_{500}$ between $8.4$ and $11.8\%$, respectively, and uses a
%normalization factor of $p_{I}/\tau_{V}\approx2.55\%$.
%Although our measurement is close to the models' predictions, 
%there is a significant difference (considering the uncertainty), suggesting 
%that models suited for specific molecular cloud conditions are needed for a reliable comparison. 
Of course we cannot draw scientific conclusions from this comparison, but we will
note that the spread in the model \per/ values is comparable to the uncertainty
in our observed value of \per/. Thus, if corresponding models for molecular cloud
dust grains become available, and if there are similar spread in \per/ values 
among the models, then with modest reduction in the observational uncertainties
it will become possible to discriminate among the models using observed \per/ values.

\subsection{Diffuse ISM far-background contamination in the determination of \per/}
\label{s:contamback}

\begin{figure}
   \centering
   \includegraphics[width=0.48\textwidth]{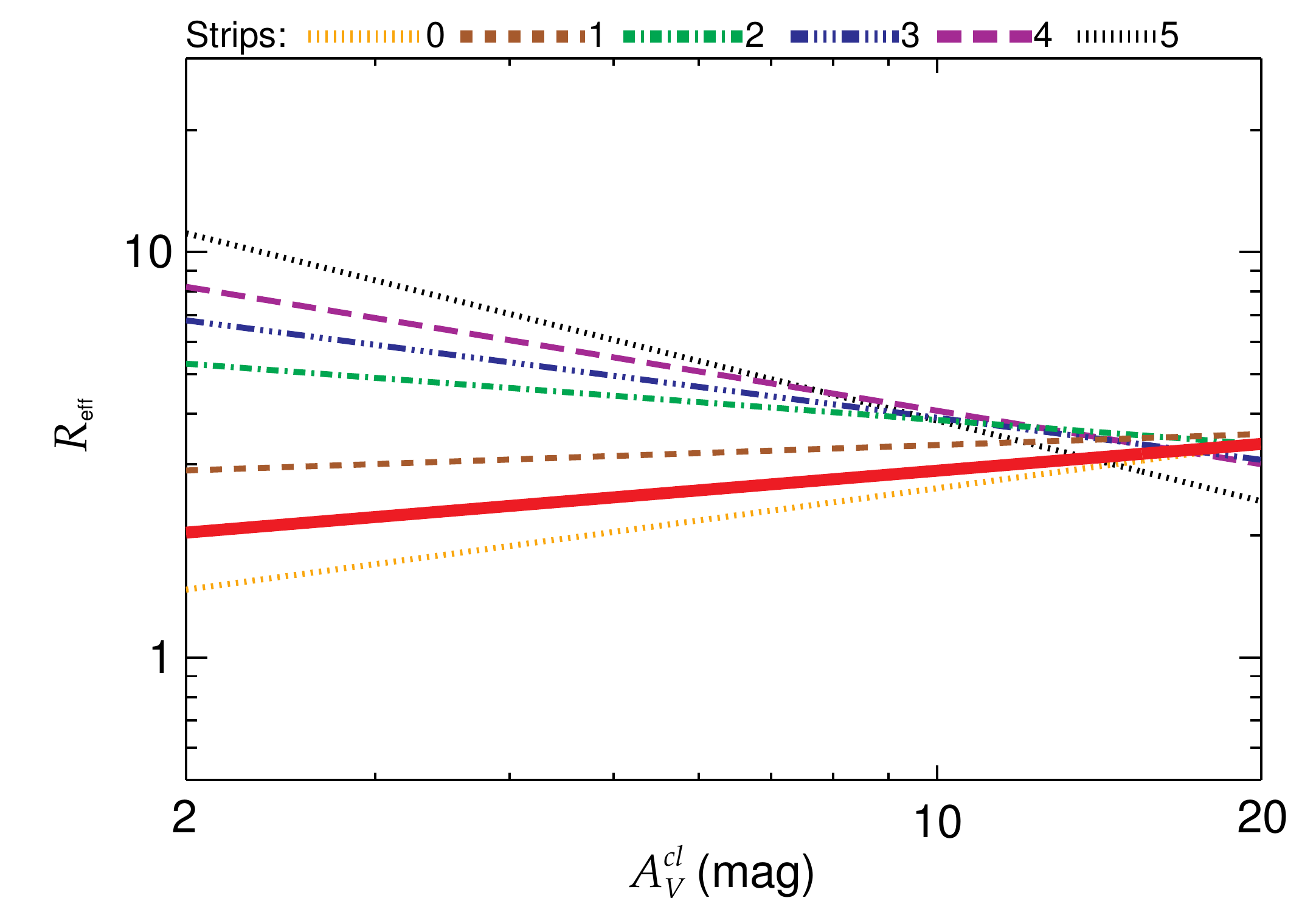} 
   \includegraphics[width=0.48\textwidth]{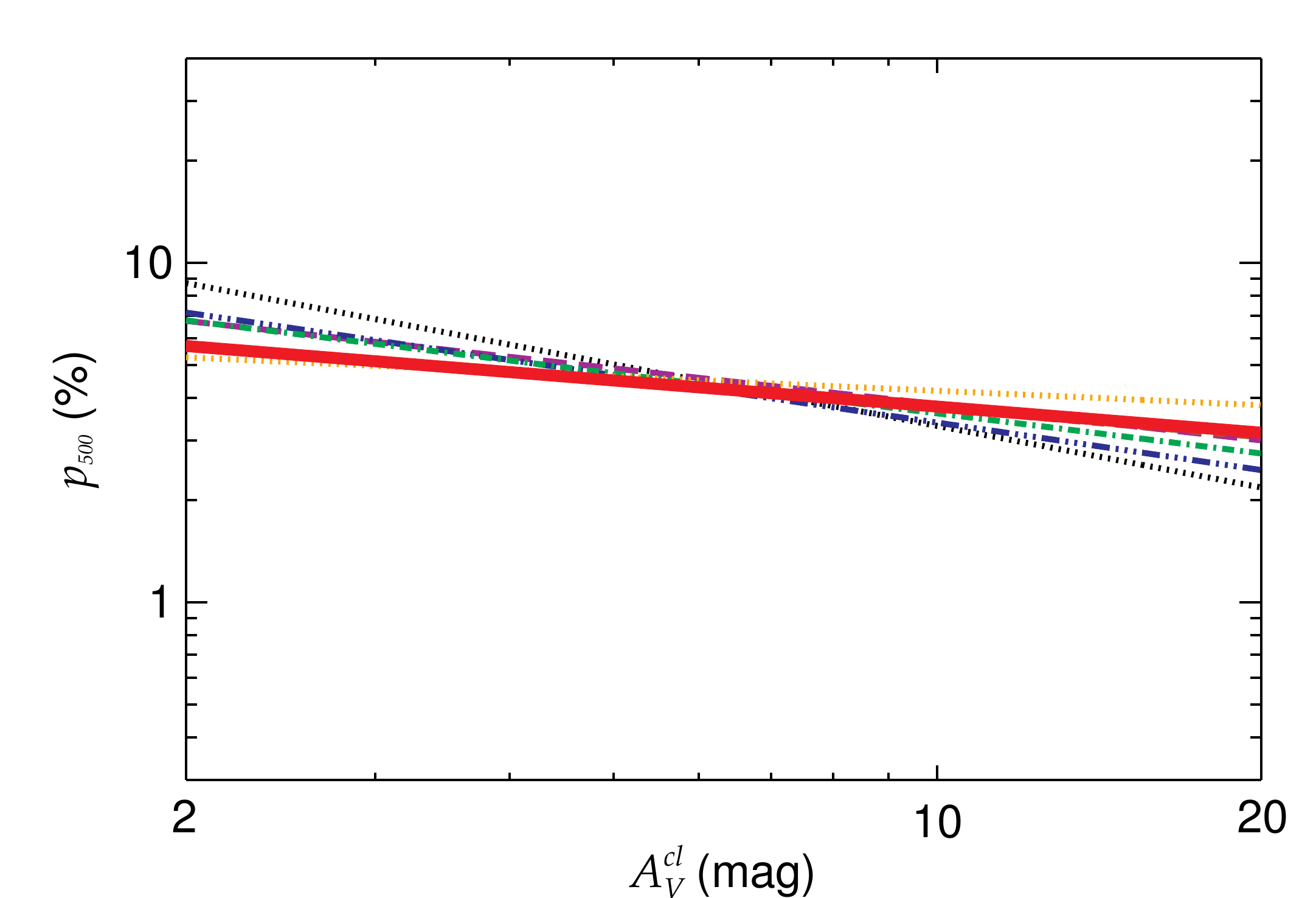} \\
   \includegraphics[width=0.48\textwidth]{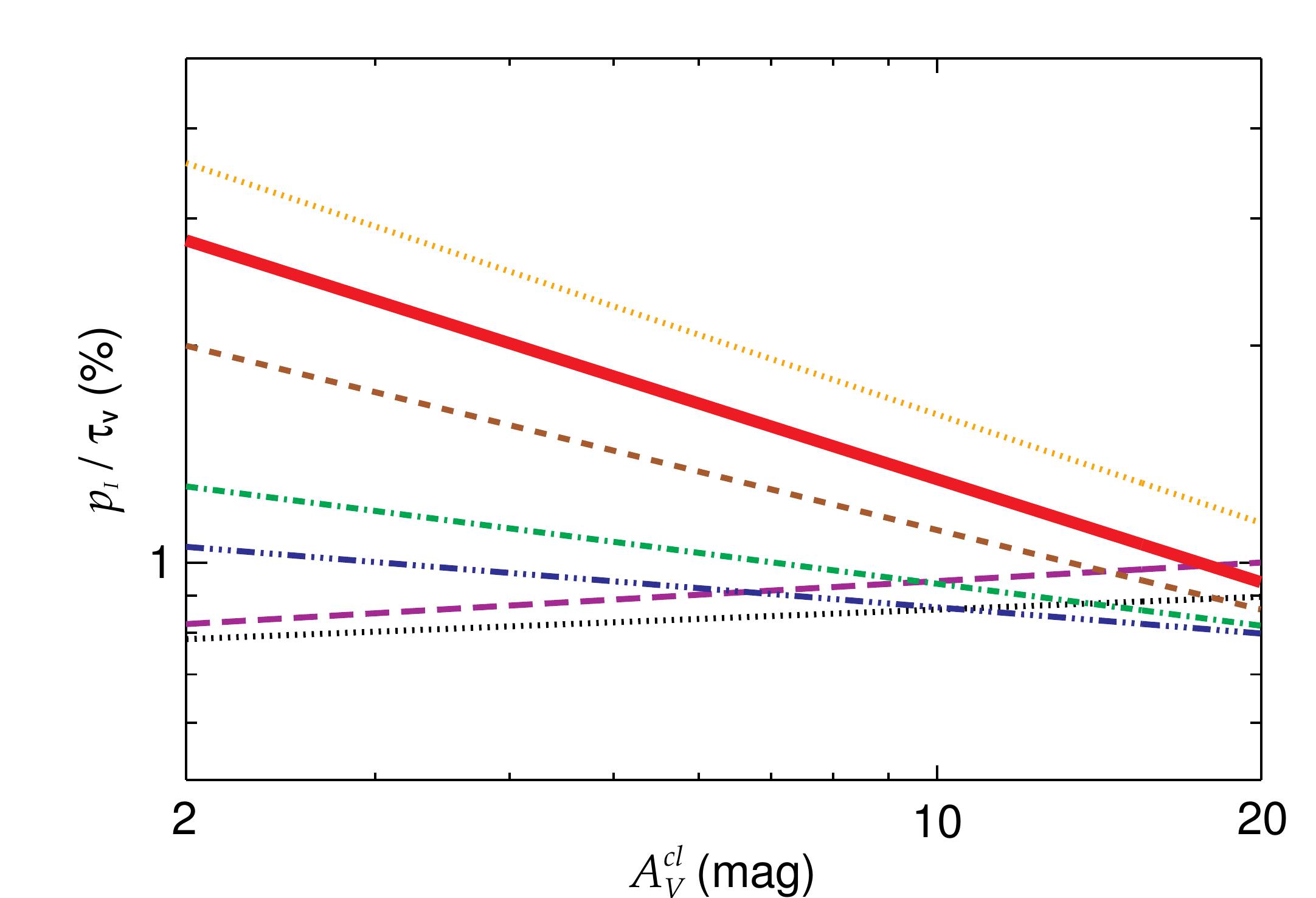} \\
      \caption{Average curves of polarization parameters as a function of cloud extinction 
	      \avcl/, for  objects located within Strips 0 to 5 (see Figure \ref{f:avavdiag2}).
	      Top left, top right and bottom panels show curves for the polarization efficiency ratio 
	      \per/=$p_{500}/(p_{I}/\tau_{V})$, $p_{500}$ and 
               $p_{I}/\tau_{V}$, respectively. The average curves in each case are obtained after 
	       applying the ``standard analysis" described in Section \ref{s:reff1} and in Appendix \ref{ap:gl}.
               The average curves for the ideal stellar locus (red) are shown for reference, and are identical
               to the ones shown in Figure \ref{f:reff1} ({\it right}).
              }
         \label{f:reffback}
   \end{figure}

The ``standard analysis" of polarization properties (described in 
Section \ref{s:reff1}) was applied to strips 0 to 5. 
As before, each variation of the GL method (see Appendix \ref{ap:gl}) leads to 
different power-law fits of \per/$\mathrm{\ vs.\ }$\avcl/, $p_{500}$$\mathrm{\ vs.\ }$\avcl/, and 
$p_{I}/\tau_{V}$$\mathrm{\ vs.\ }$\avcl/. Just as for the ideal strip, 
by taking the average of the power-law exponents
an ``average curve" is obtained for each strip.
Figure \ref{f:reffback} shows the results of this analysis, in which all curves correspond 
only to the average curve obtained for each strip (different colors and line styles are associated 
with different strips, according to the label at the top). The red curves are the same as displayed 
in Figure \ref{f:reff1} ({\it right}), corresponding to the ideal strip. 

%-------------------------------------------------------------
%-------------------------------------------------------------

Figure \ref{f:reffback} (top) shows that the inclusion of stars contaminated 
by the background material significantly affects the analysis of \per/. 
It is obvious that for strips 2 to 5, for which stars are 
increasingly contaminated by the far-background ISM, the \per/$\mathrm{\ vs.\ }$\avcl/
curves are all displaced toward higher values in comparison with the curve from the ideal 
strip (red curve). This is especially true for lower cloud extinctions. Strips 0 and 1 were defined 
to be below and above
the ideal strip (see Figure \ref{f:avavdiag2}), respectively, but also sharing a subset 
of stars located in this strip. Therefore, they appear to suffer mildly from the 
displacement effect (they can be regarded as lower and upper limits
to the ideal strip curve), and similarly to the ideal strip curve, they also
exhibit a slowly increasing trend with \avcl/.

When the curves for $p_{500}$ and $p_{I}/\tau_{V}$ are analysed separately 
(Figure \ref{f:reffback}, respectively middle and bottom), it becomes 
obvious which of these two parameters are most affected by the background 
contamination. The submillimiter polarized emission alone should not depend on 
the degree to which stars are
contaminated by the background material. As expected, 
all $p_{500}\mathrm{\ vs.\ }$\avcl/ curves overlap, with only slight 
variations. For $p_{I}/\tau_{V}\mathrm{\ vs.\ }$\avcl/, however, 
the curves corresponding to the various strips show very
significant differences in behavior. 
On the one hand, strips closer to the cloud (0, 1, and ideal) show the characteristic decrease
with \avcl/. On the other hand, objects increasingly affected by the far-background 
ISM (represented by strips 2 to 5) show lower $p_{I}/\tau_{V}$ values, 
and profiles more consistent with a flat trend, as a function of \avcl/.
If the radiation from far-background sources is being affected by extra layers 
of interstellar material behind the cloud, then while the column
density (and therefore $\tau_{V}$) is expected to be higher, $p_{I}$ 
should not necessarily increase in direct proportion, because different layers
could have different magnetic field orientations. This scenario is consistent
with the lower $p_{I}/\tau_{V}$ values observed in strips 2 to 5.

It is interesting to notice, however, that even considering 
the background contamination for strips 2 to 5, in 
Figure \ref{f:reffback} (top), all \per/$\mathrm{\ vs.\ }$\avcl/ curves 
seem to converge at the highest cloud extinctions probed by our sample.
This suggests that at increasingly higher cloud extinctions, because the relative 
amount of cloud material along the LOS is large compared to the background 
diffuse ISM, the presence of background contamination becomes negligible for 
the purposes of calculating the polarization efficiency ratio. 
In addition, this shows that for higher extinctions, even if clumping and beam 
averaging effects become important (see Appendix \ref{ap:avstuncertainties}), 
this does not impact the calculation of \per/ in a significant way.
The convergence at \per/ values close to $2.4$ for 
all curves at higher extinctions provides
extra confidence that the application of the GL method was successful 
in determining the ideal subset of stars used for this work.

\section{Conclusions}
\label{s:conclusions}

We have carried out the first large-scale quantitative comparison of near-IR and 
submillimeter polarization magnitudes measured toward the same molecular cloud.  Our aim was to 
study the polarization efficiency ratio, which provides a constraint for physical grain models. 
For the Vela C molecular cloud, we combined polarized emission data from BLASTPol at 
$500\,\mu$m with starlight polarimetry in the $I$-band. We also used 
complementary data from 2MASS, {\it Herschel} and {\it Planck}. The main
conclusions are summarized below.

\begin{enumerate}

\item The average polarization efficiency ratio ($R_\mathrm{eff} = p_{500}/(p_{I}/\tau_{V}))$
        is found to be $2.4\pm0.8$ for cloud visual extinctions between $\sim2\,$mag and $\sim20\,$mag.
This value can be used to test dust grain models designed specifically for the environment 
found inside molecular clouds;
%The value determined for \per/ is significantly different from the predictions of models
%designed for the diffuse ISM.  
%Dust grain models designed specifically for the environment found inside molecular clouds are needed
%to make a reliable comparison, including 
%the effects of varying density, temperature, exposure to radiation, and other physical conditions associated with higher 
%depths.
\item We have examined the dependence of \per/ on cloud visual extinction and we find no
statistically significant deviations from a flat trend over the range of extinctions
probed;
%Taking into account our sensitivity limitations, the dependence of \per/ with cloud
%extinction in the range between $2\,$mag and $\sim20\,$mag shows a relatively flat trend, 
%suggesting that no significant modifications in grain properties take place in the corresponding
%range of cloud depth.
%We cannot rule out the possibility that more significant changes to grain properties are taking
%place at cloud depths greater than $20\,$mag;
%Alternatively, the changes in polarization efficiency ratio might be too small for us to detect, 
%given the statistical and systematic uncertainties.
\item The polarization efficiency ratio is shown to vary significantly if far-background objects (contaminated by the
        diffuse background ISM) are included. This effect highlights the importance of selecting suitable stellar objects, such 
that the columns of material probed by polarized extinction and emission are similar. Nevertheless, 
we find that at higher cloud extinctions, the effect of the background contamination is negligible, 
since the relative contribution from the molecular cloud itself is dominant.
\end{enumerate}

The type of study conducted here would significantly benefit from more precise distance determinations.
Complementary data sets that could improve the near-IR versus sub-mm polarimetric comparison include 
products from trigonometric distance surveys such as the next {\it GAIA} data releases. 
In addition, as previously mentioned, dust grain models specifically developed for molecular clouds
are needed for a meaningful comparison. Grain models that are suitable for predicting \per/ in this environment
have yet to be developed, and would be valuable tools for understanding 
which particular changes in grain properties are taking place in molecular clouds,
thereby affecting the polarization efficiency ratio.

\vspace{0.5cm}
\acknowledgements
We are grateful to the anonymous referee for the very valuable suggestions and comments.
The BLASTPol collaboration acknowledges support from NASA (through grant numbers NAG5-12785, NAG5-13301, NNGO-6GI11G, NNX0-9AB98G, and the Illinois Space Grant Consortium), the Canadian Space Agency (CSA), the Leverhulme Trust through the Research Project Grant F/00 407/BN, Canada's Natural Sciences and Engineering Research Council (NSERC), the Canada Foundation for Innovation, the Ontario Innovation Trust, and the US National Science Foundation Office of Polar Programs. C. B. Netterfield also acknowledges support from the Canadian Institute for Advanced Research. F.P.S. was supported by the CAPES grant 2397/13-7. We thank the Columbia Scientific Balloon Facility (CSBF) staff for their outstanding work. 
F.P. thanks the European Commission under the Marie Sklodowska-Curie Actions within the H2020 program, 
Grant Agreement number: 658499 -- PolAME -- H2020-MSCA-IF-2014.
We thank the staff of OPD/LNA (Brazil) for their invaluable help during our observing runs. 
This investigation made extensive use of data products from the Two Micron All Sky Survey (2MASS), 
which is a joint project of the University of Massachusetts and the Infrared Processing and 
Analysis Center/California Institute of Technology, funded by the National Aeronautics and 
Space Administration and the National Science Foundation.  
We are grateful to Drs. A. M. Magalh\~aes and A. Pereyra for providing the polarimeter 
and software used for the near-IR data reduction.

{\it Facilities:} 
\facility{BLASTPol},
\facility{LNA: 1.6\,m, LNA: 0.6\,m}.

%%%%%%%%%%%%%%%% REFERENCES %%%%%%%%%%%%%%%%%%%%%%%%
\bibliography{astroref}
\bibliographystyle{aa}

\appendix

\section{Factors affecting the stellar extinction (\avst/) distributions}
\label{ap:avstuncertainties}

Below we list three factors that may explain the relatively wide distribution 
of \avst/ values observed toward regions having fixed cloud extinction \avcl/, 
when considering non-foreground stars in the wide photometric field data set
(e.g., Figures \ref{f:avavdiag}, \ref{f:compavall}, and \ref{f:hist}).
Our aim is to identify which factor is dominant.

%In general we may identify three main factors that affect the \avst/ values, impacting
%how we interpret \avst/ distributions in this work (e.g., Figures \ref{f:compavall} and \ref{f:hist})
%and how we compare \avst/ with the cloud extinction \avcl/.
%The purpose here is to evaluate which of these effects is dominant in the context
%of the wide photometric field data set, specially in the off-cloud positions (low \avcl/).

\begin{figure}[!t]
   \centering
   \includegraphics[width=0.50\textwidth]{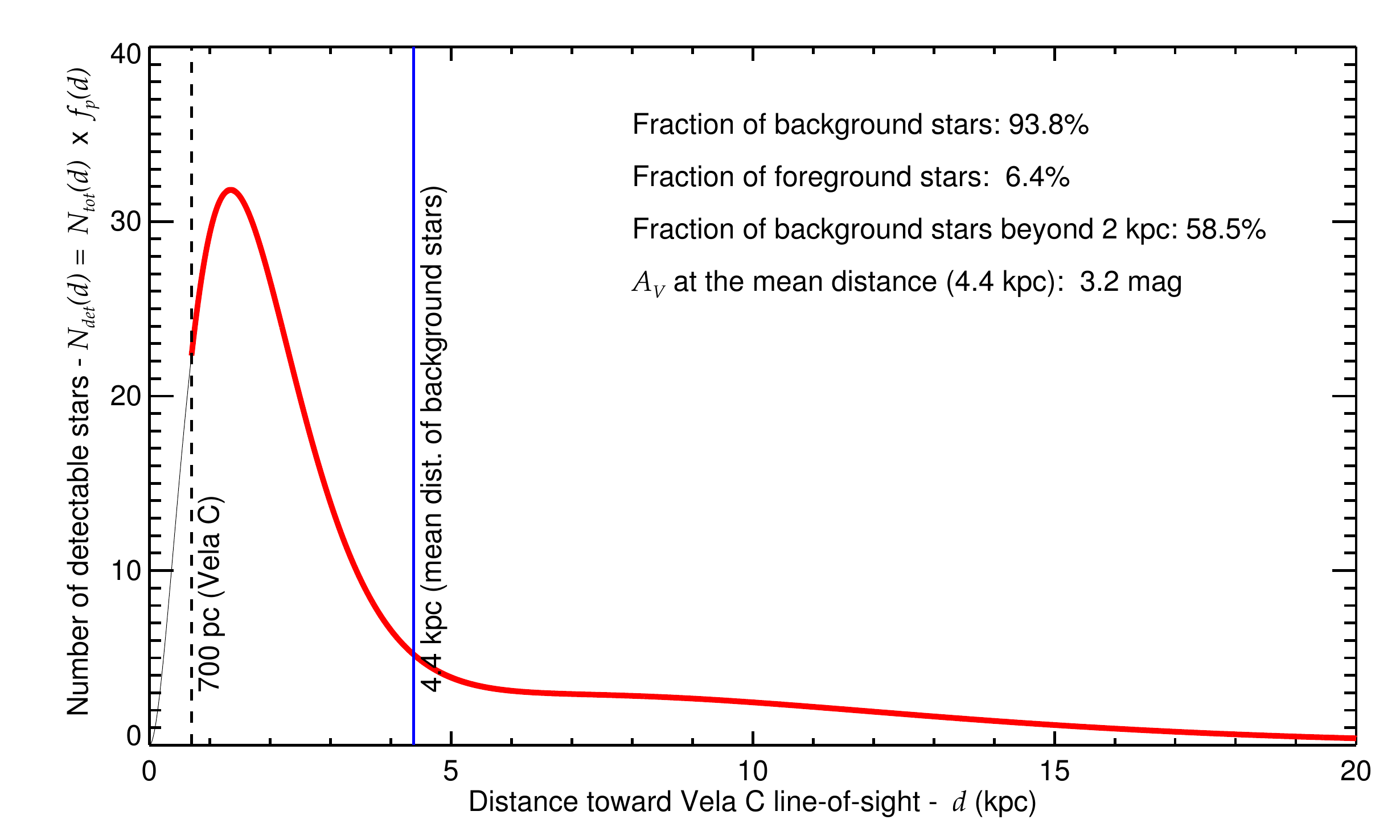} \\
   \caption{Number of detectable 2MASS stars ($N_{det}$) as a function of distance $d$
           toward Vela C, according to estimates from Galactic stellar distribution 
           models combined with the estimated photometric sensitivity of 2MASS.
           Extinction information used to obtain this curve is 
           derived from the Galactic extinction model of \citet{amores2005}
           in combination with results from {\it Planck} (see discussion in Appendix \ref{ap:avstuncertainties}).
              }
         \label{f:fracs}
   \end{figure}

\begin{enumerate}

\item Photometric errors or uncertainties associated with assumptions used in the \avst/ 
calculation method (Section \ref{s:avst}): The statistical uncertainties \eavst/ derived solely through 
propagation of errors (from 2MASS $J$, $H$ and $K_{s}$ magnitudes) typically range between $0.3$ and 
$0.8\,$mag. These correspond to lower limits for the true uncertainties, which may also be 
affected by systematic effects such as variations in grain properties (which influence the 
total-to-selective extinction ratio and consequently the conversion between $E(J-K_{s})$ and 
\avst/, as discussed in Section \ref{s:avst}) and uncertainties in intrinsic colors. 
The combined statistical and systematic uncertainties in \avst/ are $\sim1.5\,$mag, 
as discussed in Section \ref{s:gl} and in Appendix \ref{ap:gl}.

\item Stars distributed over a range of distances and thus possibly contaminated by far-background material:
It is possible to estimate statistically the fraction of stars in our wide photometric field sample 
that are located respectively in the foreground and background, by combining 
a simple model of the Galactic stellar distribution with information regarding the 2MASS sensitivity for our sample. 
A standard stellar distribution model \citep{1986bahcall} gives the 
total number density of stars $n_{tot}$ as a function of the Galactic radius $r$ 
and perpendicular distance from the Galactic plane $z$:

\begin{equation}
        n_{tot}(r,z) = n(R_{0}) e^{-z/z_{0}} e^{-(r-R_{0})/h}
\end{equation}

\noindent where $R_{0}$ is the distance from the Sun to the Galactic center ($\approx 8\,$kpc), $n(R_{0})$ 
is the number density of stars in the solar neighborhood ($\approx 0.13$ stars/pc$^3$), $z_{0}$ is the 
scale height ($\approx 250\,$pc), and $h$ is the disk scale length ($\approx 3.5\,$kpc).
This equation can be re-written in terms of $d$, defined as the distance from the Sun along the Vela C line-of-sight,
using $r^{2} = R_{0}^{2} + d^{2} - 2dR_{0}\cos(l)$ and $z = d \sin(b)$ (where $l$ and $b$ are the cloud's Galactic 
longitude and latitude, respectively taken as $266\degr$ and $1\degr$).
The total number of stars in a field-of-view of area $A(d)$ may be found by integrating the function
$N_{tot}(d) = A(d) n_{tot}(d)$ along $d$. 
Finally, the actual number of stars detected by 2MASS is given by $N_{det}(d)$ = $f_{p}(d) N_{tot}(d)$, where $f_{p}(d)$
is the total fraction of stars that are detectable given the 2MASS sensitivity, as a function of 
distance $d$ (see below).

Using standard methods \citep[e.g.,][]{santos2012}, we find that for the wide photometric field the 
photometric completeness limits are given by $(J_{\mathrm{cl}},H_{\mathrm{cl}},Ks_{\mathrm{cl}})$ = (15.3, 14.5, 14.3).
The $J$ and $H$ values are slightly smaller than the canonical 2MASS limits 
\citep[given by 15.8 and 15.1, respectively, ][]{skru2006}, due to the fact that our sample selects only 
stars with ``AAA" 2MASS photometric quality and excludes points outside the reddening band (see Section \ref{s:avst}).
Using these completeness limits (referred to as
$m_{\mathrm{cl}}$), the maximum distance $d_{\mathrm{max}}$ that a star of a given spectral type and luminosity class
can be detected may be obtained through $m_{\mathrm{cl}} - M_{\lambda} = 5\log(d_{\mathrm{max}}) - 5 + A_{\lambda}$,
where $M_\lambda$ is the intrinsic magnitude \citep{koornneef1983,carpenter2001,wegner2007} and $A_\lambda$ is the extinction
at each band \citep[converted to \av/ using canonical relations, ][]{fitzpatrick1999}.
Combining $d_{\mathrm{max}}$ with the information on the typical fractions for each stellar type 
\citep[e.g.,][]{ledrew2001} we find that the fraction function for our sample is approximately given by 
$f_{p}(d) = e^{-0.0014\,d} + 0.0018$. 
To derive this curve, we also use models of Galactic extinction
from \citet{amores2005} to estimate the extinctions
of diffuse matter as a function of distance toward Vela C.
The saturating extinction levels at high distances ($\approx 20\,$kpc) given by this 
model ($\approx3.3\,$mag) do not agree with nearby
off-cloud extinctions derived from {\it Planck} for the same 
Galactic latitudes ($\approx 6\,$mag). This difference might be due to increased sub-mm optical depth
per unit column density near the Galactic plane (see Section \ref{s:avcl})
or diffuse molecular material not be accounted for in the \citet{amores2005} model.
We scaled the extinction values from \citet{amores2005} by a factor of 1.8, so that the saturating \av/ values at high 
distances correspond to the values found by {\it Planck}.
Note that even without this scaling, 
the mean far-background \avst/ values found for the wide photometric field 
are still fairly large, as will be discussed below.

The function $N_{det}(d)$ gives the total number of detectable stars as a function
of distance, as shown in Figure \ref{f:fracs}.
Giants and Super-giants in general are bright enough to be detected at large distances ($>10\,$kpc).
The mean stellar distance of the background stars according to this model is $\sim4.4\,$kpc, with a broad 
distribution peaking at $\sim1.3\,$kpc.
At this mean distance, the \av/ according to the models is $3.2\,$mag. If the extinctions 
from \citet{amores2005} are not scaled to match {\it Planck} (as described above), 
then we find that the \av/ at the mean stellar distance is $\approx2.1\,$mag. The range
of stellar extinctions $2.1 - 3.2\,$mag is consistent with the mean \avst/ for 
the wide photometric field off-cloud stars (black histogram in Figure \ref{f:compavall}, top left).
Integrating the curve of Figure \ref{f:fracs}, we find that the fractions of expected foreground and background 
detected stars are respectively $6\%$ and $94\%$. 
It is also worth pointing out that the fraction of background stars located beyond $2\,$kpc is $\approx59\%$, 
showing that a significant fraction of the detected stellar sample is expected to be located in the far-background.
Thus, \avst/ is probably significantly affected by background material, which is shown 
to be non-negligible in the general direction of Vela C (See Section \ref{s:compavs} and 
Appendix \ref{ap:backevidence}).

\item Beam averaging over cloud ``holes" and clumps: 
The difference between the finite {\it Herschel}
beam and the stellar pencil beam introduces an additional source of spread 
in the \avst/ values (see below). 

\end{enumerate}

%-------------------------------------------------------------
%-------------------------------------------------------------

In order to evaluate the relative importance of these three factors,
consider Figure \ref{f:compavall} (top left) which shows the distribution of \avst/ 
for objects surrounding the cloud ($0 <$ \avcl/(mag) $<2$).
The distribution FWHM is about $6\,$mag, which cannot be explained by 
\avst/ photometric uncertainties, as the $\sim1.5\,$mag statistical errors in \avst/ lead to
a FWHM of $\approx3.5\,$mag. 
In addition, we find it unlikely that dense clumps would be found in this area,
since it represents a more diffuse material around the cloud, and therefore beam 
averaging (see item $3$ above) seems unlikely to play a major role.
We conclude that the dominant factor controlling the spread of \avst/ values 
is the wide range of stellar distances. This conclusion is also supported 
by evidence presented in Section \ref{s:compavs} and Appendix \ref{ap:backevidence}.
For higher \avcl/ values, even if clumping and beam averaging are more significant,
these factors do not appear to impact the \per/ calculation (see Section \ref{s:contamback}).
%We believe that beam averaging of clumps or cloud holes could be responsible for some 
%outliers in the \avst/ distribution, but the overall spread is more consistent 
%with the stars being located at a long range of distances.

%In addition, the line-of-sight averaging through the cloud (which 
%affects both the stellar pencil beam and the {\it Herschel} beam size) is possibly
%more significant than the averaging within the beam itself. 
%Assuming a cloud line-of-sight depth similar to its extension in the plane-of-sky 
%(on the order of $1\degr$), 
%then the cloud width is about $12\,$pc along the line-of-sight, while the 
%{\it Herschel} beam size is about $0.1\,$pc. Cloud substructures present 
%along the line-of-sight will be averaged in the pencil beam within the above 
%mentioned $\approx12\,$pc width estimate.

\section{Additional evidence for the existence of far-background interstellar material}
\label{ap:backevidence}

%-------------------------------------------------------------
\begin{figure}[!t]
   \centering
   \includegraphics[width=0.48\textwidth]{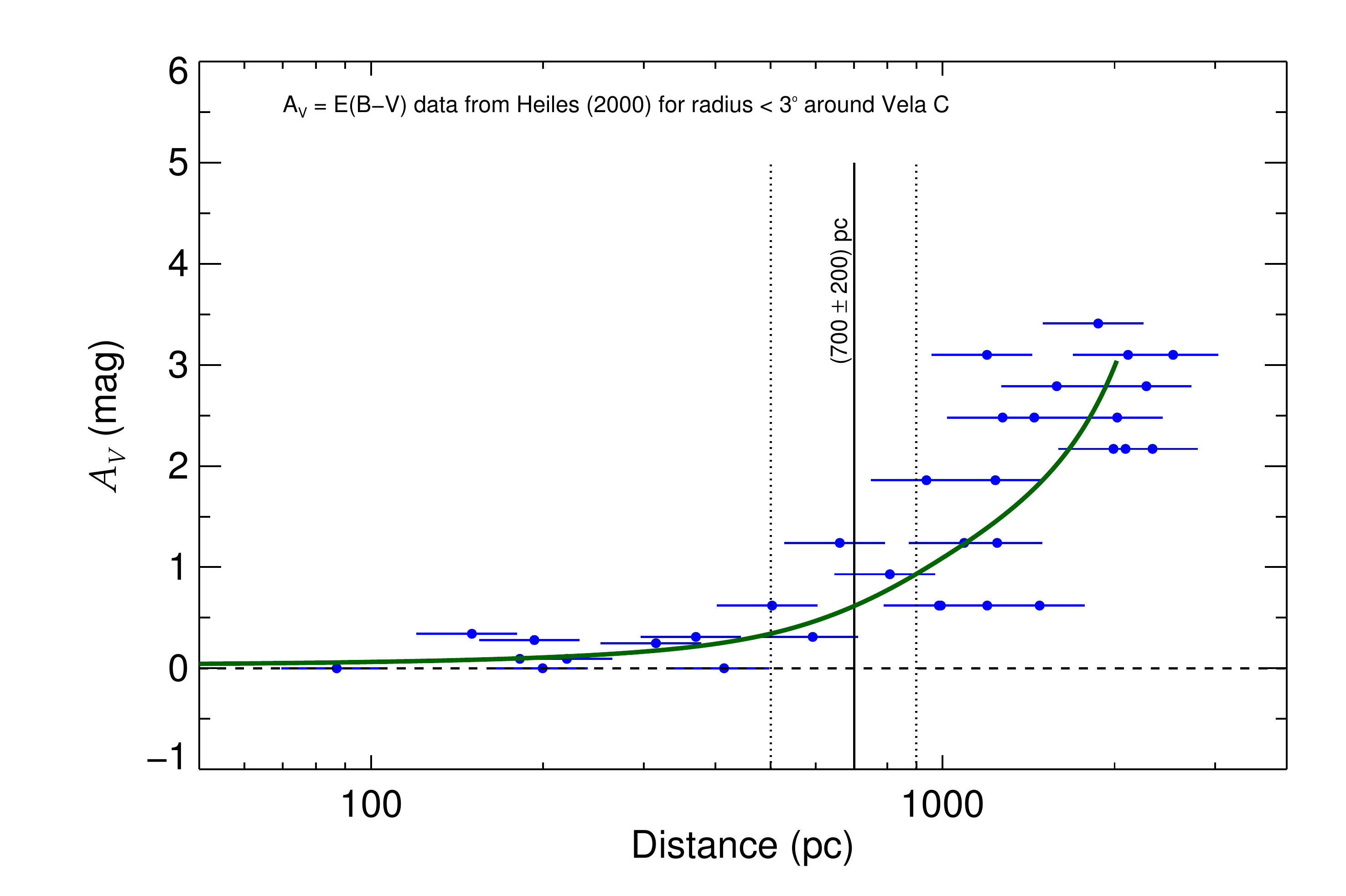} 
   \includegraphics[width=0.48\textwidth]{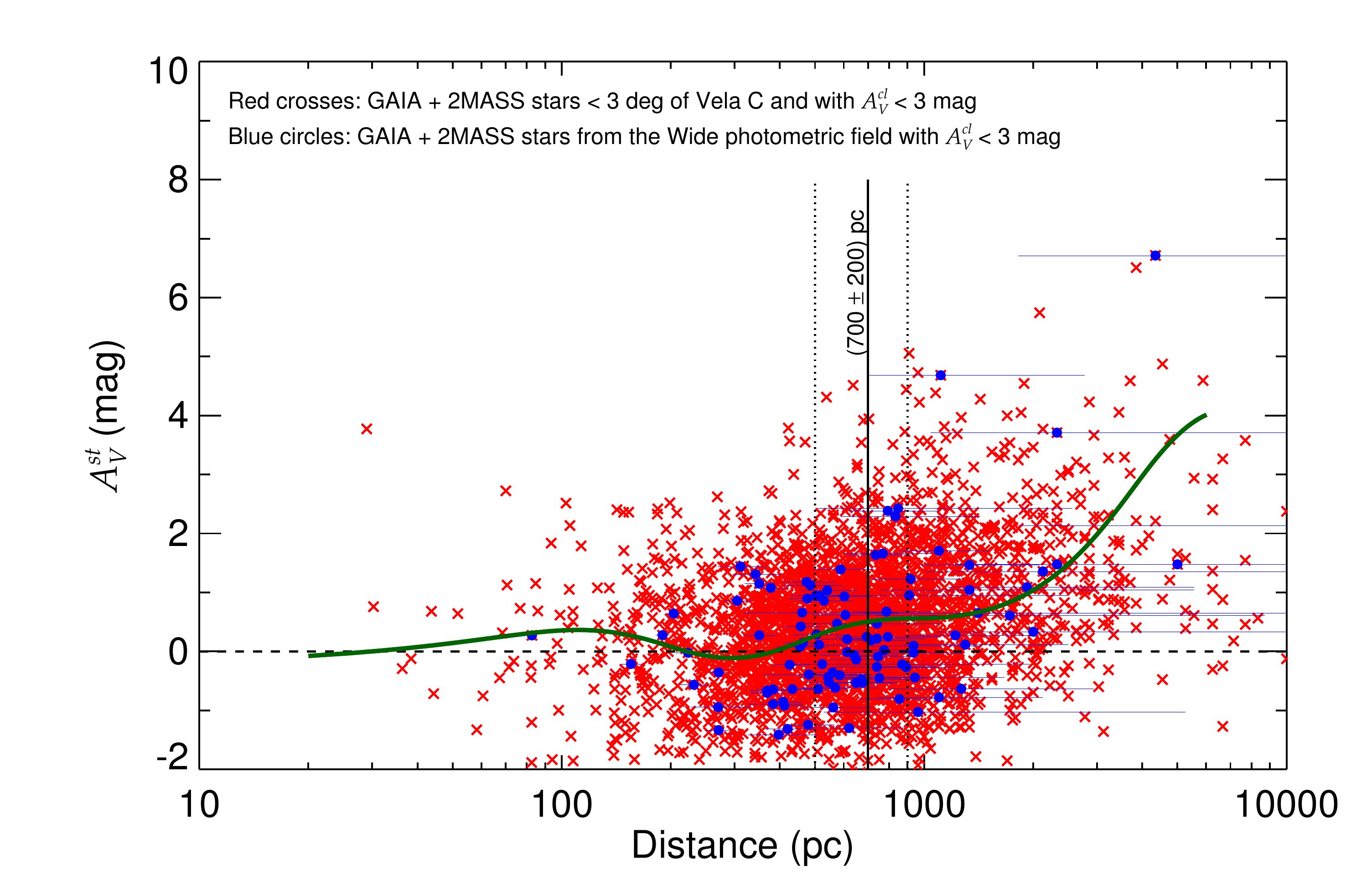} 
   \caption{
           {\it Left:} 
           visual extinction (\av/) as a function of distance
           within $3\degr$ of Vela C from \citet{heiles2000}, 
           probing the diffuse material around the cloud.
           The error bars in distance represent typical uncertainties of 20\% 
           \citep[based on agreement between different catalogs, according to][]{heiles2000}.
           {\it Right:}
           stellar extinction \avst/ as a function of distance 
           within $3\degr$ of Vela C
           for GAIA
           stars combined with 2MASS data (red crosses).
           Blue dots represent the fraction of this sample in the 
           ``wide photometric field" (see Table \ref{tab}).
           Stars in the direction of the Vela C cloud itself (\avcl/ $> 3\,$mag)
           were excluded.
           For clarity, error bars in distance are shown only for the blue dots.
           For both graphs, the solid green curve is a spline fit to the binned averaged data,
           to show the general trend.
              }
         \label{f:backevidence}
   \end{figure}
%-------------------------------------------------------------

In addition to the {\it Planck} data presented in Section \ref{s:compavs}, additional evidence 
can be gathered from the literature supporting the existence of far-background diffuse 
interstellar material. First, from the \citet{heiles2000} compilation of polarimetric
and photometric data, 33 stars are found within a radius of $3\degr$ around Vela C (centered 
on RCW 36).
These objects are spatially located in diffuse 
lines-of-sight surrounding the cloud, so they serve as adequate probes of the diffuse 
material in the disk of the Galaxy
as a function of distance. Figure \ref{f:backevidence} ({\it left}) shows a plot of 
\av/ vs. distance (pc), where \av/ is derived from color excess data $E(B-V)$ (using 
the general relation \av/$=3.1 E(B-V)$).
Visual extinctions increase as a function of distance, a trend that continues for distances 
greater than $1\,$kpc. 
The monotonic increase continues up to the maximum distance of this dataset 
($\approx 2.5\,$kpc), reaching levels around $3\,$mag, which is consistent 
with the center of the broad \avst/ distribution in Figure \ref{f:compavall} ({\it top left}).

Next, we combined trigonometric parallaxes from the GAIA early data release \citep{2016gaia}
with 2MASS, within a 3$\degr$ radius around Vela C. Stellar extinctions 
were calculated according to the method described in Section \ref{s:avst}.
In this area, 2945 GAIA stars with valid parallax detections are found,
after selecting only objects toward more diffuse lines-of-sight (\avcl/$\,<\,3\,$mag). 
Among this total, 102 objects are part of the wide photometric field data set defined 
in Table \ref{tab}.
Figure \ref{f:backevidence} ({\it right}) shows a plot of \avst/ as a function of distance.
Red crosses consist in the 2945 objects within 3$\degr$ of Vela C, while blue dots 
represent the fraction of these points corresponding to the wide photometric field. 
The stellar distribution (including wide photometric field objects) reaches very 
large distances in the far-background, up to approximately $10\,$kpc, consistent with the 
discussion of Appendix \ref{ap:avstuncertainties}.
The GAIA{\texttt+}2MASS combination shows a trend similar to the one found from the Heiles (2000) data:
For distances smaller than the cloud's location ($700\,$pc) the foreground stars 
show a distribution of stellar extinctions close to $0\,$mag, but
a clear increase in \avst/ is found for higher distances.
Notice that for large distances, the distribution reaches values as high as $6\,$mag for some objects.
This reinforces the idea that a significant number of far-background 
stars contaminated by diffuse material behind the cloud are present within our sample.

\section{Determination of the foreground levels of extinction and polarization}
\label{ap:fore}

\citet{franco2012} investigated the extinction levels in the general
direction of Vela and Puppis within the 0 -- 1000~pc distance range, using 
$uvby$H$\beta$ Str\"omgren photometry. Although some ISM features are found in this direction
(such as the edge of the Local Bubble, the Gum Nebula, and the Vela Supernova Remnant), 
the overall color excess levels suggest that the material out to 700pc is 
very diffuse and has very low density. In particular, areas labeled as SA173 and SA171 from 
\citet{franco2012} are located respectively above and below the Galactic plane,
with an angular separation of a few degrees relative to the Vela Molecular Ridge.
The median $E(b-y)$ values out to 700 pc for these areas are 
$0.05$ and $0.03\,$mag, which correspond to  $A_{V}=0.22$ and 
$0.13\,$ mag \citep[assuming $A_{V}=4.3 E(b-y)$,][]{1976crawford}, respectively. 
Similarly, \citet{reis2011} used $uvby$H$\beta$ photometry to
map stellar distances and extinctions in the local ISM. 
For sky positions within $3\degr$ of Vela C, 17 stars distributed out
to $500\,$pc were found from their sample, with a mean $A_{V}$ of $0.10\,$mag.
These estimates show that the foreground ISM in the direction of the VMR 
is typically very diffuse, consistent with ``tunnels" observed in this 
direction from maps of the local ISM \citep{lallement2003,welsh2010,reis2011}. 
Based on the above-mentioned foreground extinction values, we estimate an average foreground 
extinction level of approximately $0.15\pm0.09\,$mag toward the Vela LOS.

%-------------------------------------------------------------
   \begin{figure*}[!t]
   \centering
   \includegraphics[width=\textwidth]{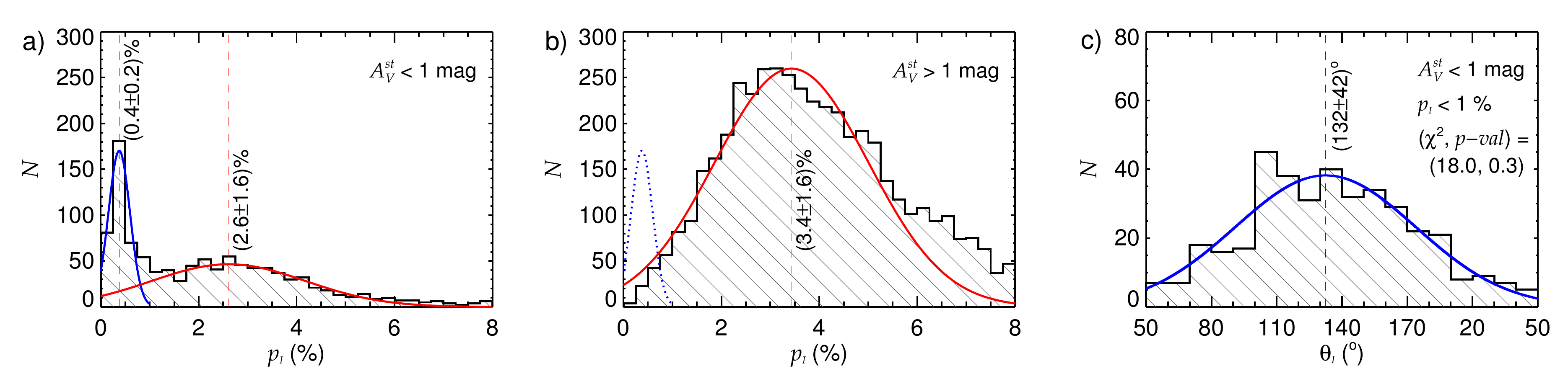} \\
      \caption{Analysis of the foreground polarization component using the $I$-band-2MASS combination 
	       data set (see Table \ref{tab}): panels {\it a} and {\it b},
               respectively, show the distributions of polarization fraction in the $I$-band 
               for ranges of stellar extinctions defined by \avst/ $<1\,$mag and \avst/ $>1\,$mag. 
               Solid red and blue curves are Gaussian fits to the sample. Panel {\it c} shows 
               a histogram of equatorial polarization angles for 
               objects with \avst/ $<1\,$mag and $p_{I} < 1\%$ 
               (corresponding to the peak shown in blue in panel {\it a}).
               The chi-square (and associated p-value) is shown only for the Gaussian 
               fit in panel {\it c} (in panel {\it a} the red Gaussian contaminates the chi-square
               calculation for the blue Gaussian).
               %(therefore mostly those under the blue Gaussian fit of panel {\it a}).
               %The distributions in panels {\it a} and {\it c}
               %contain all objects with low extinctions, including those for which the procedure
               %described in Section \ref{s:avst} would render negative \avst/ values.
               %correct pi to lower case
              }
         \label{f:fore}
   \end{figure*}
%-------------------------------------------------------------

%As an independent check, we used galaxy models 
%by \citet{amores2005}, which takes into account the spiral arms of the Galaxy, 
%to estimate $A_{V}$ in this direction. We find that up to 1 kpc, $A_{V}\approx0.75$ mag, 
%which is approximately consistent with the upper levels found by \citet{franco2012}.
%Based on this analysis, 

In order to estimate the foreground polarization component, we 
first define a conservative upper limit for the foreground stellar extinction of $\approx1\,$mag,
based on the observation that in Figure \ref{f:avavdiag} (top) most of the stars in the ``band" that 
defines the foreground objects are below this limit. 
The results from this analysis remain essentially unaltered if this choice is 
varied within reasonable limits (see below).
Subsequently, we analyse the distribution of $p_{I}$ for objects from the $I$-band-2MASS combination 
data set (see Table \ref{tab}) possessing 
stellar extinction values below this upper limit. 
This is shown in Figure \ref{f:fore}a.
We compare this histogram
with the one for \avst/$ > 1$ mag, shown in Figure \ref{f:fore}b. In the first histogram, we notice 
a peak centered on low polarization values, around $\approx 0.4\%$ 
(blue Gaussian curve). This peaked distribution is obviously absent in the second 
histogram (for reference, it is shown as a dotted blue line). Instead, it shows a broad distribution
centered on much higher polarization values (around $3.4\%$). This indicates that the 
stars within the peak shown in blue are mostly foreground objects. 
%since these polarization levels are clearly distinct for higher visual extinctions.
%The broad distribution of higher polarization values
%is also seen in the first histogram (centered on $2.6\%$), showing that the $A_{V} < 1$ mag 
%criteria ends up selecting a small population of background sources as well.
In addition to the analysis of color excess as a function of distance, \citet{franco2012}
also studied linear polarization in the $B$ band toward this general direction, showing that 
levels between 0 and 0.60\% may be found out to $700$ pc. This corresponds to a range of values between 0 and 
$\approx 0.54\%$ in the $I$-band, \citep[assuming the general spectral relation by][]{serkowski1975}, 
which is consistent with the distribution of values seen within the peak shown in blue in Figure \ref{f:fore}a.

In order to find the mean foreground polarization orientation, we use the histogram of position angles
in Figure \ref{f:fore}c, which includes only stars with \avst/$ < 1$ mag and $p_{I} < 1\%$ 
(these criteria are used to select only the stars within the peak shown in blue in  
Figure \ref{f:fore}a). We find a broad distribution of polarization angles peaked at 
$\theta_{I} = 132\degr$. Although a large spread is expected for such low polarization levels 
(the typical $p_{I}$ signal-to-noise for this particular sample is just above the threshold of $3$, so the angle uncertainties
are $\approx10\degr$), this distribution suggests that the intervening diffuse ISM 
features located in the foreground might have a wide range of magnetic field orientations.
However, the peak of the distribution is a reasonable estimate since it represents the most common
orientation found in this distance range.
We therefore adopt $p_{I}=0.4\%$ and \thetai/$=132\degr$ as the foreground polarization 
fraction and angle toward Vela C. 

These results are robust 
with respect to the choice of the \avst/ upper limit. If instead of $1\,$mag, levels
of $0.5$ or $2\,$mag are chosen, the estimated $p_{I}$ and \thetai/ from the foreground
remain fixed, although the spreads in the Gaussian distributions from which they are derived 
vary slightly.

\section{Detailed description of the Gaussian-logistic method and its sources of systematic uncertainties}
\label{ap:gl}

\begin{figure*}
   \centering
   \includegraphics[width=0.300\textwidth]{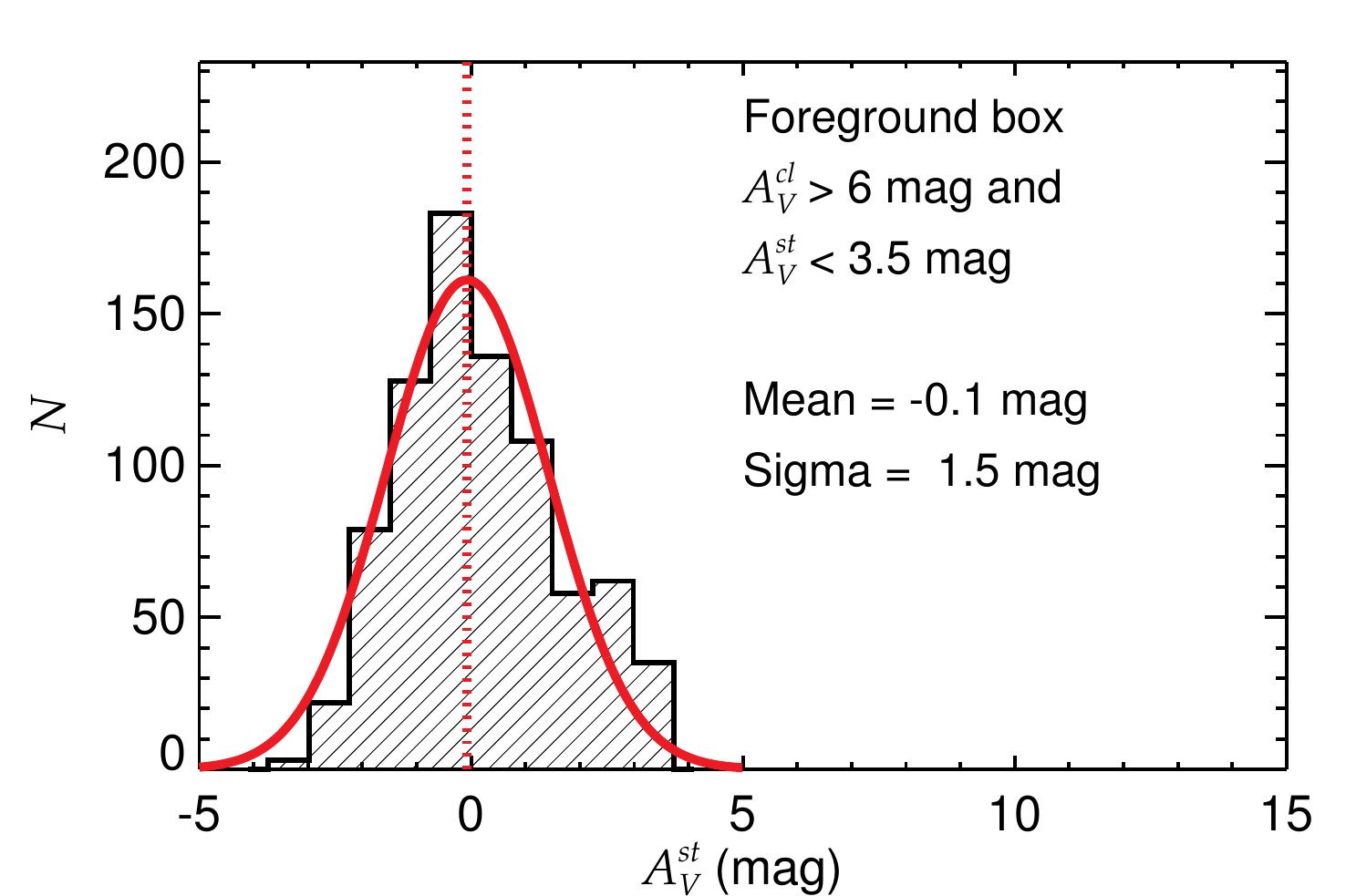}
   \includegraphics[width=0.300\textwidth]{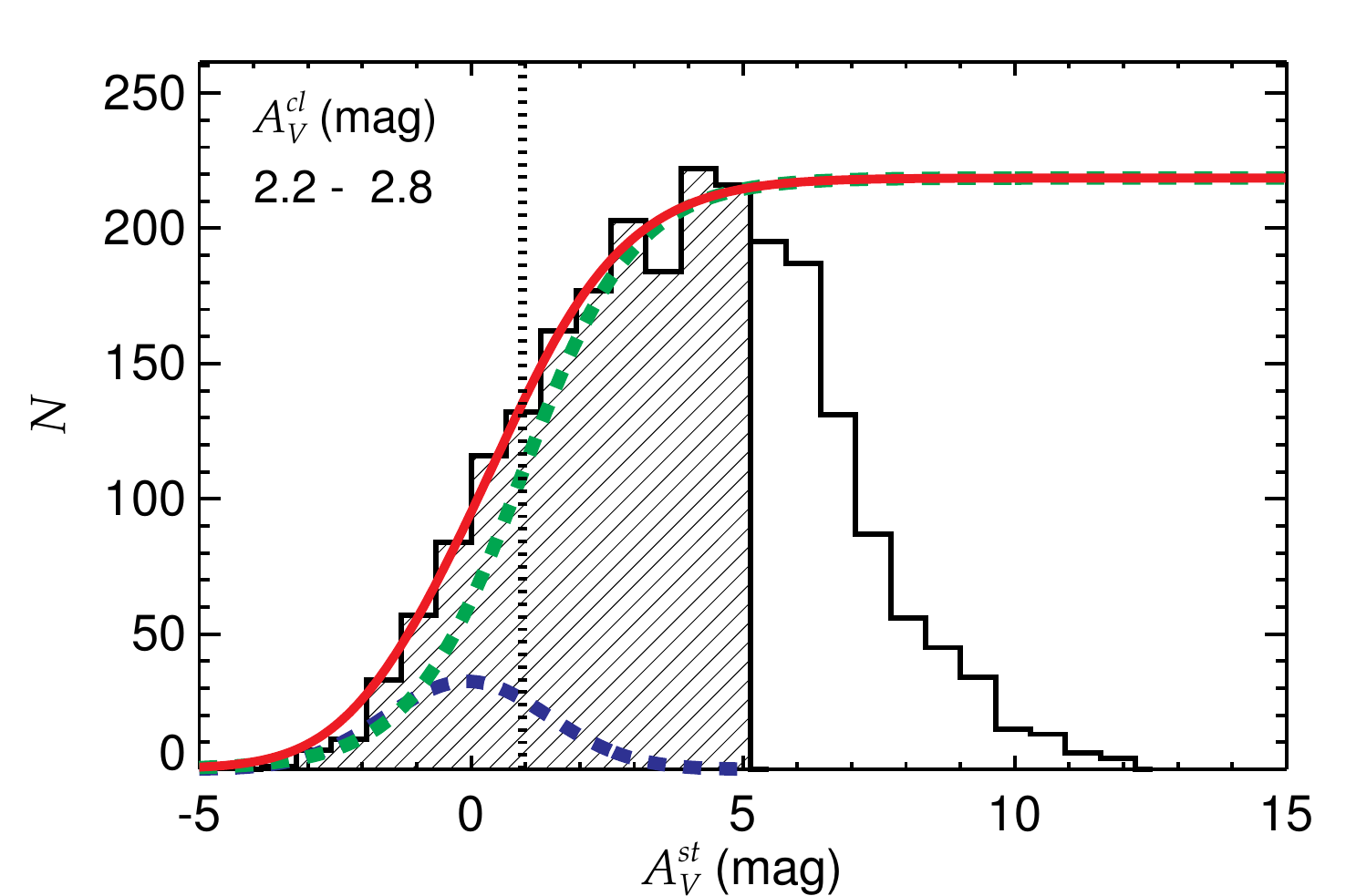} 
   \includegraphics[width=0.300\textwidth]{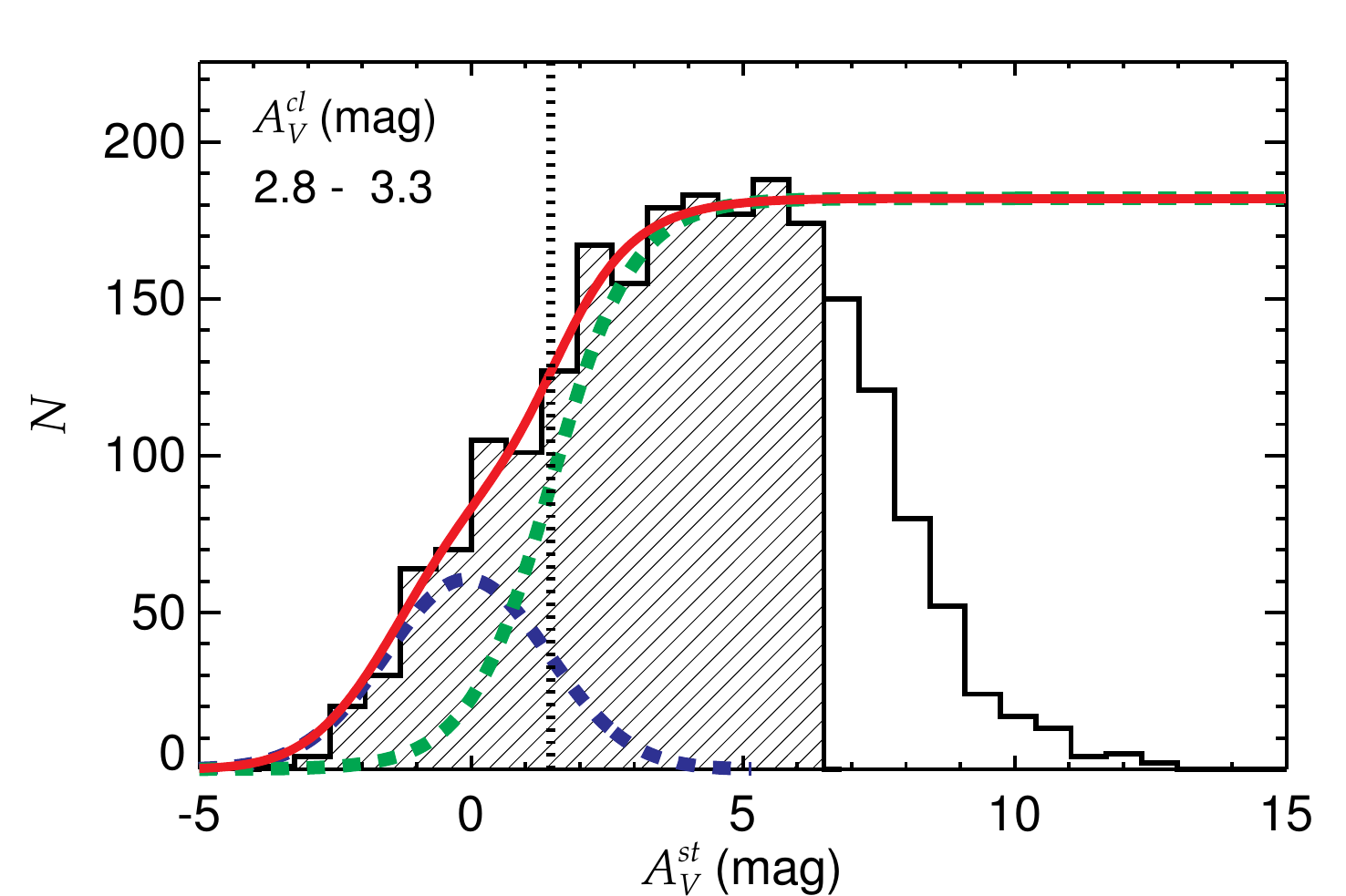} \\ 
   \includegraphics[width=0.300\textwidth]{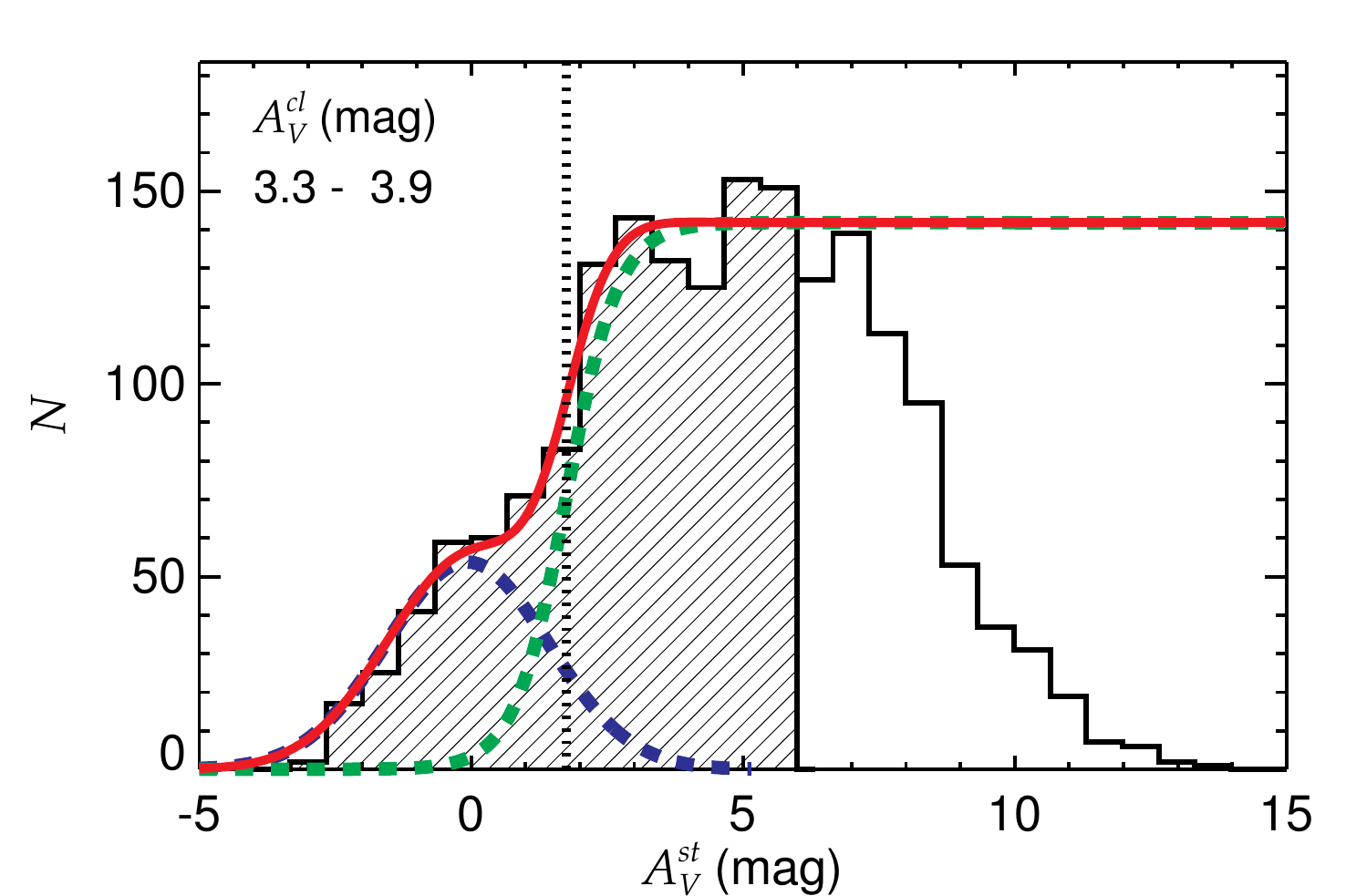} 
   \includegraphics[width=0.300\textwidth]{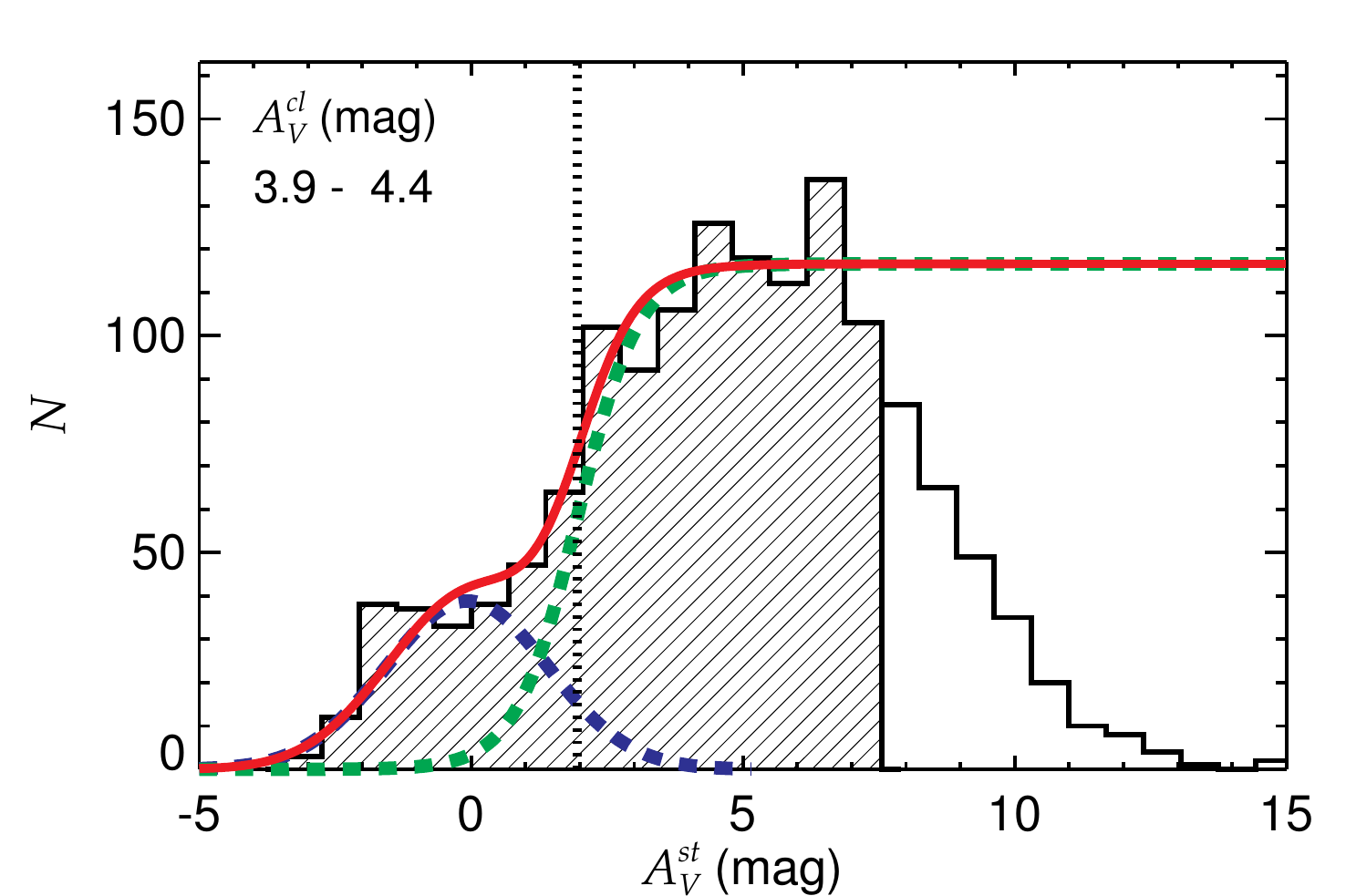} 
   \includegraphics[width=0.300\textwidth]{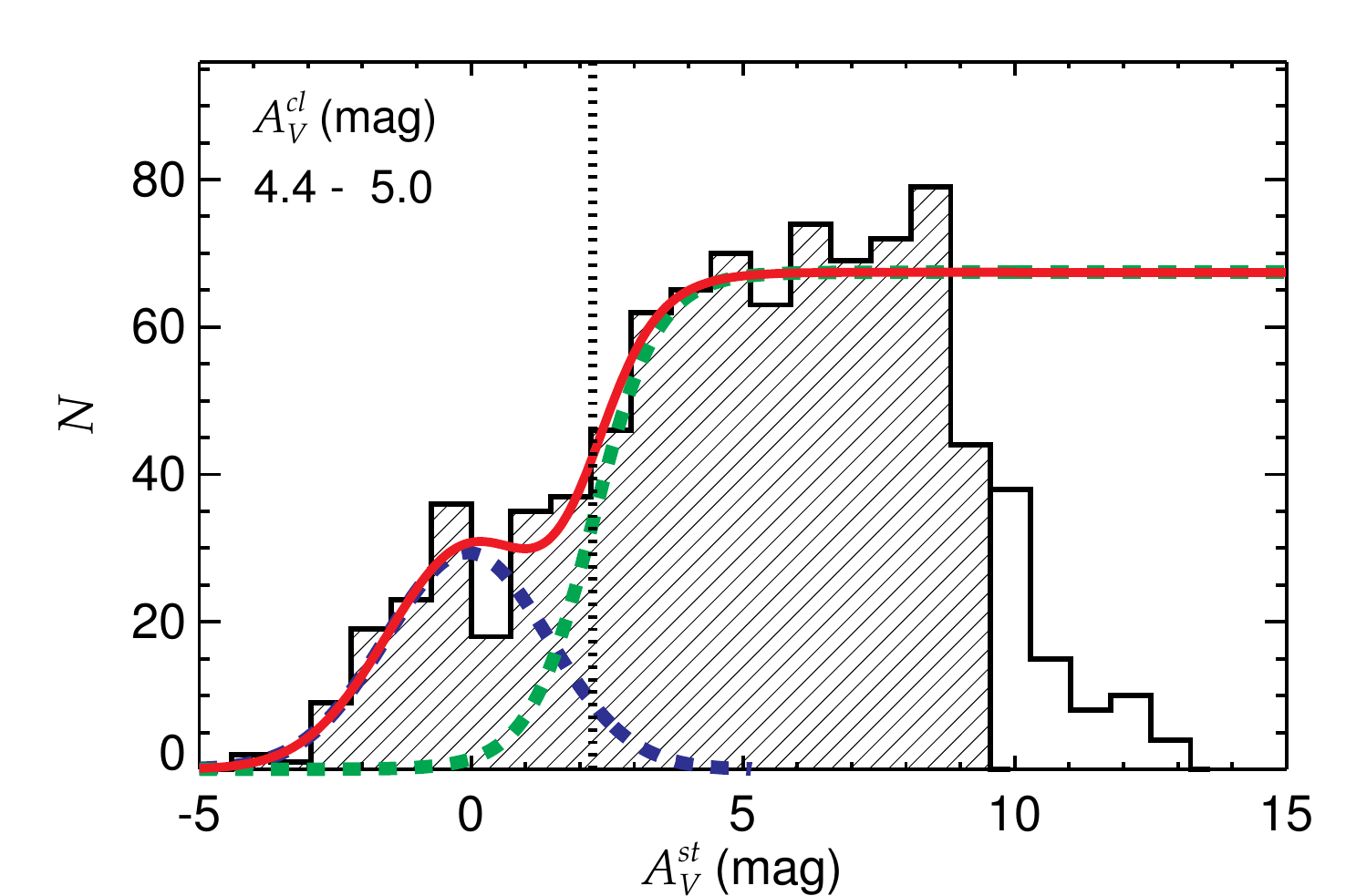} \\
   \includegraphics[width=0.300\textwidth]{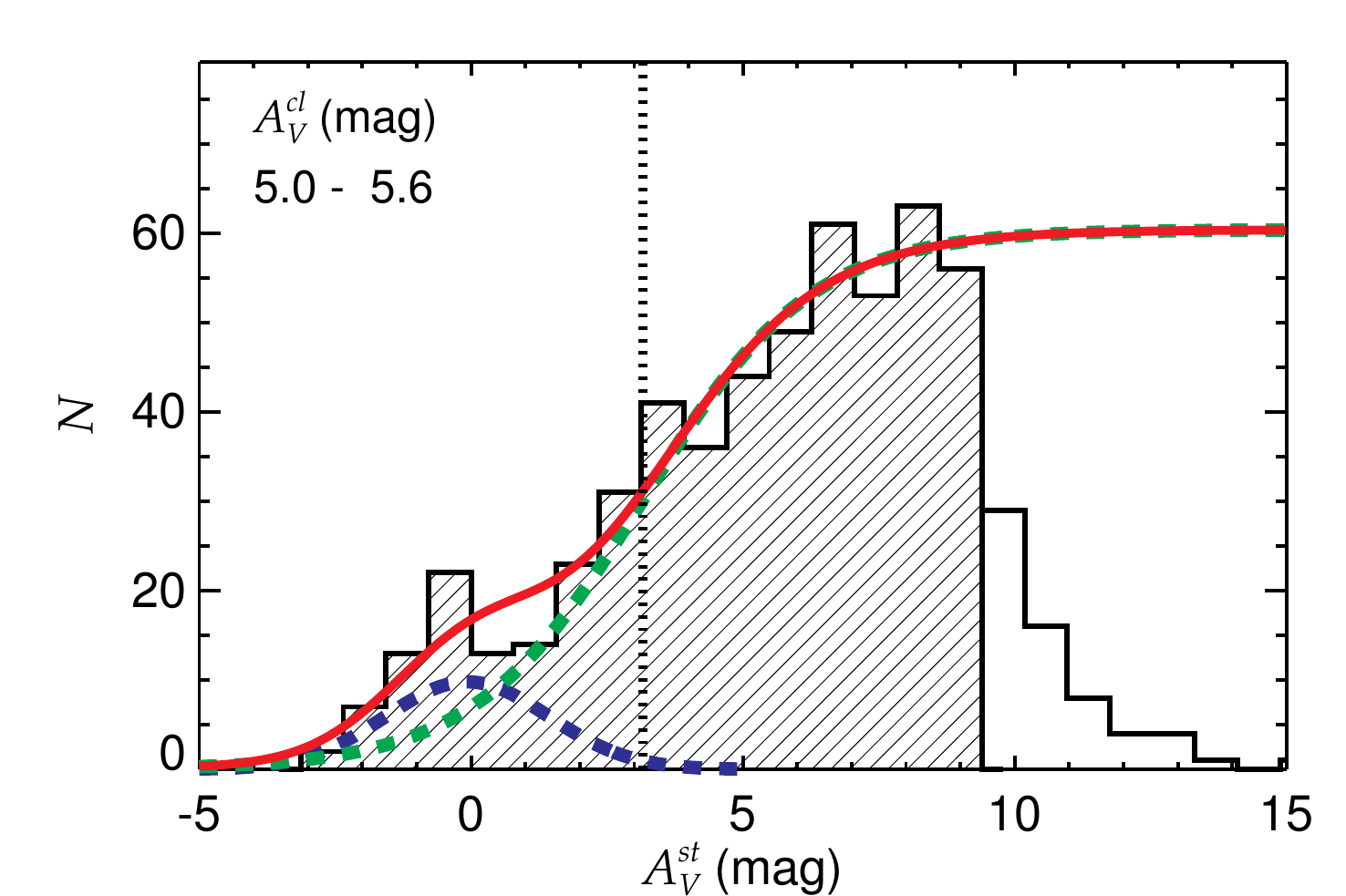} 
   \includegraphics[width=0.300\textwidth]{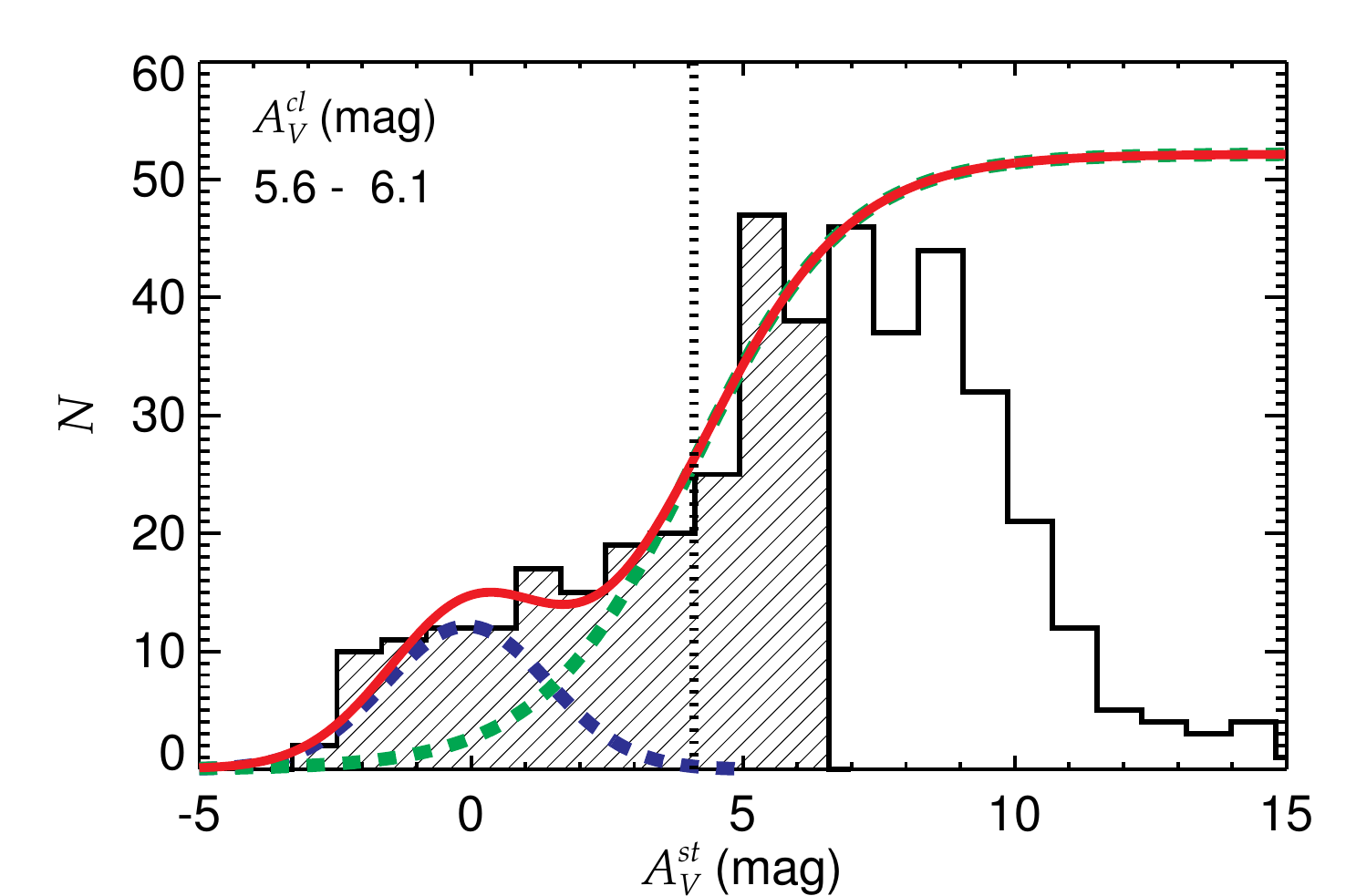} 
   \includegraphics[width=0.300\textwidth]{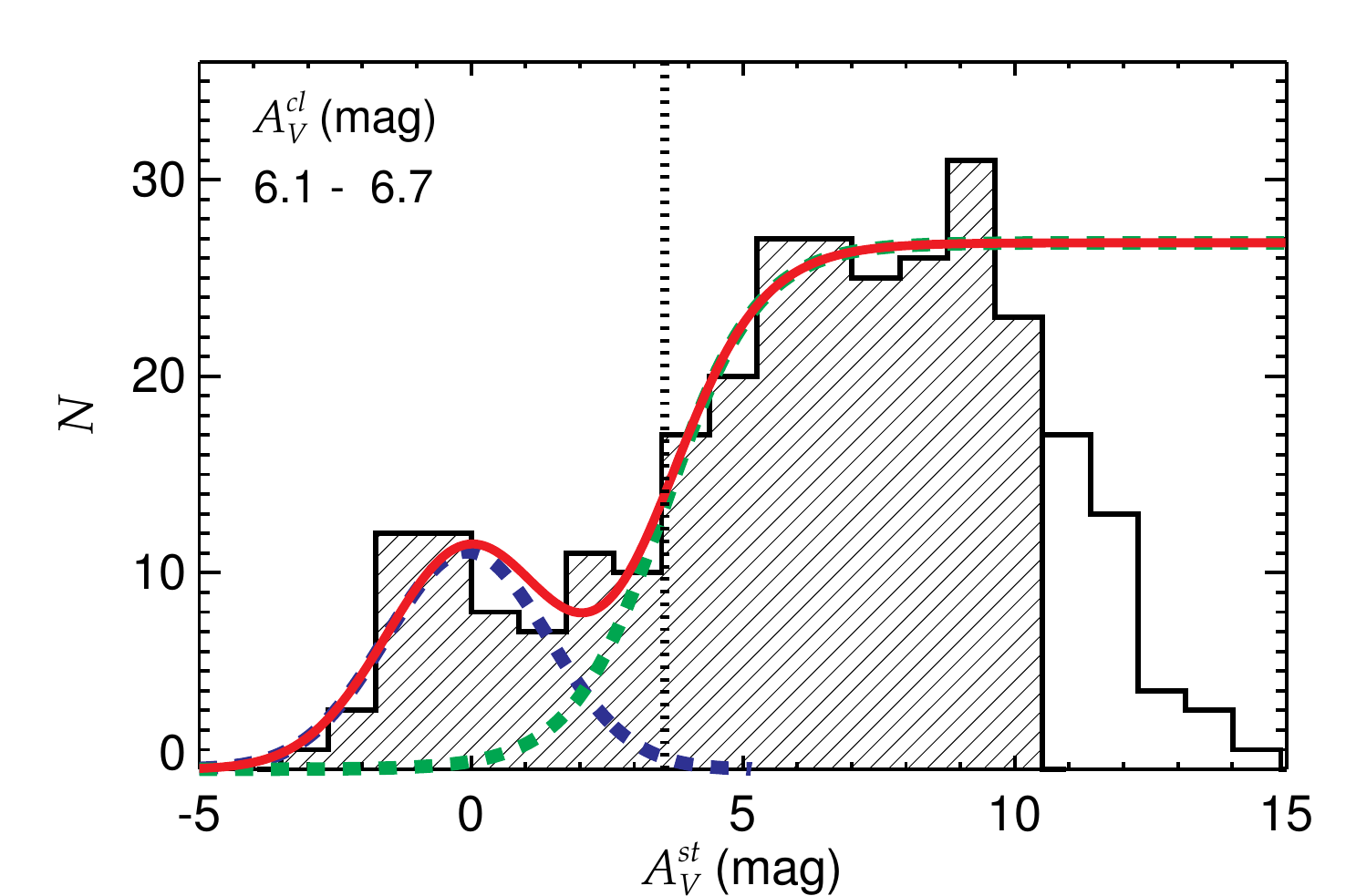} \\
   \includegraphics[width=0.300\textwidth]{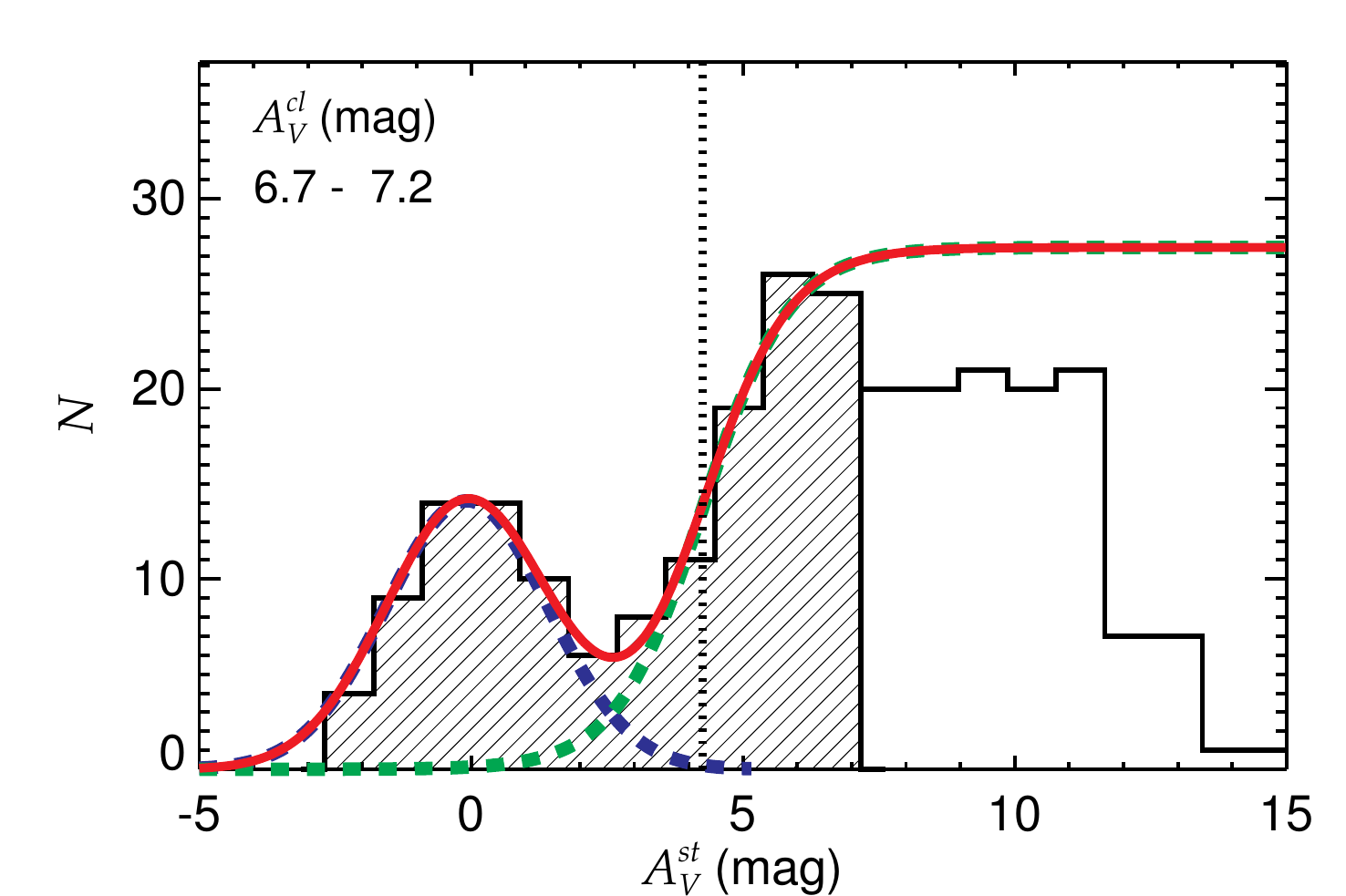} 
   \includegraphics[width=0.300\textwidth]{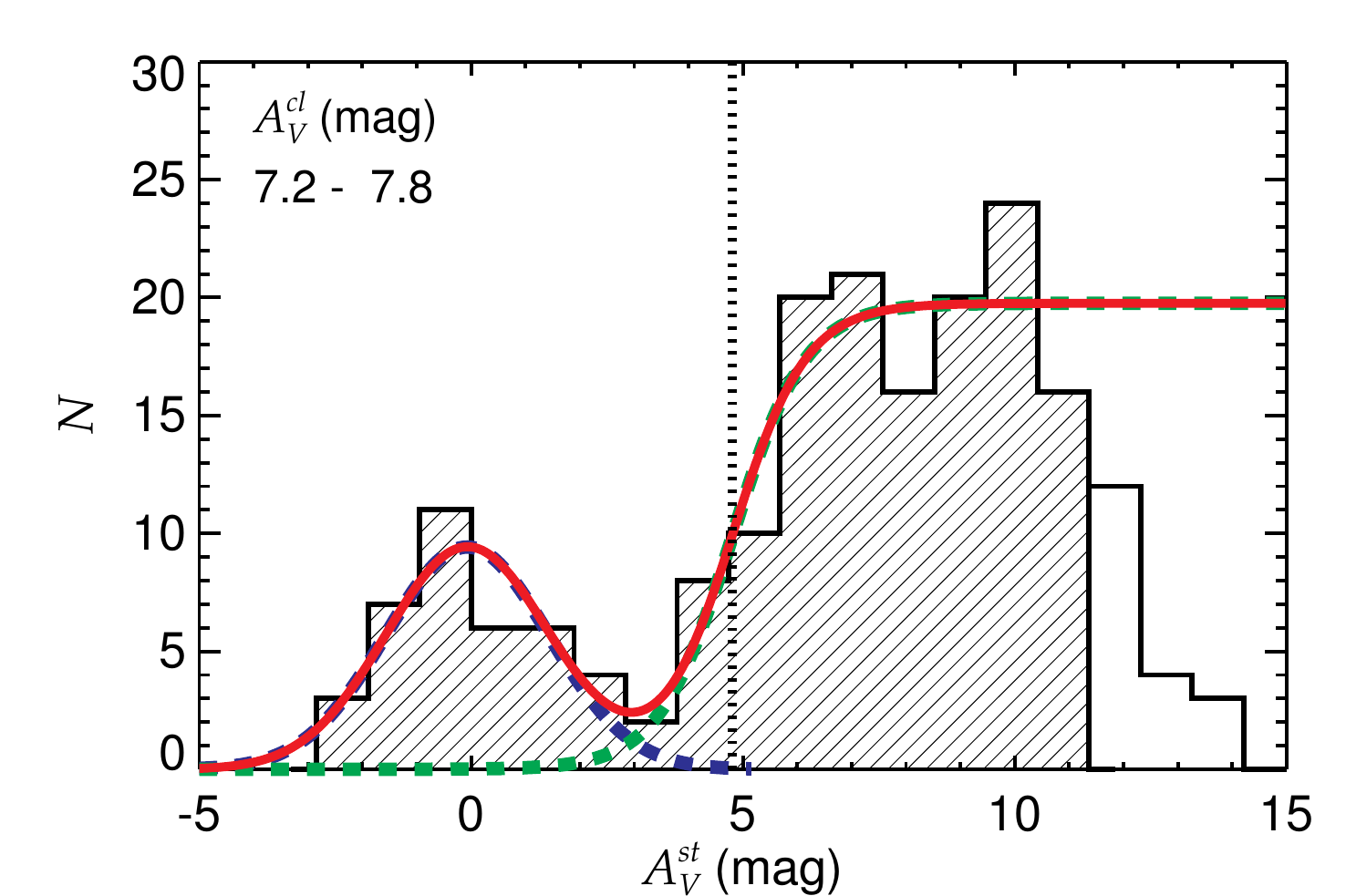} 
   \includegraphics[width=0.300\textwidth]{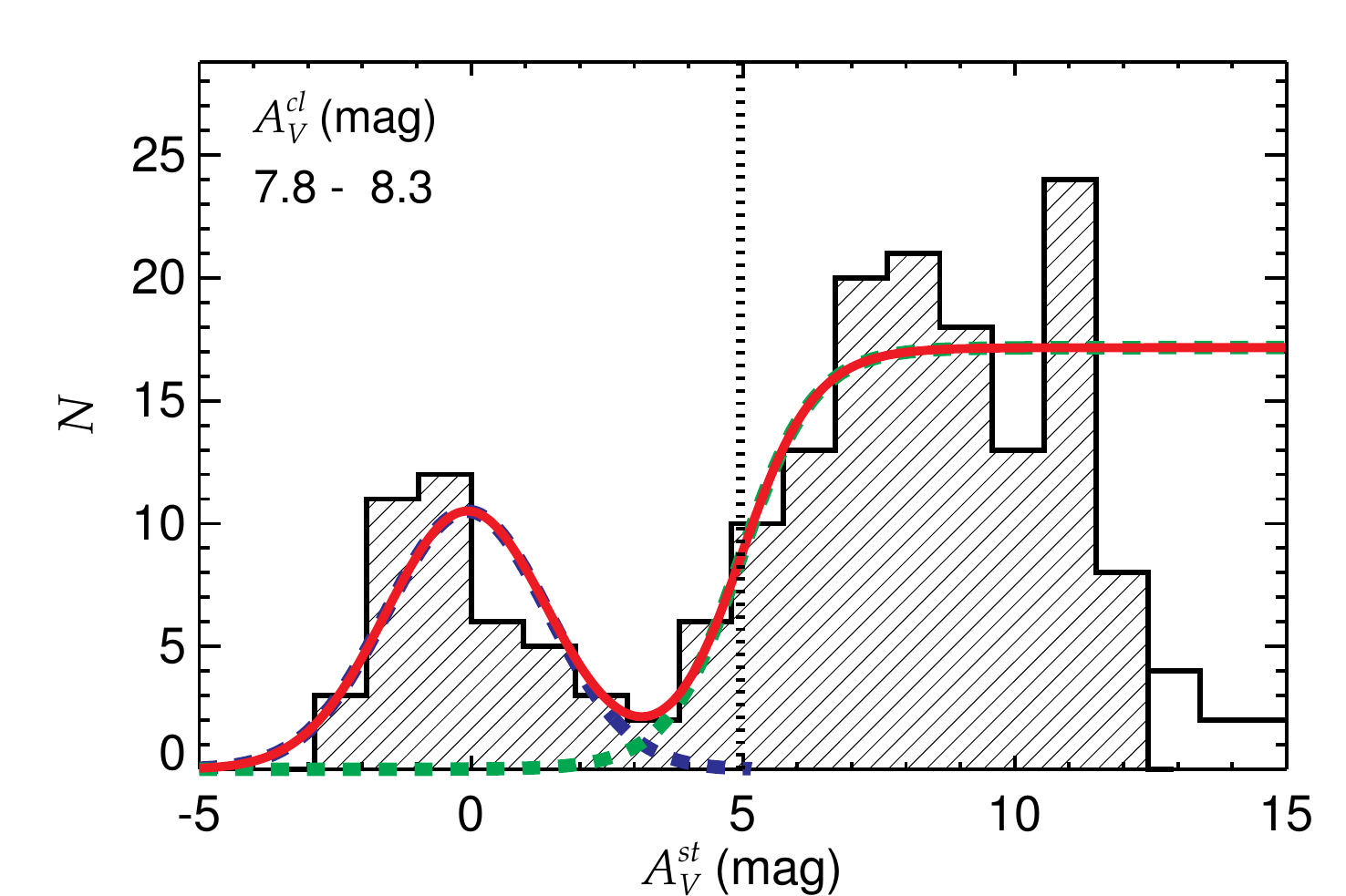} \\
   \includegraphics[width=0.300\textwidth]{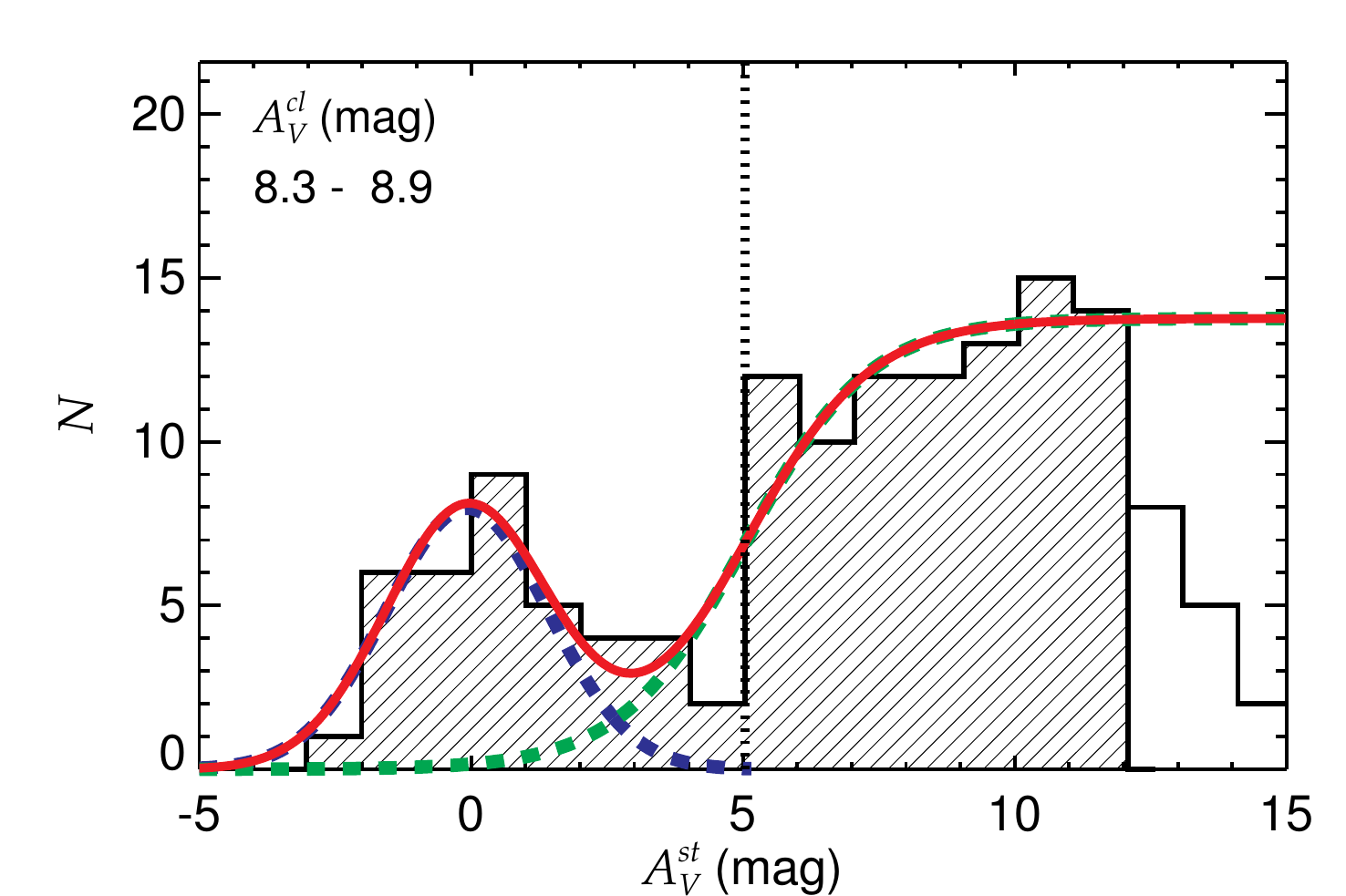} 
   \includegraphics[width=0.300\textwidth]{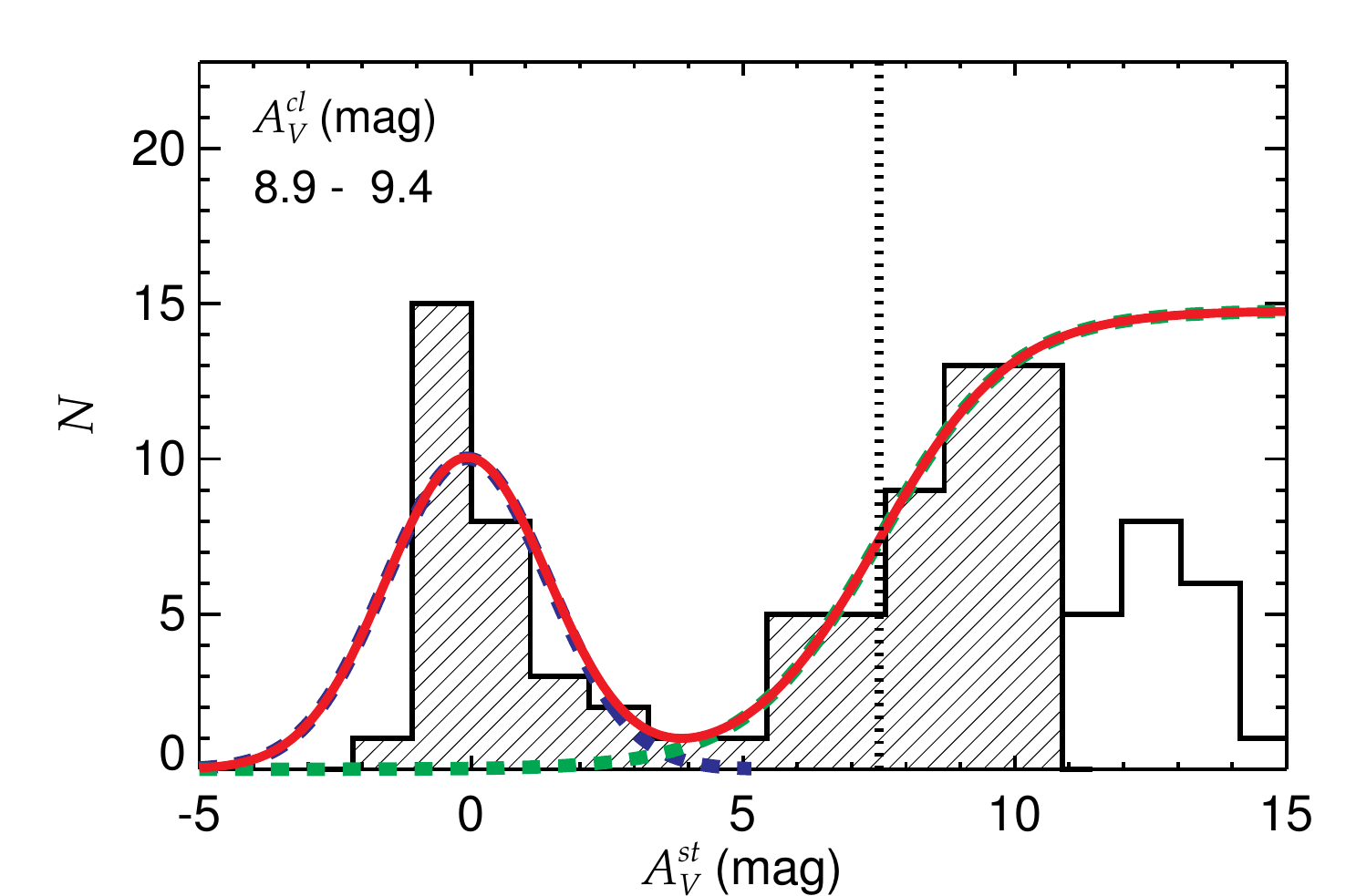} 
   \includegraphics[width=0.300\textwidth]{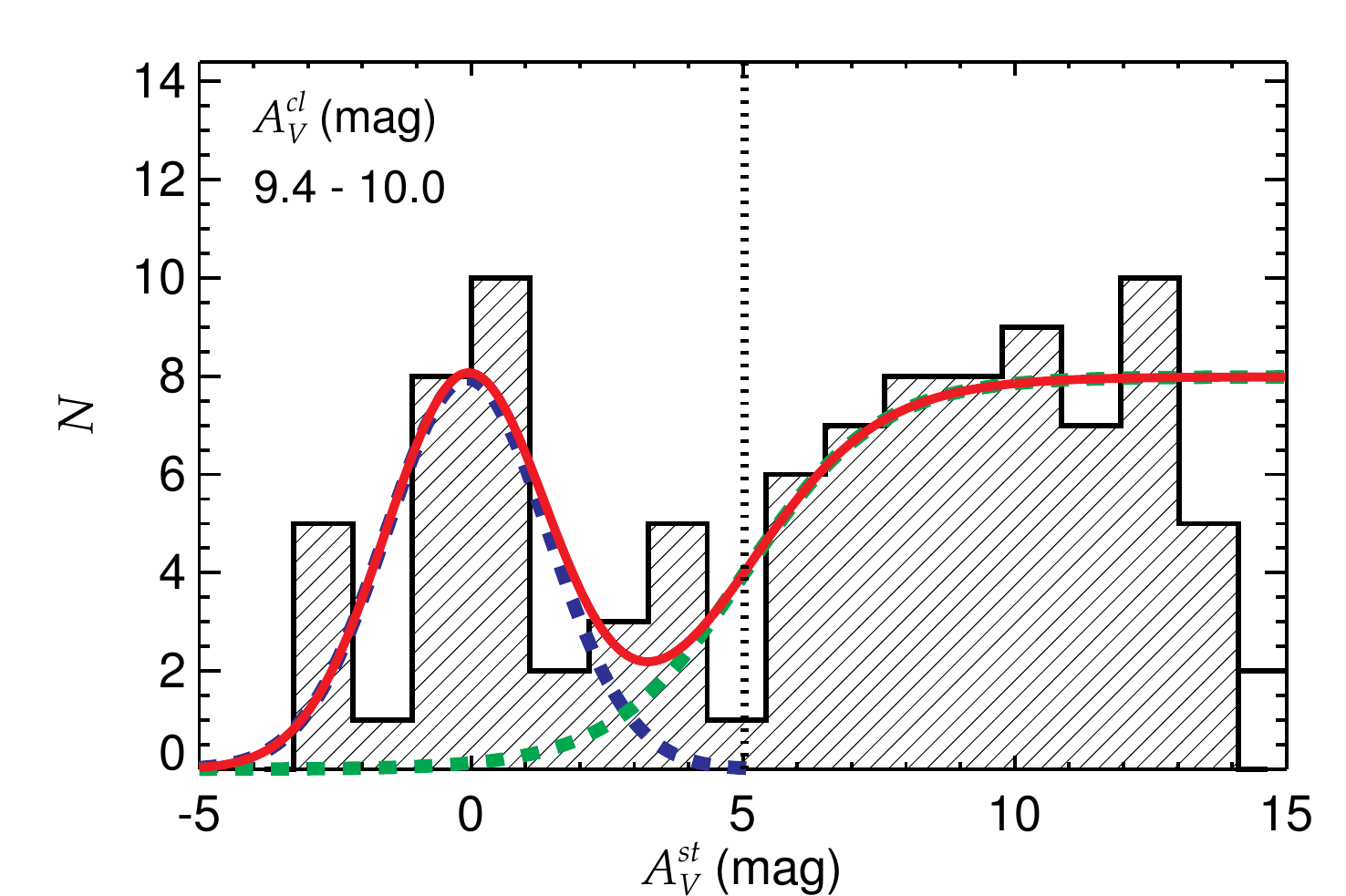} \\
      \caption{Distributions of stellar extinctions (\avst/) for different cloud extinction (\avcl/) bins, 
               using objects from the wide photometric field (see Table \ref{tab}), 
	       illustrating the GL procedure. The first histogram
               at the top left represents the distribution of foreground stars 
               located inside the gray box defined in Figure \ref{f:avavdiag2}, from which the 
               Gaussian parameters $\beta$ and $\sigma$ are obtained and used as inputs for the GL fits.
               All the other distributions are defined within narrow \avcl/ bins, 
               and used to fit the GL function (the red curve, with the Gaussian and logistic 
               components separately defined by the blue and green curves). The shaded areas 
               of each histogram represent the points effectively used in the fits (the drop
               in the number of stars at higher extinctions is ignored). The vertical dotted 
               black lines are the mid-points of the logistic functions fit in each case, used 
               to define the position of the yellow circles in Figure \ref{f:avavdiag2}.
               For this particular run of the GL method we use $N_{\mathrm{d}}=18$, 
	       $R_{\mathrm{bin}}=5\,$mag, and $N_{\mathrm{max}}=1$, which corresponds to the standard example 
	       (see description in the text).
              }
         \label{f:hist}
   \end{figure*}

As described in Section \ref{s:gl}, the first term of Equation \ref{e:gl} corresponds to the foreground stellar population, 
previously identified as the ``band" of points roughly parallel to the \avst/$=0$ line
in Figure \ref{f:avavdiag2}. For a given \avcl/ bin, parameters 
$\alpha$, $\beta$, and $\sigma$ (the height, displacement and width, respectively) define a Gaussian curve. 
The displacement and width of the 
Gaussian distribution should be independent of \avcl/, since the extinction
of foreground objects is not affected by the cloud. Therefore, the first step before applying 
GL fits is to obtain single values of $\beta$ and $\sigma$ to be used for all \avcl/ bins.
We define a population of foreground objects in Figure \ref{f:avavdiag2}
as the objects inside the gray box (delimited by \avcl/$>6\,$mag and \avst/$<3.5\,$mag), and show the 
\avst/ distribution for the corresponding stars as the first histogram in Figure \ref{f:hist} (top left). 
A simple Gaussian fit to this distribution gives $\beta=-0.1\,$mag and 
$\sigma=1.5\,$mag. 
%These two parameters can therefore be held fixed in the GL fits. 
For each \avcl/ bin, the number of foreground stars obviously changes significantly,
and therefore the only Gaussian parameter allowed vary in the following 
analysis is $\alpha$ (the height of the Gaussian curve).
It is interesting to note that 
the $1.5\,$mag spread in the foreground stellar extinction provides a good estimate 
of the total uncertainty in \avst/ (which includes both statistical and systematic
errors, see Section \ref{s:foreground}).

%-------------------------------------------------------------
%-------------------------------------------------------------

The next step is to build histograms of \avst/ for different 
\avcl/ bins; one can imagine dividing the diagram in Figure \ref{f:avavdiag2}
into a series of vertical bands with fixed $\Delta$\avcl/ widths, and  
a \avst/ histogram is drawn for each of these vertical bands. 
For \avcl/$>10\,$mag the number of points is usually insufficient to apply GL fits 
so the GL method is not applied to these points.
The histograms are shown in
Figure \ref{f:hist} (all panels except for the first one). 
As \avcl/ increases, it is easy to distinguish the foreground (Gaussian-like) population at low 
extinction, followed by a ``gap", and finally a steep rise defining the ideal stars, 
as previously depicted in the schematic of Figure \ref{f:scheme}. The GL function 
(Equation \ref{e:gl}, shown by the red curves) is then fit to each of these distributions; 
as described above, while $\beta$ and $\sigma$ are held constant, parameters $\alpha$ (the height of the Gaussian function), 
$a$ (the height of the logistic function), $b$ (the steepness of the curve) and $A_{V}^\mathrm{0}$ (the mid-point
of the logistic function) are allowed to vary independently for each distribution.
For cloud extinctions between \avcl/$=0$ and $2.2\,$mag the distributions lead to bad fits and therefore
are not shown. For such low \avcl/ the foreground and ideal-star population are merged.
The quantity $A_{V}^\mathrm{0}$ is shown as a vertical dotted line in each 
histogram of Figure \ref{f:hist}. 

As mentioned in Section \ref{s:reff1}, the results from the GL method depend on certain 
choices of parameters because these affect the selection of ideal stars.
The goal here is to identify the sources of systematics and vary them within reasonable values, 
repeating the entire GL analysis in each case.
There are three parameters that can significantly affect the 
fits of the GL function to the histograms of Figure \ref{f:hist}.

\begin{enumerate}
\item The number of \avst/ distributions ($N_\mathrm{d}$) between \avcl/$=0$ and $10\,$mag,
which naturally affects the \avcl/ bin widths for each histogram of Figure \ref{f:hist}.
For larger $N_\mathrm{d}$, the number of points available for the GL fits inside each histogram 
decreases. For the particular case shown in Figure \ref{f:avavdiag2}, we chose $N_\mathrm{d} = 18$ (and therefore 
$\Delta$\avcl/$\approx0.56\,$mag for each distribution).  
To account for systematics, values of $N_\mathrm{d}$  $= 18$, $14$, and $10$
were used.

\item The bin sizes $\Delta$\avst/ for each distribution, which are defined according to 
the following relation: $\Delta$\avst/=$R_{\mathrm{bin}}/(\log{N_{\mathrm{hist}}})$, where
$N_{\mathrm{hist}}$ is the total number of elements in a given histogram and 
$R_{\mathrm{bin}}$ is a proportionality factor that may be varied. This allows the bin sizes
to decrease or increase if the number of elements is, respectively, higher or lower. 
In Figure \ref{f:avavdiag2}, we used $R_{\mathrm{bin}}=5\,$mag. Here, values of
$R_{\mathrm{bin}}$  $=3$, $5$, and $7\,$mag are used.
 
\item The maximum \avst/ to truncate the distributions in order to apply the 
GL fits. In each histogram of Figure \ref{f:hist}, only the shaded area is used in the 
GL fits, because the drop in the number of stars for higher \avst/ is not accounted for
in Equation \ref{e:gl} (the precise position where this drop occurs is not important for 
our purposes). Therefore, a maximum \avst/ needs to be chosen. 
In Figure \ref{f:avavdiag2}, we set this maximum limit to be the position of the first 
bin after the highest peak of the histogram ($N_{\mathrm{max}}=1$). 
Here, values of $N_{\mathrm{max}}$ $=0$, $1$, and $2$ are used. 
\end{enumerate}

As described above, the analysis shown in Figures \ref{f:avavdiag2}, \ref{f:reff1} (left), 
and \ref{f:hist} corresponds to the ``standard example", in which we used the intermediate
diffuse emission subtraction method along with $N_\mathrm{d} = 18$, $R_{\mathrm{bin}}=5\,$mag,
and $N_{\mathrm{max}}=1$. 
Variations of these three quantities (allowing 27 different combinations) result in 
slight changes in the determinations 
of $A_{V}^\mathrm{0}$ for each histogram. When considered as a whole, these variations also change 
the linear fit shown in Figure \ref{f:avavdiag2}, which therefore impacts
the determination of the ideal stellar locus. 
%(the blue straight linear fit defined a range 
%of possible inclinations between 0.6 and 1.0). 
In addition, the entire process is repeated separately using BLASTPol data sets with 
aggressive, conservative and intermediate diffuse emission subtraction, 
leading to results given in Figure \ref{f:reff1} ({\it right}).

Another parameter that should be mentioned is the width chosen for the strips of Figure 
\ref{f:avavdiag2}. We set the width of the strips to $1.5\,$mag, based on the estimated \avst/ uncertainties.  
Varying this width effectively increases or reduces the number of points inside each strip. 
We verified that such variation causes only minor changes 
in our final results. Because these changes are smaller than those caused by varying the three
parameters discussed above, changes in strip width are not included in the formal 
systematic error analysis described here.

\end{document}